\documentclass[preprint, aps,showpacs]{revtex4}
\usepackage{graphicx}
\usepackage{psfrag}
\usepackage{color}
\usepackage{amssymb}
\usepackage{amsmath}
\usepackage{epstopdf}

\usepackage{multirow}
\usepackage{booktabs}

\newcommand{\be}{\begin{eqnarray}}
\newcommand{\ee}{\end{eqnarray}}

\newcommand{\bfq}{{\bf q}_{\perp}}

\newcommand{\bfp}{{\bf p}_{\perp}}

\begin{document}
\title{A light front quark-diquark model for the  nucleons}
\author{ Tanmay Maji}
\author{Dipankar Chakrabarti}
\affiliation{ Department of Physics, Indian Institute of Technology Kanpur, Kanpur 208016, India}

\date{\today}

\begin{abstract}

We present a quark-diquark model for the nucleons   where the light front wave functions are constructed from the soft-wall AdS/QCD prediction. The model is consistent with quark counting rule and Drell-Yan-West  relation.  The scale  evolution of unpolarized PDF  of proton is simulated by making the parameters in the PDF   scale dependent. The  evolution of the DPFs are reproduced for a wide range of evolution scale. Helicity and transversity distributions  for the  proton predicted in this model agree with phenomenological fits. The axial and tensor charges are also shown to agree with the experimental data. The model can be used to evaluate distributions like GPDS, TMDs etc. and their scale evolutions.  

\end{abstract}

\maketitle

\section{Introduction}

 In recent years, there have been a lot of activities to investigate the three dimensional structure of proton.  Different model investigations gave many interesting insights into the nucleon structure and inter-relations among different distribution functions like TMDs, GPDs, Wigner distribution etc and their properties.  There are many model calculations for integrated PDFs but for TMDs and Wigner distributions, only few such model calculations are available.  Since different experiments produce data at different energy scales, predictions of the scale evolutions of these different distributions are also  very important. In this paper, we like to build a simple but phenomenological quark-diquark model for the nucleons which can be used to calculate all these distributions and their inter-relations and at the same time we can also evaluate the scale evolutions of different distributions.
 
 The quark-diquark  model  describes a nucleon as a composite of a diquark spectator with a definite mass  and an active quark.  The model assumes the factorization of short(hard) and long distance(soft) dynamics  in the high energy scattering, and assumes that the lepton basically scatters off a single quark in a nucleon, the other two quarks can be treated as a composite diquark spectator. The diquarks are the effective degrees of freedom and the nonperturbative gluon exchanges between the two spectator quarks are taken into  account by considering an invariant mass of the diquark. 
 This simple model of nucleons is very successful in describing many interesting phenomena. There are many many different variations or parameterizations of the quark-diquark model in the literature\cite{kss,jkss,Jakob97,Bacc08}.   Here we want to construct a quark-diquark model for proton with light front wave functions which
  should not just include the valence structure but also include some nonperturbative ingredients into it. Light-front AdS/QCD  provides one such choice.  Light front AdS/QCD predicts a general form of two particle bound state wave function which can not be derived from just valence quarks \cite{BT}. One needs infinite number of Fock states to have that wave function and thus includes nonperturbative informations in it.
  Recently, the light front wave functions for the nucleons in  quark- scalar diquark models\cite{valery0,valery1} have been constructed from the light-front AdS/QCD prediction.  These  models have been applied to evaluate many interesting properties of proton , e.g, GPDs, Wigner distributions, TMDs etc\cite{CM,valery2}.  Recently, some interesting relations among the GPDs and TMDs  have been investigated\cite{MMCT} in the scalar diquark model\cite{valery0} with the light front wave functions modeled  from soft-wall AdS/QCD wave function. Though the models successfully describe many nucleon properties,   they include only the scalar diquark state.
   In the quark model, the nucleons consists of three quarks of two different flavors $u$ and $d$ ($p=|uud\rangle, ~n=|udd\rangle$).  In the quark-diquark picture, we can schematically write,  for example the proton state, $p=|u (ud)\rangle +|d (uu)\rangle$, where $(ud)$ and $(uu)$ are the diquark states.    With spin-flavor symmetry,
  the diquark   can be  either scalar or axial vector, and hence  both of them are required to build a model. Scalar diquark alone cannot give the complete picture of a nucleon.

In this work, we construct a quark-diquark model with $SU(4)$ spin-flavor structure  for  the nucleons with the light front wave functions modeled from AdS/QCD prediction  including both the scalar and axial vector  diquarks.  
  The model is consistent with the quark counting rule and Drell-Yan-West relation. The parameters  are fitted to proton form factors and unpolarized PDF data  at the initial scale $\mu_0$. We consider the leading order QCD evolution of the unpolarized  PDF for the proton and set the initial scale to $\mu_0=0.313 $ GeV \cite{Bron08,valery0}.  The scale evolution of the PDFs are simulated by introducing scale dependence in the parameters.
  The model reproduces the PDF scale evolution up to a very high scale. The helicity distribution  $g_1(x,\mu)$ and transversity distribution $h_1(x,\mu)$ are predicted in this model at different scales $\mu$. We can also get numerical estimation of different physical quantities to match with the available data. We show that  the predictions of tensor and axial charges  in this model are in good agreement  with the observed data.
  
In Sec.\ref{model}, we describe the quark-diquark model with the detail expressions of the light front wave functions. The parameters in the model are fitted to the proton form factors and the details are discussed in Sec.\ref{FF_fit}. In Sec.\ref{PDF_fit}, we discuss the scale evolution of the unpolarized PDFs. In Sec.\ref{predicts}, we discuss the polarized PDFs, axial and tensor charges predicted in our model. Finally, we present a brief summary and conclusion in Sec.\ref{concl}.

\section{ Diquark Model}\label{model}
In the diquark model, we assume that the virtual incoming photon is interacting with a valence and other two valence quark form a diquark of definite mass with spin-0, called scalar diquark, or with spin-1, called vector diquark. The spin-0 diquarks are in in a flavor singlet state and spin-1 diquarks are in flavor triplet state. The proton state is written
as a sum of isoscalar-scalar diquark singlet state $|u~ S^0\rangle$, isoscalar-vector diquark state $|u~ A^0\rangle$ and isovector-vector diquark $|d~ A^1\rangle$ state\cite{Jakob97,Bacc08}. The proton state is written in the spin-flavor $SU(4)$ structure as
\be 
|P; \pm\rangle = C_S|u~ S^0\rangle^\pm + C_V|u~ A^0\rangle^\pm + C_{VV}|d~ A^1\rangle^\pm \label{PS_state}
\ee
Where $S$ and $A$ represent the scalar and vector diquark having isospin at their superscript.  Under the isospin symmetry, the neutron  state is given by the above formula with $u\leftrightarrow d$.

We use the light-cone convention $x^\pm=x^0 \pm x^3$. We choose a frame where the transverse momentum of proton vanishes i,e. $P \equiv \big(P^+,\frac{M^2}{P^+},\textbf{0}_\perp\big)$. Where the momentum of struck quark  
$p\equiv (xP^+, \frac{p^2+|\bfp|^2}{xP^+},\bfp)$
 and  that of diquark   $P_X\equiv ((1-x)P^+,P^-_X,-\bfp)$. Here $x=p^+/P^+$ is the longitudinal momentum fraction carried by the struck quark. 
The two particle Fock-state expansion for $J^z =\pm1/2$ with spin-0 diquark  is given by
\be
|u~ S\rangle^\pm & =& \int \frac{dx~ d^2\bfp}{2(2\pi)^3\sqrt{x(1-x)}} \bigg[ \psi^{\pm(u)}_{+}(x,\bfp)|+\frac{1}{2}~s; xP^+,\bfp\rangle \nonumber \\
 &+& \psi^{\pm(u)}_{-}(x,\bfp)|-\frac{1}{2}~s; xP^+,\bfp\rangle\bigg],\label{fock_PS}
\ee
and the LF wave functions with spin-0 diquark, for $J=\pm1/2$, are given by\cite{Lepa80}
\be 
\psi^{+(u)}_+(x,\bfp)&=& N_S~ \varphi^{(u)}_{1}(x,\bfp),\nonumber \\
\psi^{+(u)}_-(x,\bfp)&=& N_S\bigg(- \frac{p^1+ip^2}{xM} \bigg)\varphi^{(u)}_{2}(x,\bfp)\label{LFWF_S}\\
\psi^{-(u)}_+(x,\bfp)&=& N_S \bigg(\frac{p^1-ip^2}{xM}\bigg) \varphi^{(u)}_{2}(x,\bfp),\nonumber \\
\psi^{-(u)}_-(x,\bfp)&=&  N_S~ \varphi^{(u)}_{1}(x,\bfp),\nonumber
\ee
where $|\lambda_q~\lambda_S; xP^+,\bfp\rangle$ is the two particle state having struck quark of helicity $\lambda_q$ and a scalar diquark having helicity $\lambda_S=s$(spin-0 singlet diquark helicity is denoted by s to distinguish from triplet diquark). The state with spin-1 diquark is given as \cite{Ellis08}
\be
|\nu~ A \rangle^\pm & =& \int \frac{dx~ d^2\bfp}{2(2\pi)^3\sqrt{x(1-x)}} \bigg[ \psi^{\pm(\nu)}_{++}(x,\bfp)|+\frac{1}{2}~+1; xP^+,\bfp\rangle \nonumber\\
 &+& \psi^{\pm(\nu)}_{-+}(x,\bfp)|-\frac{1}{2}~+1; xP^+,\bfp\rangle +\psi^{\pm(\nu)}_{+0}(x,\bfp)|+\frac{1}{2}~0; xP^+,\bfp\rangle \nonumber \\
 &+& \psi^{\pm(\nu)}_{-0}(x,\bfp)|-\frac{1}{2}~0; xP^+,\bfp\rangle + \psi^{\pm(\nu)}_{+-}(x,\bfp)|+\frac{1}{2}~-1; xP^+,\bfp\rangle \nonumber\\
 &+& \psi^{\pm(\nu)}_{--}(x,\bfp)|-\frac{1}{2}~-1; xP^+,\bfp\rangle  \bigg].\label{fock_PS}
\ee
Where $|\lambda_q~\lambda_D; xP^+,\bfp\rangle$ represents a two-particle state with a quark of helicity $\lambda_q=\pm\frac{1}{2}$ and a vector diquark of helicity $\lambda_D=\pm 1,0(triplet)$.
The LFWFs are, for $J=+1/2$ 
\be 
\psi^{+(\nu)}_{+~+}(x,\bfp)&=& N^{(\nu)}_1 \sqrt{\frac{2}{3}} \bigg(\frac{p^1-ip^2}{xM}\bigg) \varphi^{(\nu)}_{2}(x,\bfp),\nonumber \\
\psi^{+(\nu)}_{-~+}(x,\bfp)&=& N^{(\nu)}_1 \sqrt{\frac{2}{3}} \varphi^{(\nu)}_{1}(x,\bfp),\nonumber \\
\psi^{+(\nu)}_{+~0}(x,\bfp)&=& - N^{(\nu)}_0 \sqrt{\frac{1}{3}} \varphi^{(\nu)}_{1}(x,\bfp),\label{LFWF_Vp}\\
\psi^{+(\nu)}_{-~0}(x,\bfp)&=& N^{(\nu)}_0 \sqrt{\frac{1}{3}} \bigg(\frac{p^1+ip^2}{xM} \bigg)\varphi^{(\nu)}_{2}(x,\bfp),\nonumber \\
\psi^{+(\nu)}_{+~-}(x,\bfp)&=& 0,\nonumber \\
\psi^{+(\nu)}_{-~-}(x,\bfp)&=&  0, \nonumber 
\ee
and for $J=-1/2$
\be 
\psi^{-(\nu)}_{+~+}(x,\bfp)&=& 0,\nonumber \\
\psi^{-(\nu)}_{-~+}(x,\bfp)&=& 0,\nonumber \\
\psi^{-(\nu)}_{+~0}(x,\bfp)&=& N^{(\nu)}_0 \sqrt{\frac{1}{3}} \bigg( \frac{p^1-ip^2}{xM} \bigg) \varphi^{(\nu)}_{2}(x,\bfp),\label{LFWF_Vm}\\
\psi^{-(\nu)}_{-~0}(x,\bfp)&=& N^{(\nu)}_0\sqrt{\frac{1}{3}} \varphi^{(\nu)}_{1}(x,\bfp),\nonumber \\
\psi^{-(\nu)}_{+~-}(x,\bfp)&=& - N^{(\nu)}_1 \sqrt{\frac{2}{3}} \varphi^{(\nu)}_{1}(x,\bfp),\nonumber \\
\psi^{-(\nu)}_{-~-}(x,\bfp)&=& N^{(\nu)}_1 \sqrt{\frac{2}{3}} \bigg(\frac{p^1+ip^2}{xM}\bigg) \varphi^{(\nu)}_{2}(x,\bfp),\nonumber
\ee
having flavour index $\nu=u,d$.
The LFWFs $\varphi^{(\nu)}_i(x,\bfp)$ are  a modified form of the  soft-wall AdS/QCD prediction\cite{valery0}
\be
\varphi_i^{(\nu)}(x,\bfp)=\frac{4\pi}{\kappa}\sqrt{\frac{\log(1/x)}{1-x}}x^{a_i^\nu}(1-x)^{b_i^\nu}\exp\bigg[-\delta^\nu\frac{\bfp^2}{2\kappa^2}\frac{\log(1/x)}{(1-x)^2}\bigg].
\label{LFWF_phi}
\ee
The wave functions $\varphi_i^\nu ~(i=1,2)$ reduce to the AdS/QCD prediction\cite{BT} for the parameters $a_i^\nu=b_i^\nu=0$  and $\delta^\nu=1.0$.
 We use the AdS/QCD scale parameter $\kappa =0.4~GeV$ as determined in \cite{CM1} and the quarks are  assumed  to be  massless. 
\section{Form Factor Fitting}\label{FF_fit}
In the light-front formalism, for a spin-$\frac{1}{2}$  composite particle system the Dirac and Pauli form factors are defined as \cite{BD80}
\be 
\langle P+q;+|\frac{J^+(0)}{2P^+}|P;+\rangle &=& F_1(q^2)\\
\langle P+q;+|\frac{J^+(0)}{2P^+}|P;-\rangle &=& -(q^1-iq^2)\frac{F_2(q^2)}{2M}
\ee 
Where the $q^2$ is square of the momentum transferred to the nucleon of mass $M$. The normalization of form factors for proton and neutron are given as $F^p_1(0)=1,~F^p_2(0)=\kappa^p=1.793$ and $F^n_1(0)=0,~F^n_2(0)=\kappa^n=-1.913$ respectively. Considering the charge and isospin symmetry the nucleon form factors are decomposed into flavour form factors as\cite{Cates11} $F^{p(n)}_i=e_uF^{u(d)}_i+e_dF^{d(u)}_i$.
 
In the SU(4) structure, flavored form factors are written in terms of scalar and vector diquarks as\cite{Bacc08}
\be 
F^{(u)}_i(Q^2)&=&C^2_S F^{(S)}_i(Q^2) + C^2_V F^{(V)}_i(Q^2),\label{Fiu}\\
F^{(d)}_i(Q^2)&=&C^2_{VV} F^{(VV)}_i(Q^2).\label{Fid}
\ee
In the quark-diquark model Dirac and Pauli form factors for quarks can be written in terms of LFWFs as
\be 
F^{(S)}_1(Q^2)&=&\int^1_0\int\frac{d^2\bfp}{16\pi^3}[\psi^{+(u)\dagger}_+(x,\bfp^\prime)\psi^{+(u)}_+(x,\bfp)+\psi^{+(u)\dagger}_-(x,\bfp^\prime)\psi^{+(u)}_-(x,\bfp)],\\
F^{(S)}_2(Q^2)&=&-\frac{2M}{q^1-iq^2}\int^1_0\int\frac{d^2\bfp}{16\pi^3}[\psi^{+(u)\dagger}_+(x,\bfp^\prime)\psi^{-(u)}_+(x,\bfp)+\psi^{+(u)\dagger}_-(x,\bfp^\prime)\psi^{-(u)}_-(x,\bfp)]\nonumber\\
\ee
for scalar diquark and 
\be 
F^{(A)}_1(Q^2)&=&\int^1_0\int\frac{d^2\bfp}{16\pi^3}[\psi^{+(\nu)\dagger}_{++}(x,\bfp^\prime)\psi^{+(\nu)}_{++}(x,\bfp)+\psi^{+(\nu)\dagger}_{-+}(x,\bfp^\prime)\psi^{+(\nu)}_{-+}(x,\bfp)\nonumber\\
&&+\psi^{+(\nu)\dagger}_{+0}(x,\bfp^\prime)\psi^{+(\nu)}_{+0}(x,\bfp)+\psi^{+(\nu)\dagger}_{-0}(x,\bfp^\prime)\psi^{+(\nu)}_{-0}(x,\bfp)],\\
F^{(A)}_2(Q^2)&=&-\frac{2M}{q^1-iq^2}\int^1_0\int\frac{d^2\bfp}{16\pi^3}[\psi^{+(\nu)\dagger}_{+0}(x,\bfp^\prime)\psi^{-(\nu)}_{+0}(x,\bfp)+\psi^{+(\nu)\dagger}_{-0}(x,\bfp^\prime)\psi^{-(\nu)}_{-0}(x,\bfp)]\nonumber\\
\ee
for vector diquark.
Where $\bfp^\prime = \bfp+(1-x)\bfq$. The superscript  $A=V,VV$ for isoscalar-vector diquark and isovector-vector diquark respectively. We consider the frame where $q=(0,0,\textbf{q}_\perp)$ and $Q^2=-q^2=\bfq^2$.

In this model the Dirac and Pauli form factors read as
\be 
F^{(S)}_1(Q^2)&=&N^2_S  R_1^{(u)}(Q^2)\label{F1s}\\
F^{(S)}_2(Q^2)&=& N^2_S R_1^{(u)}(Q^2) \label{F2s}\\
F^{(V)}_1(Q^2)&=& (\frac{1}{3} N^{(u)2}_0+ \frac{2}{3} N^{(u)2}_1) R_1^{(u)}(Q^2) \label{F1v}\\
F^{(V)}_2(Q^2)&=& - \frac{1}{3} N^{(u)2}_0 R_2^{(u)}(Q^2) \label{F2v}\\
F^{(VV)}_1(Q^2)&=&(\frac{1}{3} N^{(d)2}_0+ \frac{2}{3} N^{(d)2}_1) R_1^{(d)}(Q^2)\label{F1vv}\\
F^{(VV)}_2(Q^2)&=& - \frac{1}{3} N^{(d)2}_0 R_2^{(d)}(Q^2) \label{F2vv}
\ee
Where superscript $S,V$ and $VV$ represent the contributions with isoscalar-scalar diquark, isoscalar-vector diquark and isovector-vector diquarks respectively. $R_1^{(\nu)}(Q^2)$ and $R_2^{(\nu)}(Q^2)$ are defined as 
\be 
 R_1^{(\nu)}(Q^2)&=&\int dx \bigg[T^{(\nu)}_1(x)\frac{(1-x)^2}{\delta^\nu}+T^{(\nu)}_2(x)\frac{(1-x)^4}{(\delta^\nu)^2} \frac{\kappa^2}{M^2\log(1/x)} \nonumber\\
&&\times \bigg(1-\frac{\delta^\nu Q^2}{4\kappa^2} \log(1/x)\bigg)\bigg]\exp\bigg[-\delta^\nu\frac{Q^2}{4\kappa^2} \log(1/x)\bigg],\label{F1s}\\
 R_2^{(\nu)}(Q^2)&=&\int dx~ 2T^{(\nu)}_3(x)\frac{(1-x)^3}{\delta^\nu} \exp\bigg[-\delta^\nu\frac{Q^2}{4\kappa^2} \log(1/x)\bigg].
\ee
Where
\be 
T^{(\nu)}_1(x)&=& x^{2a^{\nu}_1}(1-x)^{2b^{\nu}_1-1},\\
T^{(\nu)}_2(x)&=& x^{2a^{\mu}_2-2}(1-x)^{2b^{\nu}_2-1}, \label{F2}\\
T^{(\nu)}_3(x)&=& x^{a^{\nu}_1+a^{\nu}_2-1}(1-x)^{b^{\nu}_1+b^{\nu}_2-1}.
\ee

\begin{figure}[htbp]
\begin{minipage}[c]{0.98\textwidth}
\small{(a)}\includegraphics[width=7.5cm,clip]{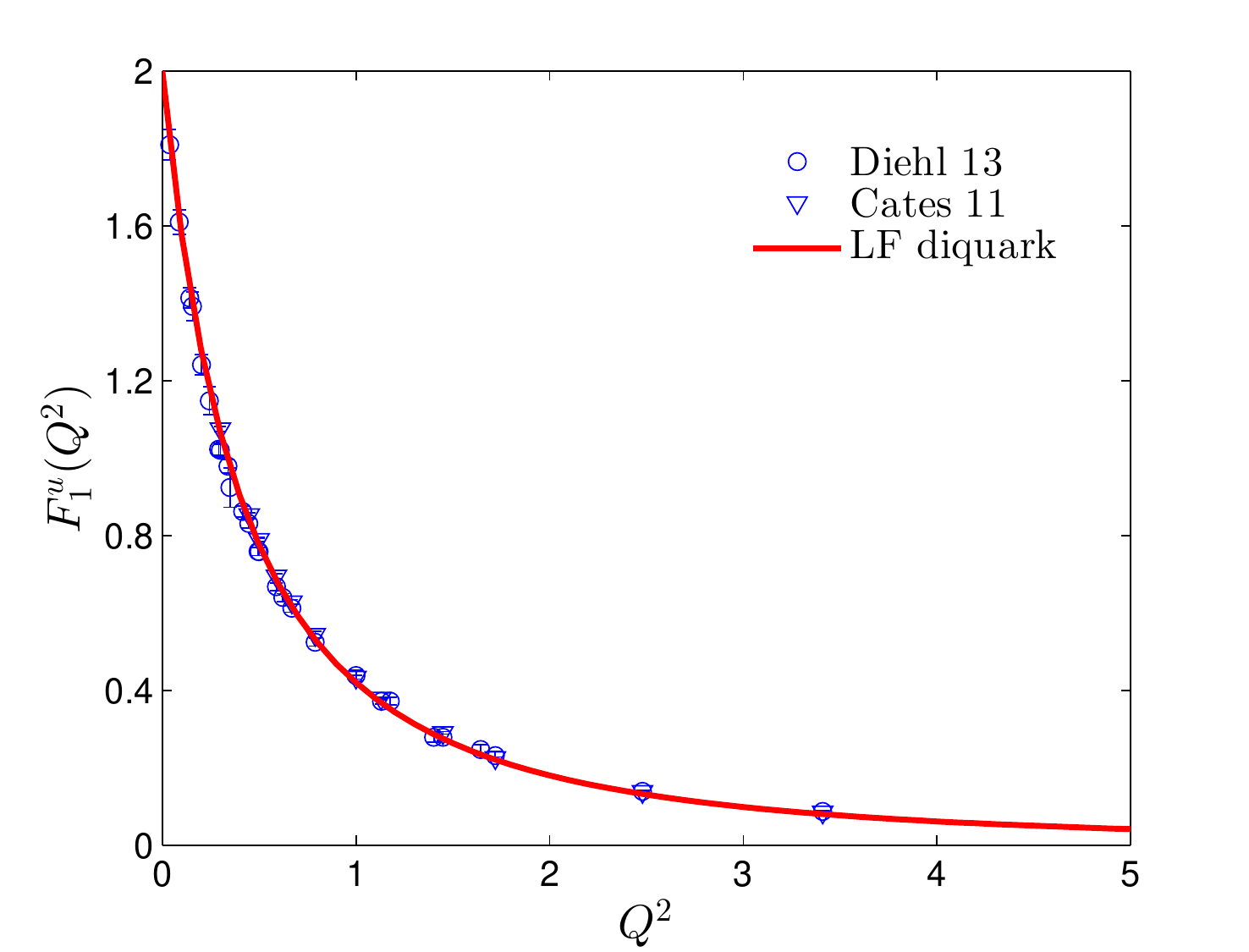}
\small{(b)}\includegraphics[width=7.5cm,clip]{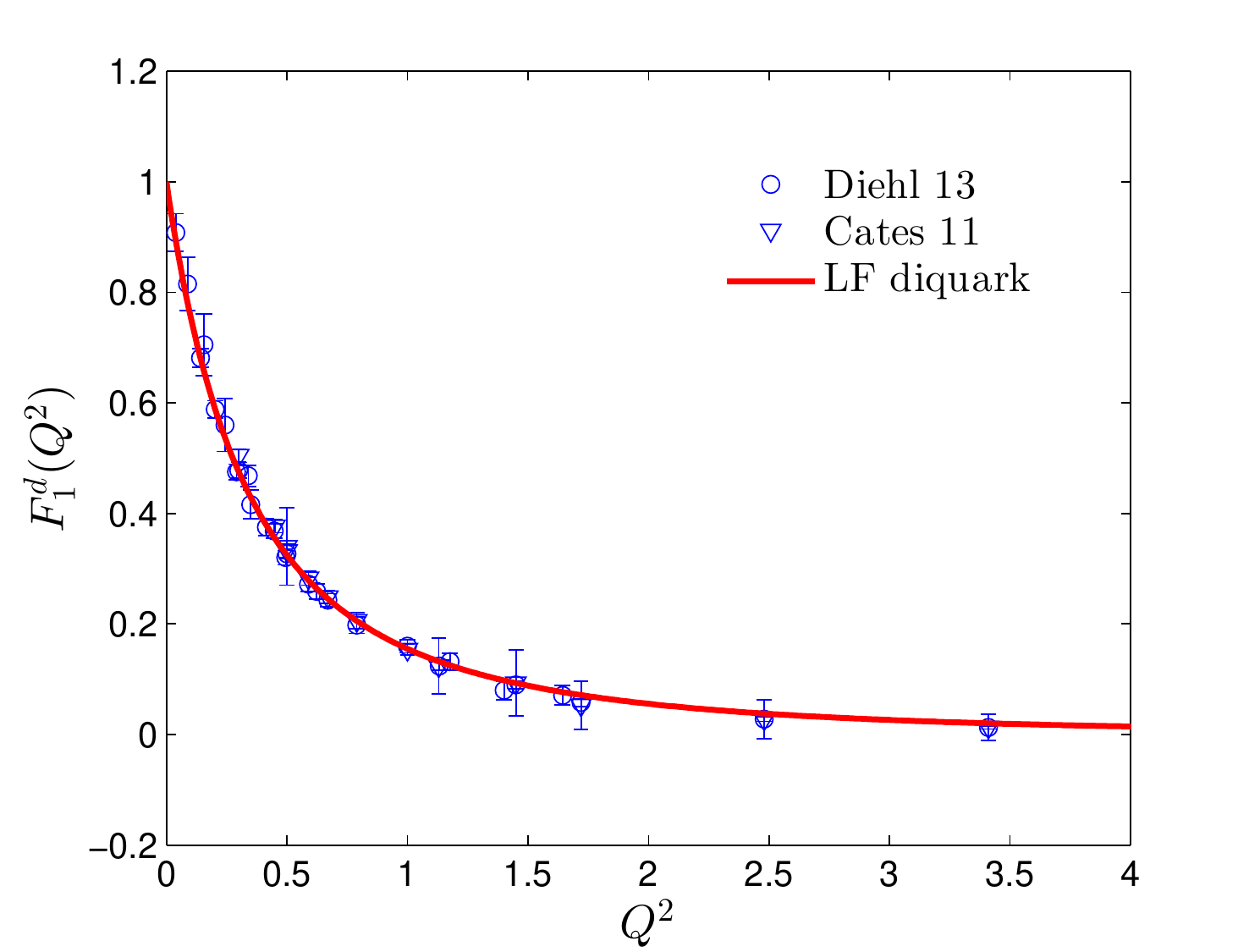}
\end{minipage}
\begin{minipage}[c]{0.98\textwidth}
\small{(c)}\includegraphics[width=7.5cm,clip]{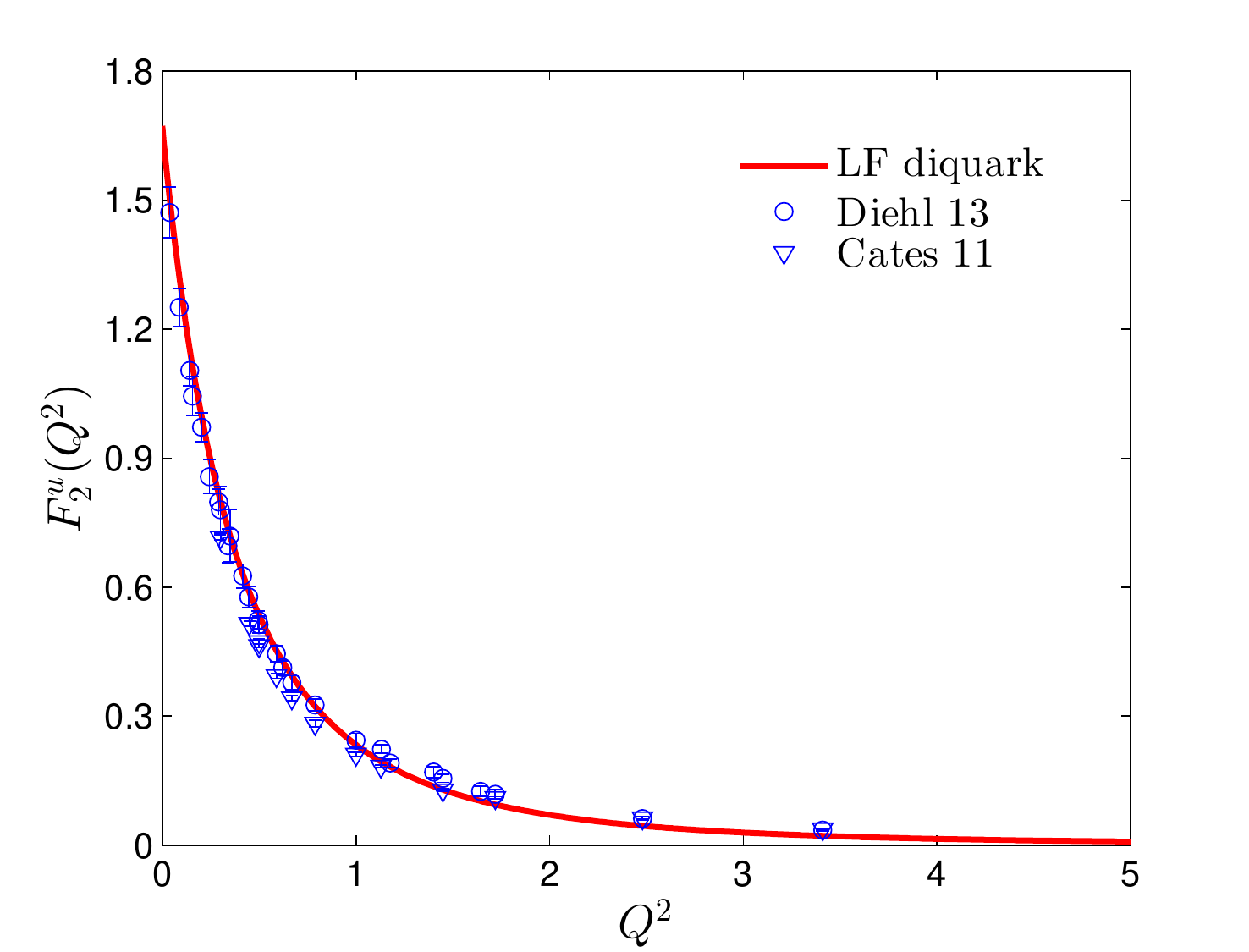}
\small{(d)}\includegraphics[width=7.5cm,clip]{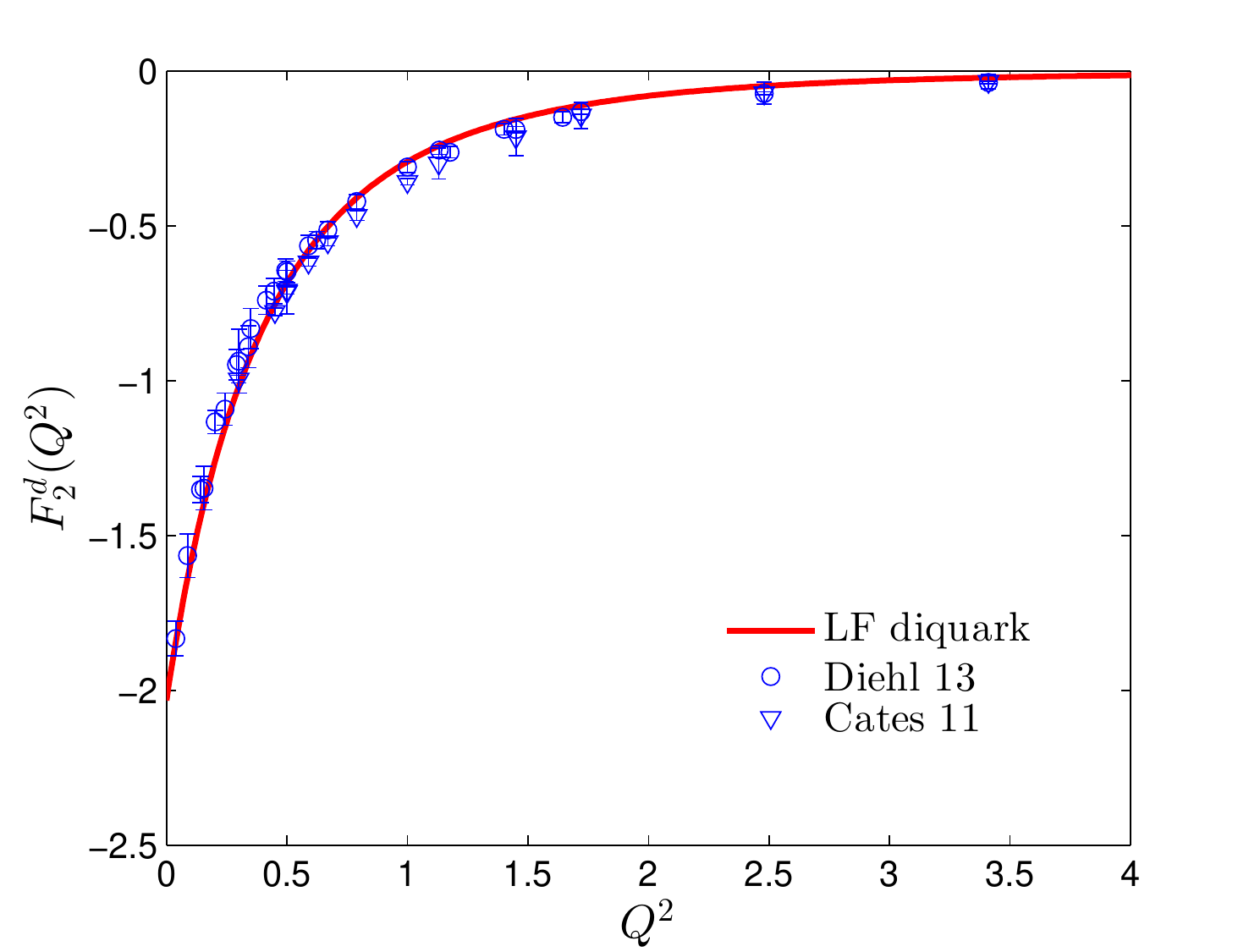} 
\end{minipage}
\caption{\label{fig_FF} Flavour form factors fitting in  the light-front diquark model. Data are taken from ref.\cite{Diehl13,Cates11}. 
} 
\end{figure} 
\begin{figure}[htbp]
\begin{minipage}[c]{0.98\textwidth}
\small{(a)}\includegraphics[width=7.5cm,clip]{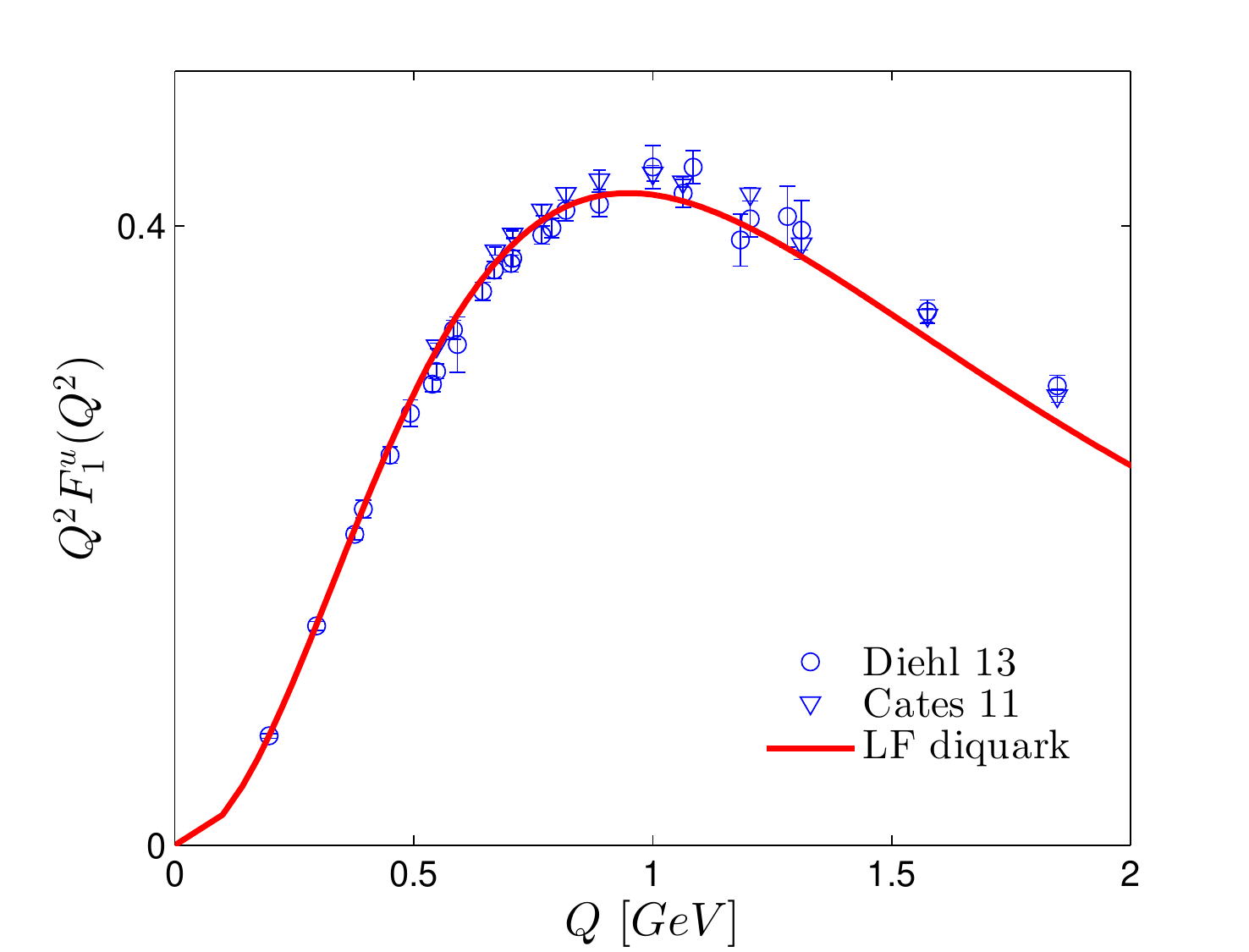}
\small{(b)}\includegraphics[width=7.5cm,clip]{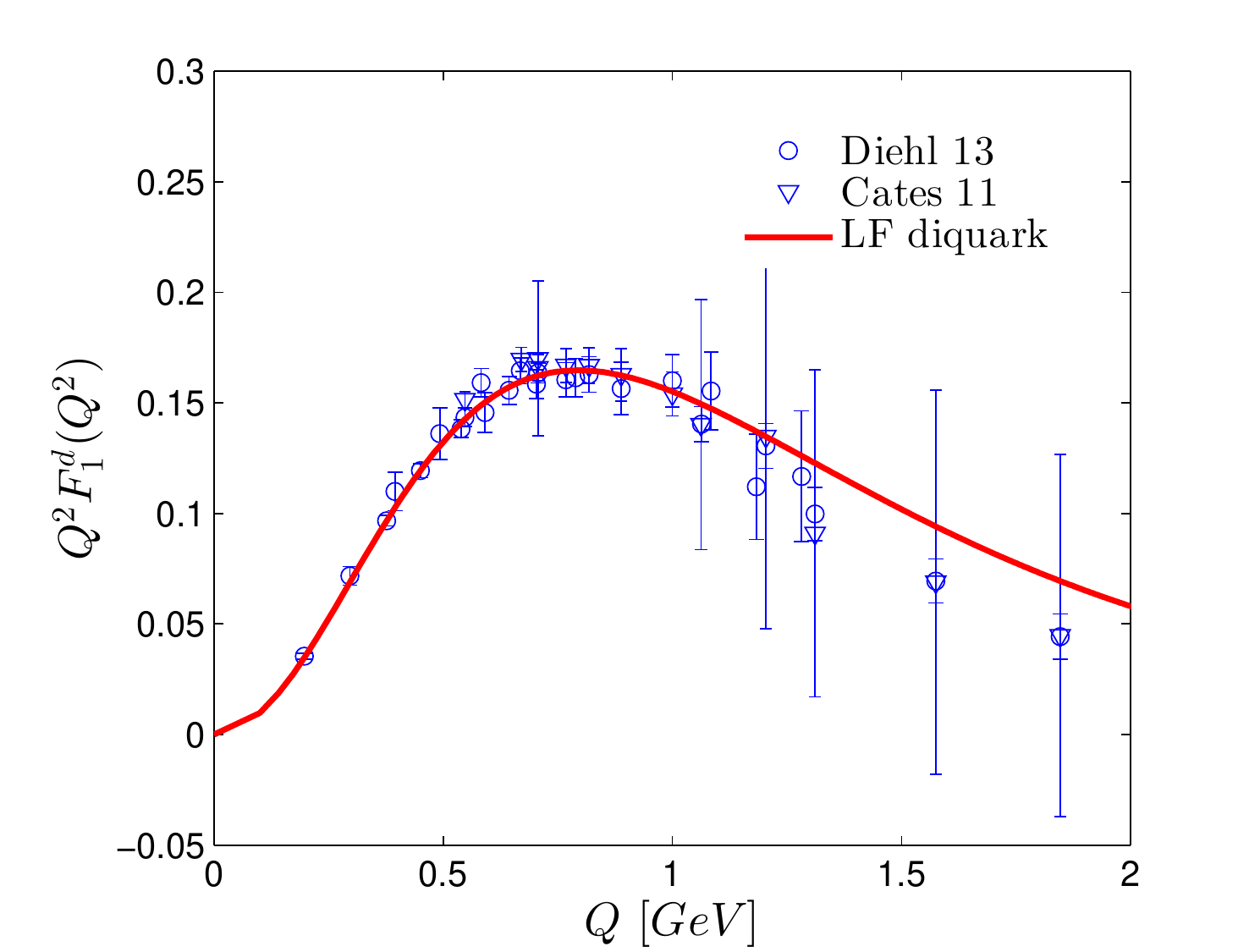}
\end{minipage}
\begin{minipage}[c]{0.98\textwidth}
\small{(c)}\includegraphics[width=7.5cm,clip]{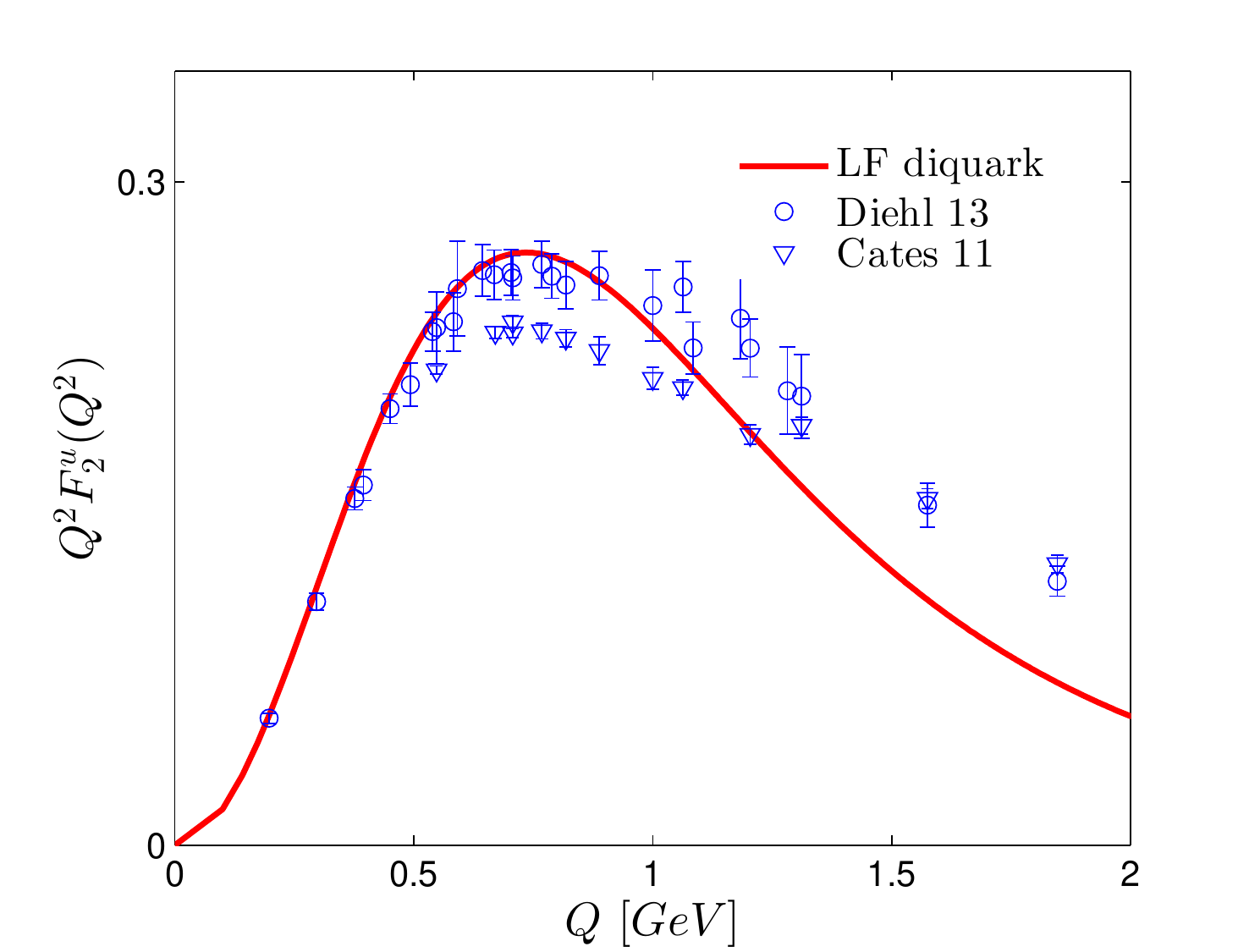}
\small{(d)}\includegraphics[width=7.5cm,clip]{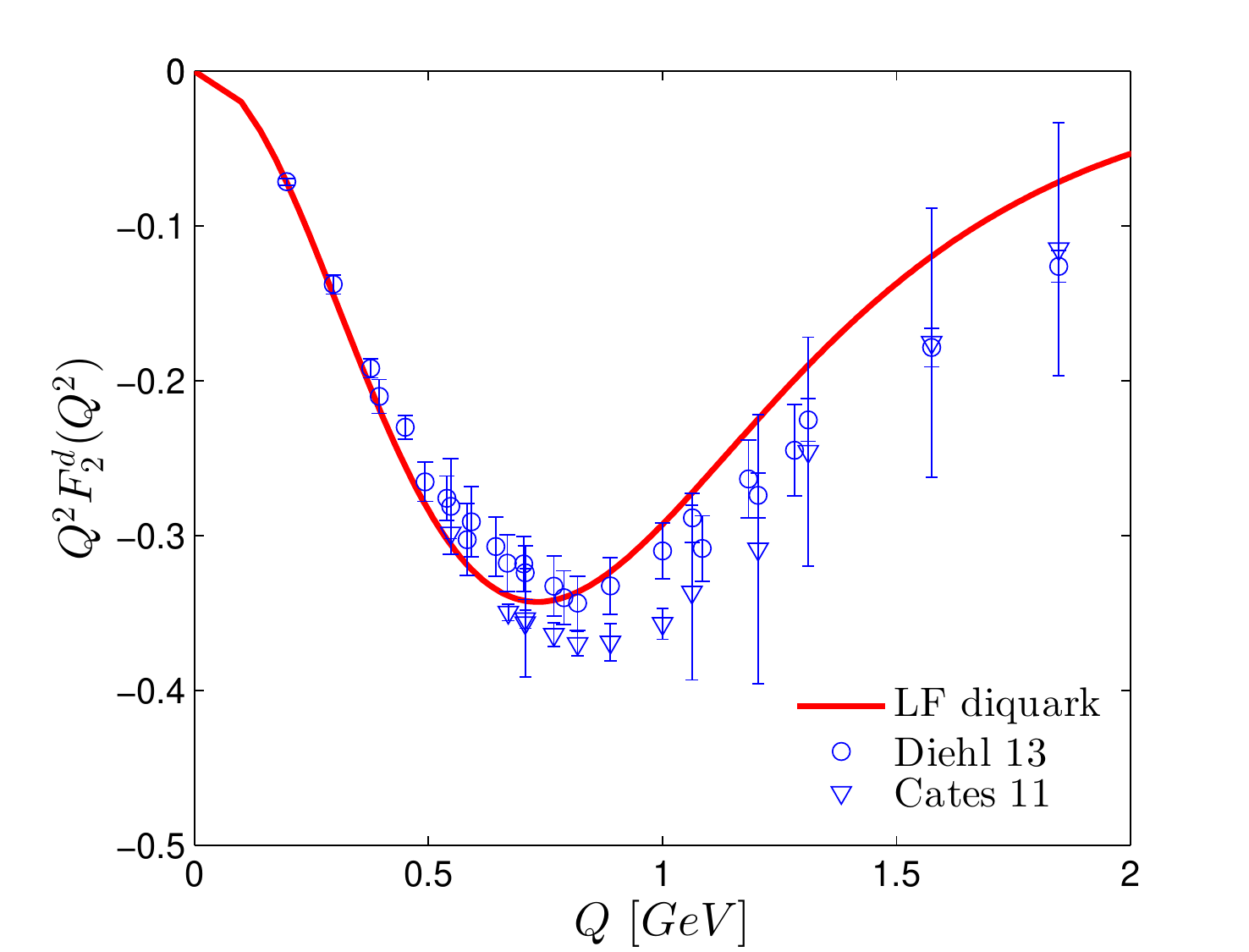}
\end{minipage}
\caption{\label{fig_Q2F} Dirac and pauli form factors multiplied by $Q^2$ for $u$ and $d$ quarks and compared with the data\cite{Diehl13,Cates11}.}
\end{figure} 
\begin{figure}[htbp]
\begin{minipage}[c]{0.98\textwidth}
\small{(a)}\includegraphics[width=7.5cm,clip]{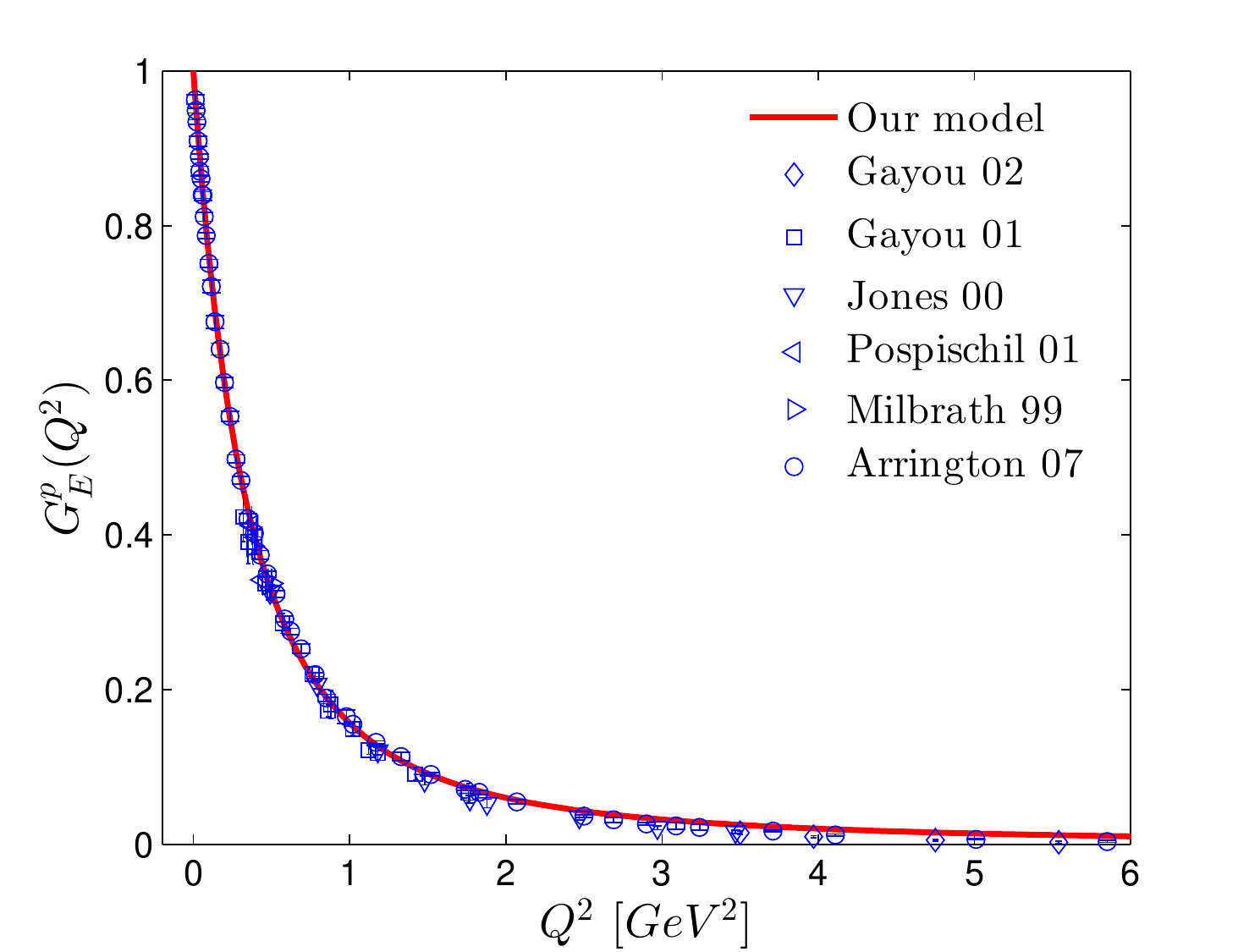}
\small{(b)}\includegraphics[width=7.5cm,clip]{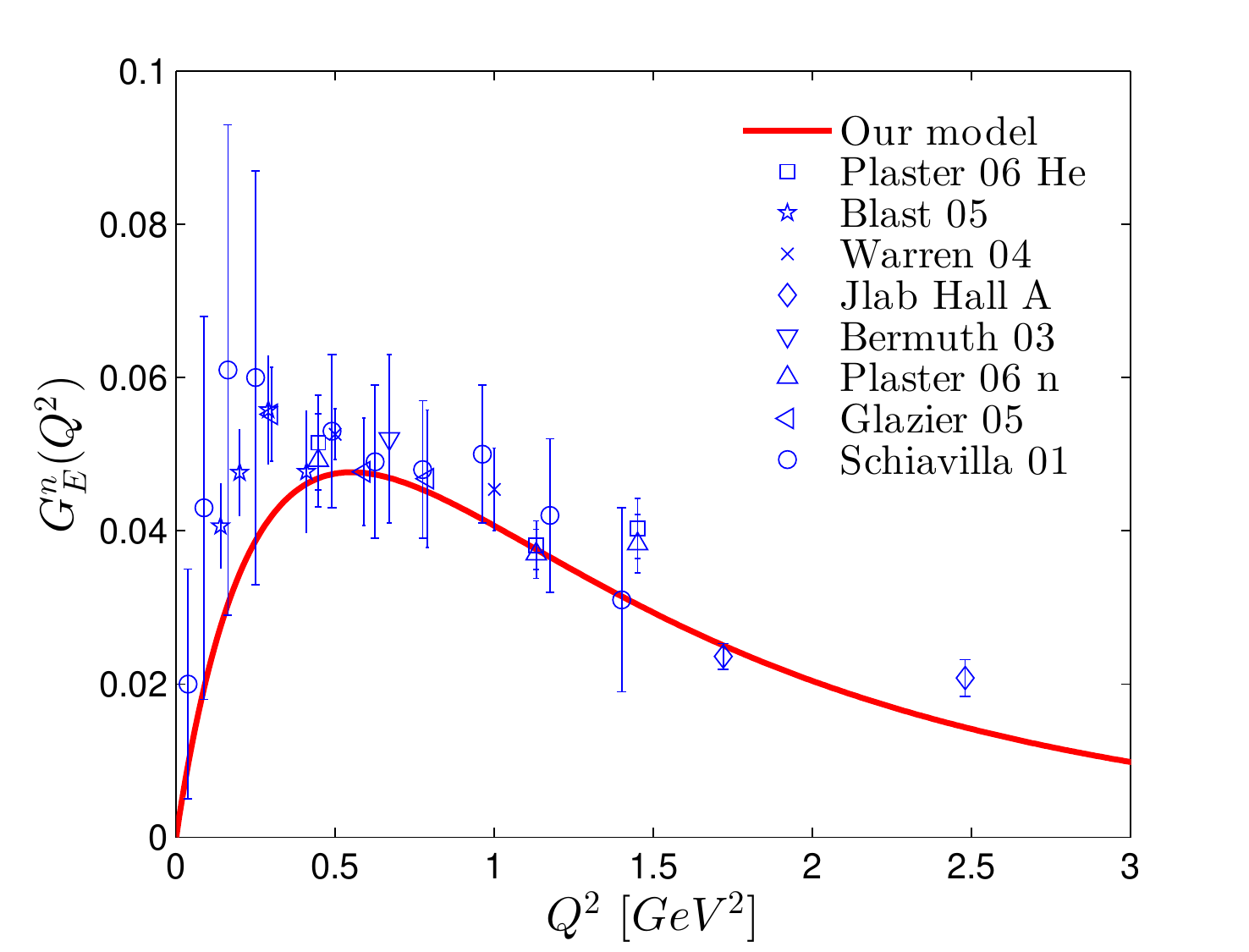}\\
\small{(c)}\includegraphics[width=7.5cm,clip]{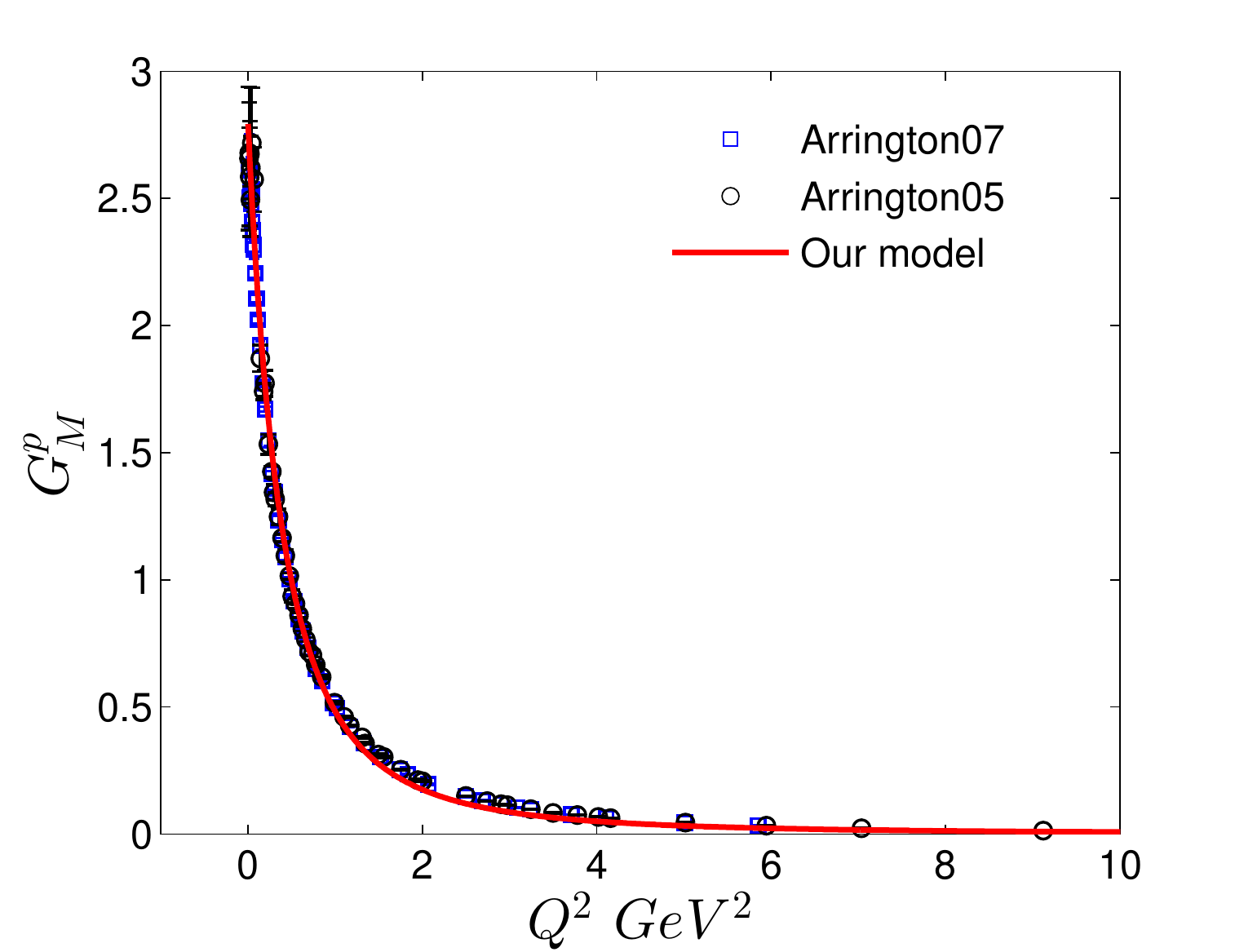}
\small{(d)}\includegraphics[width=7.5cm,clip]{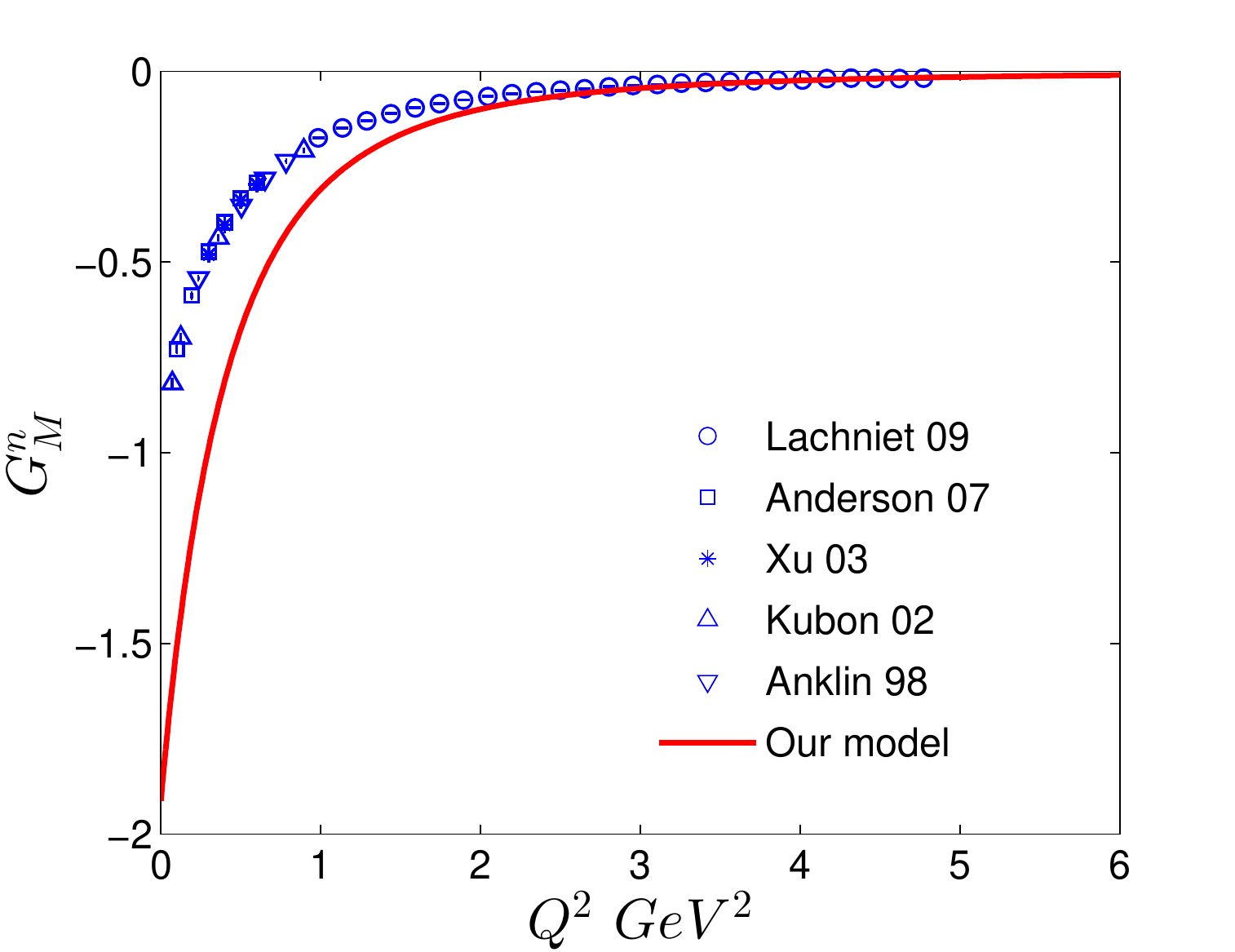}
\end{minipage}
\caption{\label{fig_SachsFF} Sachs form factors $G^{p(n)}_E(Q^2)$  and $G^{p(n)}_M(Q^2)$ for proton(a,c) and neutron(b,d) respectively. The data are taken from ref.\cite{Gayou02,Gayou01,Jones00,Posp01,Milb99,Arri07} for $G^{p}_E(Q^2)$, ref.\cite{Blast05,Warr04,JlabHA,Berm03,Plas06,Glaz05,Schi01} for $G^{n}_E(Q^2)$, ref.\cite{Arri07,Arri05} for $G^{p}_M(Q^2)$ and ref.\cite{Lach09,Ande07,Xu03,Kubo02,Ankl98} for $G^{n}_M(Q^2)$.}
\end{figure} 
\begin{figure}[htbp]
\small{(a)}\includegraphics[width=7.5cm,clip]{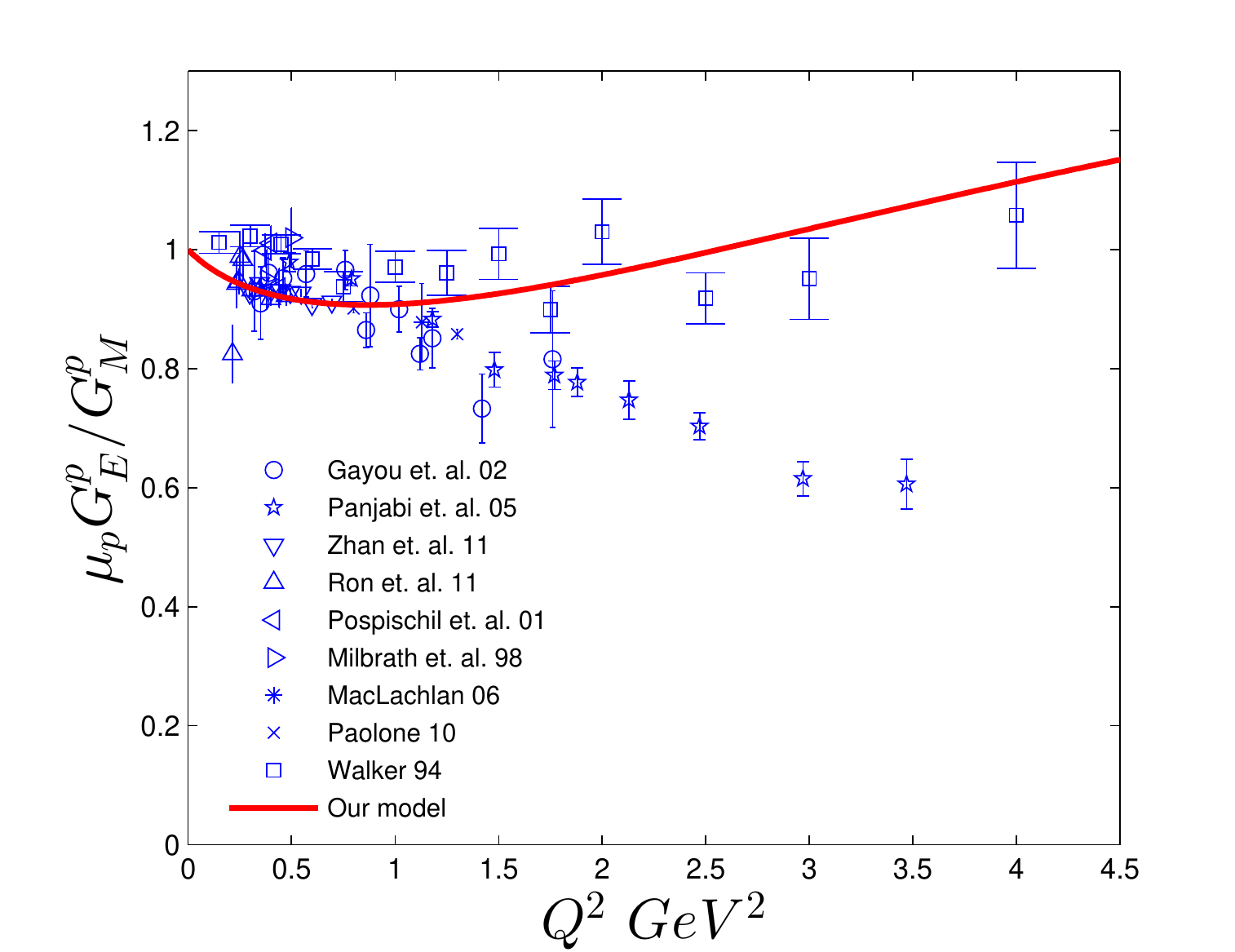}
\small{(b)}\includegraphics[width=7.5cm,clip]{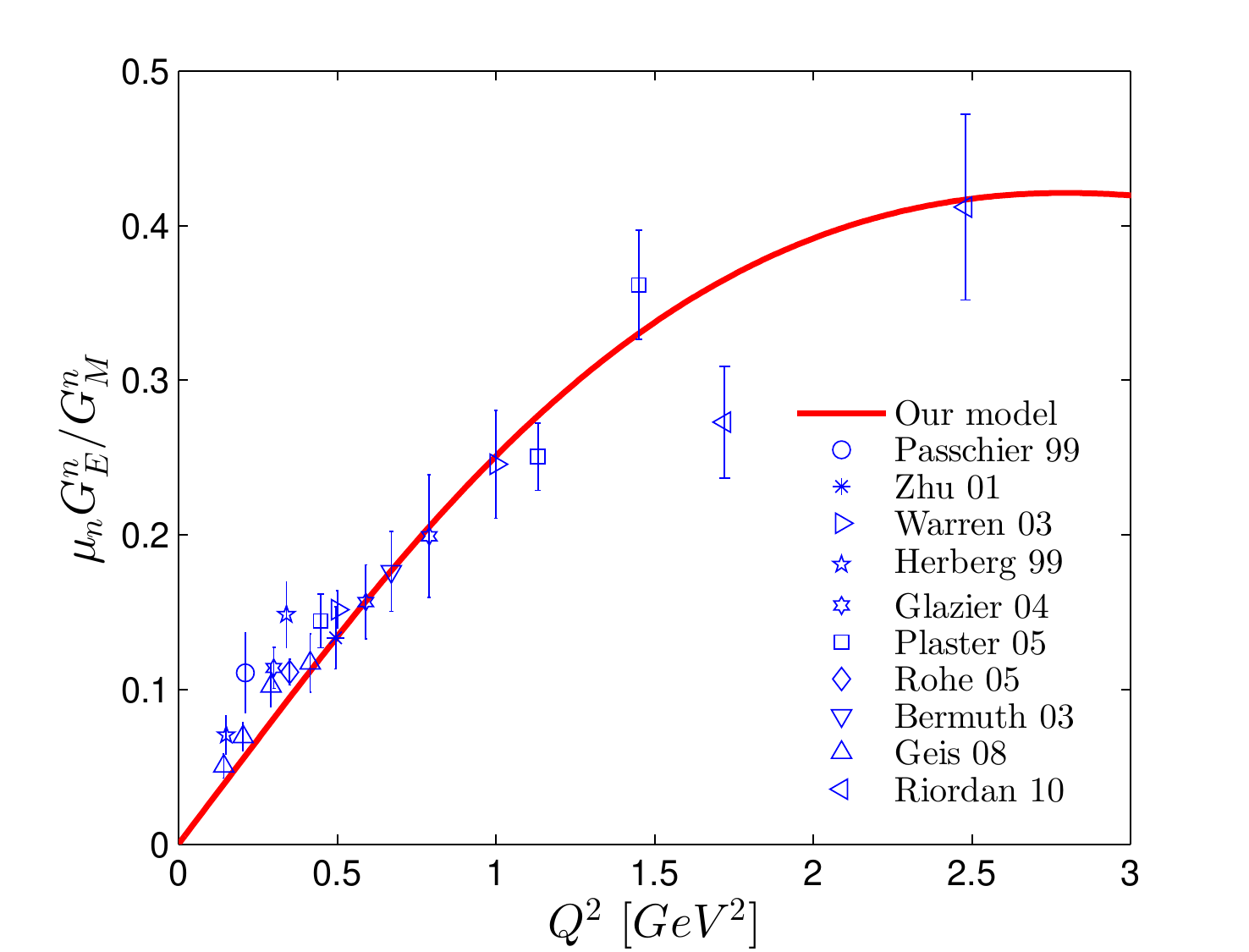}
\caption{\label{fig_RSachsFF} Ratio of Sachs form factor $R^i=\mu_i G^i_E/G^i_M$ for proton \cite{Gayou02,Posp01,Milb99,Punj05,Ron11,Walk94,Zhan11,Paol10,MacL06} and neutron \cite{Pass99,Zhu01,Warr04,Herb99,Glaz05, Plas06,Berm03,Blast05,Rior10}.}
\end{figure} 

We find value of the parameters $a_i$ and $b_i$ by fitting Dirac and Pauli form factors data taken form Ref.\cite{Cates11,Diehl13}. Fig.\ref{fig_FF} shows the form factor fittings in our model. \begin{table}[ht]
\centering 
\begin{tabular}{|c|c|c|c|c|c|}
 \hline
$ \nu $~&~$a_1^\nu$~ & ~$b_1^\nu$~ & ~$a_2^\nu$~ & ~$b_2^\nu$~&~~$\delta^\nu$~~\\ \hline
 ~~ $u$ ~~&~~ $0.280\pm 0.001 ~$~&~~ $0.1716\pm 0.0051$ ~~&~~ $0.84 \pm 0.02$ ~~&~~ $0.2284 \pm 0.0035$&1.0 \\ 
  $d$ & $0.5850 \pm 0.0003$ & $0.7000 \pm 0.0002$ &  $0.9434^{+0.0017}_{-0.0013}$ & $0.64^{+0.0082}_{-0.0022}$ & 1.0 \\ \hline
 \end{tabular} 
\caption{The fitted parameters  for $u$ and $d$ quarks.} 
\label{tab_para_mu0} 
\end{table} 

The normalization conditions 
are defined as
\be 
\int_0^1 dx f^{(u)}_1(x)&=&F^{(u)}_1(Q^2=0)= n_u,\\
\int_0^1 dx E^{(u)}_1(x,0)&=&F^{(u)}_2(Q^2=0)= \kappa_u,
\ee 
\be 
\int_0^1 dx f^{(d)}_1(x)&=&F^{(d)}_1(Q^2=0)= n_d ,\\
\int_0^1 dx E^{(d)}_1(x,0)&=&F^{(d)}_2(Q^2=0)= \kappa_d.
\ee  
Where $f_1^{\nu}(x)$ is the unpolarized PDF and $E^{(\nu)}(x,Q^2)$ is the helicity flip GPD corresponding to the valence quark of flavour $\nu=u,d$ and according to the quark counting rules $n_u=2$ and $n_d=1$ for proton.
From isospin symmetry, the anomalous magnetic moments for $u$ and $d$ quarks are $\kappa_u=1.673$ and $\kappa_d=-2.033$ respectively.
The  coefficients $C^2_i$  are then determined as
\be 
C^2_S&=& 1.3872,\nonumber\\
C^2_V&=& 0.6128,\label{Ci2}\\
C^2_{VV}&=& 1.\nonumber
\ee
The flavour decomposition of any distribution function follows the Eq.(\ref{Fiu},\ref{Fid}) with the $C^2_i$ given above in Eq.(\ref{Ci2}).

The normalized constants $N^2_i$ are found considering the following normalizations \cite{Bacc08} 
$$ 
\int dx f^{(S)}_1(x)=F^{(S)}_1(0)=1,~
\int dx f^{(V)}_1(x)=F^{(V)}_1(0)=1,~
\int dx f^{(VV)}_1(x)=F^{(VV)}_1(0)=1;
$$
and the  values  are 
$N_S = 2.0191, ~
N_0^{(u)} = 3.2050, ~ N_0^{(d)} = 5.9423,~
N_1^{(u)} = 0.9895, ~ N_1^{(d)} = 1.1616.$
To demonstrate the accuracy of the model, in Fig.\ref{fig_Q2F}, the flavor form factors multiplied with $Q^2$ are compared with the available data. Even at large $Q^2$, the model predictions are within error bars of the experimental data. 

The Sachs form factors for nucleons $(i=p,n)$ are defined as
\be
G^i_E(Q^2)=F^i_1(Q^2)-\frac{Q^2}{4M^2_i}F^i_2(Q^2),\\
G^i_M(Q^2)=F^i_1(Q^2)+F^i_2(Q^2).
\ee
In Fig.\ref{fig_SachsFF}, the Sachs form factors $G_E$ and $G_M$ for proton and neutron in this model are shown to have excellent agreement with the experimental data, except for $G^n_M$. The ratio $R^i=\mu_i G^i_E/G^i_M$ for proton and neutron are also shown in Fig.\ref{fig_RSachsFF}.  They agree with  the experimental data quite well. 
We also calculate the electromagnetic radii of nucleons from
\be 
\langle r^2_E\rangle^i=-6\frac{dG^i_E(q^2)}{dQ^2}\bigg|_{Q^2=0},\\
\langle r^2_M\rangle^i=-\frac{6}{G^i_M(0)}\frac{dG^i_M(q^2)}{dQ^2}\bigg|_{Q^2=0}
\ee
in this model. The radii are given in Table.\ref{tab_radii} show quit well agreement with measured data.
\begin{table}[ht]
\centering 
\begin{tabular}{|c|c|c|c|c|c|}
 \hline
 ~~ Quantity ~~&~~ Our ~result ~~&~~ Measured Data \cite{Ber12}\\ \hline
 ~~ $r^p_E~(fm)$ ~~&~~ $0.830\pm 0.025$ ~~&~~ $0.877\pm0.005$ \\ 
 ~~ $r^p_M~(fm)$ ~~&~~ $0.779\pm 0.007$ ~~&~~ $0.777\pm0.016$ \\
  ~~ $\langle r^2_E\rangle^n ~(fm^2)$ ~~&~~ $-0.064\pm 0.018$ ~~&~~ $-0.1161\pm0.0022$ \\ 
 ~~ $r^n_M~(fm)$ ~~&~~ $0.758 \pm 0.005$ ~~&~~ $0.862^{+0.009}_{-0.008}$ \\ \hline
 \end{tabular} 
\caption{Electromagnetic radii of nucleon in this model compared with measured data\cite{Ber12} }.
\label{tab_radii} 
\end{table}

\section{Unpolarized PDF evolution}\label{PDF_fit}
The parton distribution function is defined as
\be 
\Phi^{\Gamma(\nu)}(x)=\frac{1}{2}\int \frac{d z^-}{2(2\pi)} e^{ip^+z^-/2} \langle P;S|\bar{\psi}^{(\nu)}(0)\Gamma\psi^{(\nu)}(z^-)|P; S\rangle \bigg|_{z^+=z_T=0}.
\ee
which depends only on the light-cone momentum fraction $x=p^+/P^+$. Where the proton state $|P; S\rangle$, having spin $S$, is given in Eq.(\ref{PS_state}).  For different  Dirac structures we get different PDFs, e.g., for $\Gamma=\gamma^+, \gamma^+ \gamma^5, i\sigma^{j+}\gamma^5$ we have the unpolarized PDF $f_1(x)$, helicity distribution $g_1(x)$ and transversity distribution $h_1(x)$ respectively.

 The leading order QCD evolution of the unpolarized PDF is given as  \cite{Bron08}
\be 
\int^1_0 dx x^n f_1(x,\mu)=\bigg(\frac{\alpha_s(\mu)}{\alpha_s(\mu_0)} \bigg)^{\gamma^{(0)}_n/2\beta_0} \int^1_0 dx x^n f_1(x,\mu_0). \label{DGLAP_Eq}
\ee  
Where the anomalous dimension is given as
\be 
\gamma^{(0)}_n=-2C_F\bigg(3+\frac{2}{(n+1)(n+2)}-4\sum_{k=1}^{n+1}\frac{1}{k} \bigg)
\ee 
with $C_F=4/3$ and $\beta_0=9$.
The strong coupling constant, at the leading order, is given as
\be 
\alpha_s(\mu)=\frac{4\pi}{\beta_0 \ln(\mu^2/\Lambda^2_{QCD})}
\ee
with $\Lambda_{QCD}=0.226~ GeV$. 
In \cite{Bron08}, for  pion PDF evolution, the initial scale in leading order evolution was found to be $\mu_0=0.313$ GeV. For proton, we use the same initial scale.



The LFWFs are independent of the hard evolution scale $\mu$. Generally, the models of LFWFs are defined at the lowest scale\cite{Bacc08} and then DGLAP equation determines the PDF scale evolution.
Thus, in the LFWF overlap representation, the unpolarised PDFs in the light-front diquark model  at the initial scale $\mu_0$  are obtained as\\
for scalar diquark:
\be 
f^{(S)}_1(x)&=& \int d^2\bfp \frac{1}{16\pi^3}\bigg[|\psi^{+(u)}_+(x,\bfp )|^2+|\psi^{+(u)}_-(x,\bfp)|^2\bigg],\nonumber\\
&=&  N^{2}_S\bigg[\frac{1}{\delta^u} x^{2a_1^u}(1-x)^{2b_1^u+1}\nonumber\\
&&~~~~~~~~~~~ + x^{2a_2^u-2}(1-x)^{2b_2^u+3}\frac{\kappa^2}{(\delta^u)^2 M^2\ln(1/x)}\bigg],
\ee
for vector diquark:
\be 
f^{(A)}_1(x)&=& \int d^2\bfp \frac{1}{16\pi^3}\bigg[|\psi^{+(\nu)}_{++}(x,\bfp)|^2+|\psi^{+(\nu)}_{-+}(x,\bfp)|^2 \nonumber\\
&&~~~~~~~+ |\psi^{+(\nu)}_{+0}(x,\bfp )|^2+|\psi^{+(\nu)}_{-0}(x,\bfp)|^2\bigg]\nonumber\\
&=&  \bigg(\frac{1}{3}N^{(\nu)2}_0+\frac{2}{3}N^{(\nu)2}_1\bigg)\nonumber\\
&&\times \bigg[ \frac{1}{\delta^\nu}x^{2a_1^\nu}(1-x)^{2b_1^\nu+1}+ x^{2a_2^\nu-2}(1-x)^{2b_2^\nu+3}\frac{\kappa^2}{(\delta^\nu)^2 M^2\ln(1/x)}\bigg].
\ee
Where $A$ represents the isoscalar-vector($V$) diquark corresponding to $u$ quark and isovector-vector($VV$) diquark corresponding to $d$ quark.

We simulate the scale evolution of the PDF  by making the parameters in the PDF  scale dependent where the values of the parameters at $\mu_0$ are  the same as in the LFWFs.  Thus, at a scale $\mu$, we parameterize the expressions for the PDFs as
\be 
f^{(S)}_1(x,\mu) &=&  N^{2}_S(\mu)\bigg[\frac{1}{\delta^u(\mu)} x^{2a_1^u(\mu)}(1-x)^{2b_1^u(\mu)+1}\nonumber\\
&&~~~~~~~~~~~ + x^{2a_2^u(\mu)-2}(1-x)^{2b_2^u(\mu)+3}\frac{\kappa^2}{(\delta^u(\mu))^2 M^2\ln(1/x)}\bigg],\\
f^{(A)}_1(x,\mu)&=&  \bigg(\frac{1}{3}N^{(\nu)2}_0(\mu)+\frac{2}{3}N^{(\nu)2}_1(\mu)\bigg)\nonumber\\
&&\!\!\! \times \bigg[ \frac{1}{\delta^\nu(\mu)}x^{2a_1^\nu(\mu)}(1-x)^{2b_1^\nu(\mu)+1}+ x^{2a_2^\nu(\mu)-2}(1-x)^{2b_2^\nu(\mu)+3}\frac{\kappa^2}{(\delta^\nu(\mu))^2 M^2\ln(1/x)}\bigg].
\ee
 The assumption is that, not only at every scale there exists a set of parameters to reproduce the desired PDFs but we can also define an evolution formula for each of  these parameters consistent with PDF evolution starting from the initial scale $\mu_0$. 
 
The flavour decomposed PDFs are given as, from Eq.(\ref{Fiu},\ref{Fid}) 
\be 
f^{u}_1(x,\mu)&=&C_S^2 f^{(S)}_1(x,\mu)+C_V^2 f^{(V)}_1(x,\mu),\\
f^{d}_1(x,\mu)&=&C_{VV}^2 f^{(VV)}_1(x,\mu).
\ee


\begin{figure}[htbp]
\begin{minipage}[center]{0.98\textwidth}
\small{(a)}\includegraphics[width=7.5cm,clip]{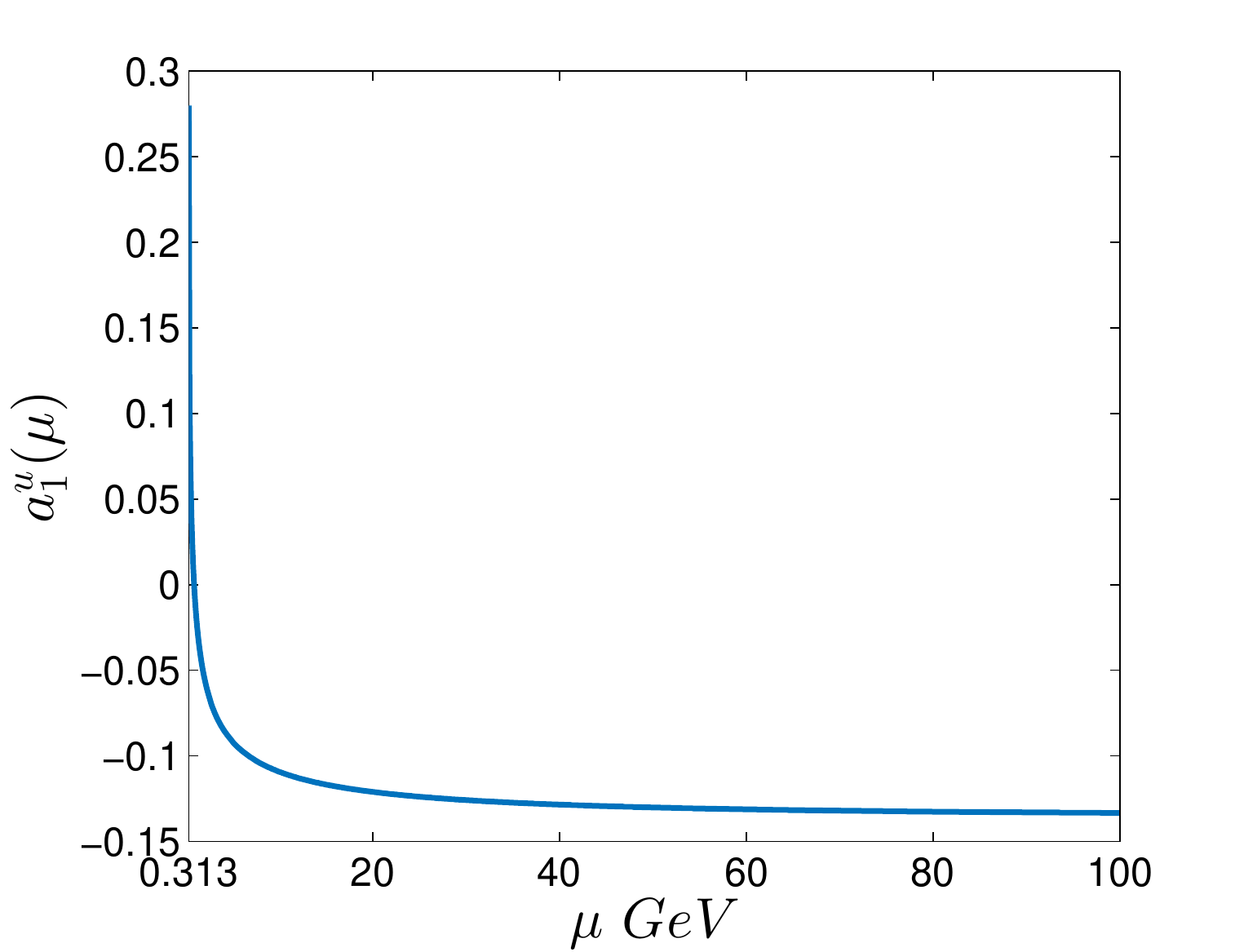}
\small{(b)}\includegraphics[width=7.5cm,clip]{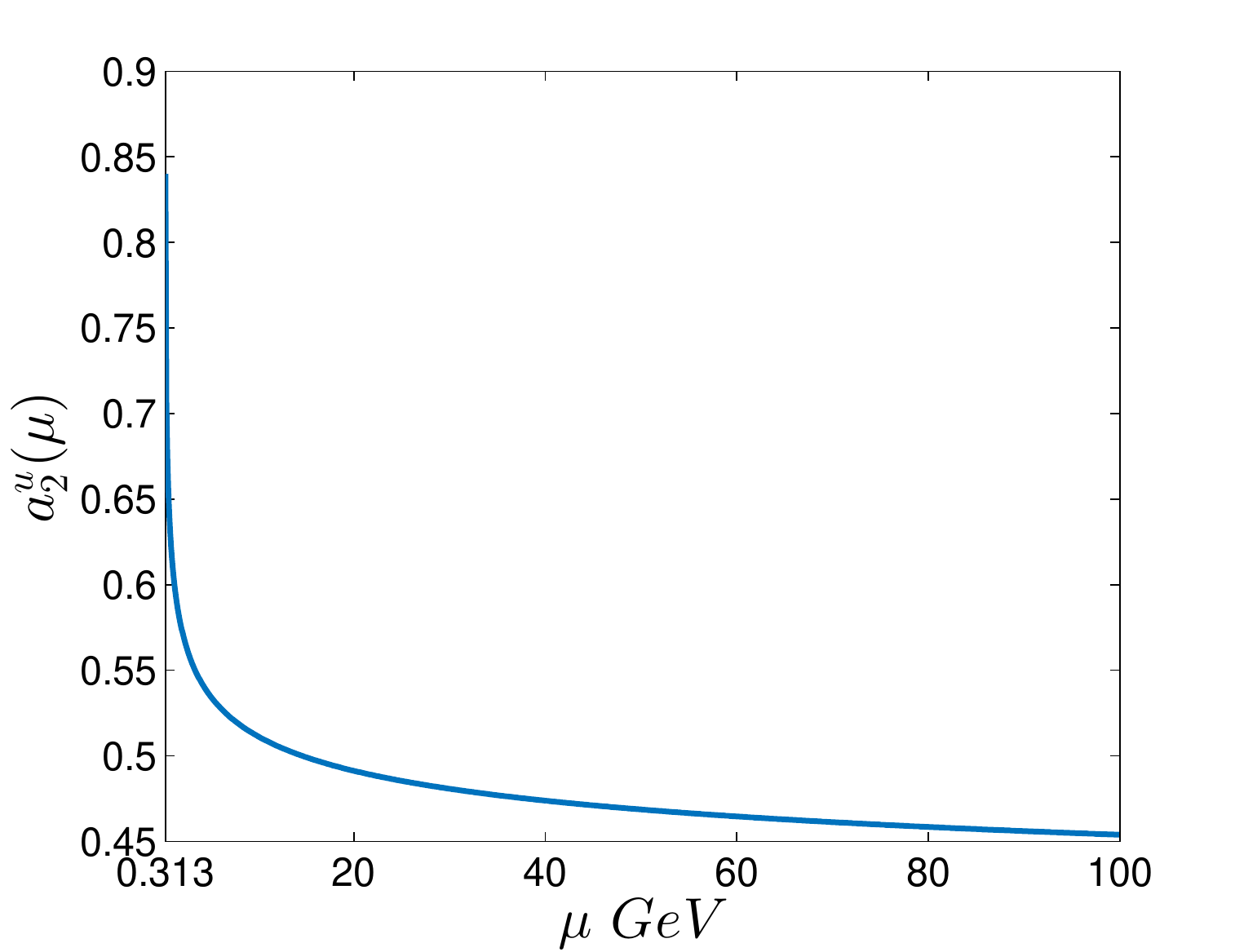}
\end{minipage}
\begin{minipage}[center]{0.98\textwidth}
\small{(c)}\includegraphics[width=7.5cm,clip]{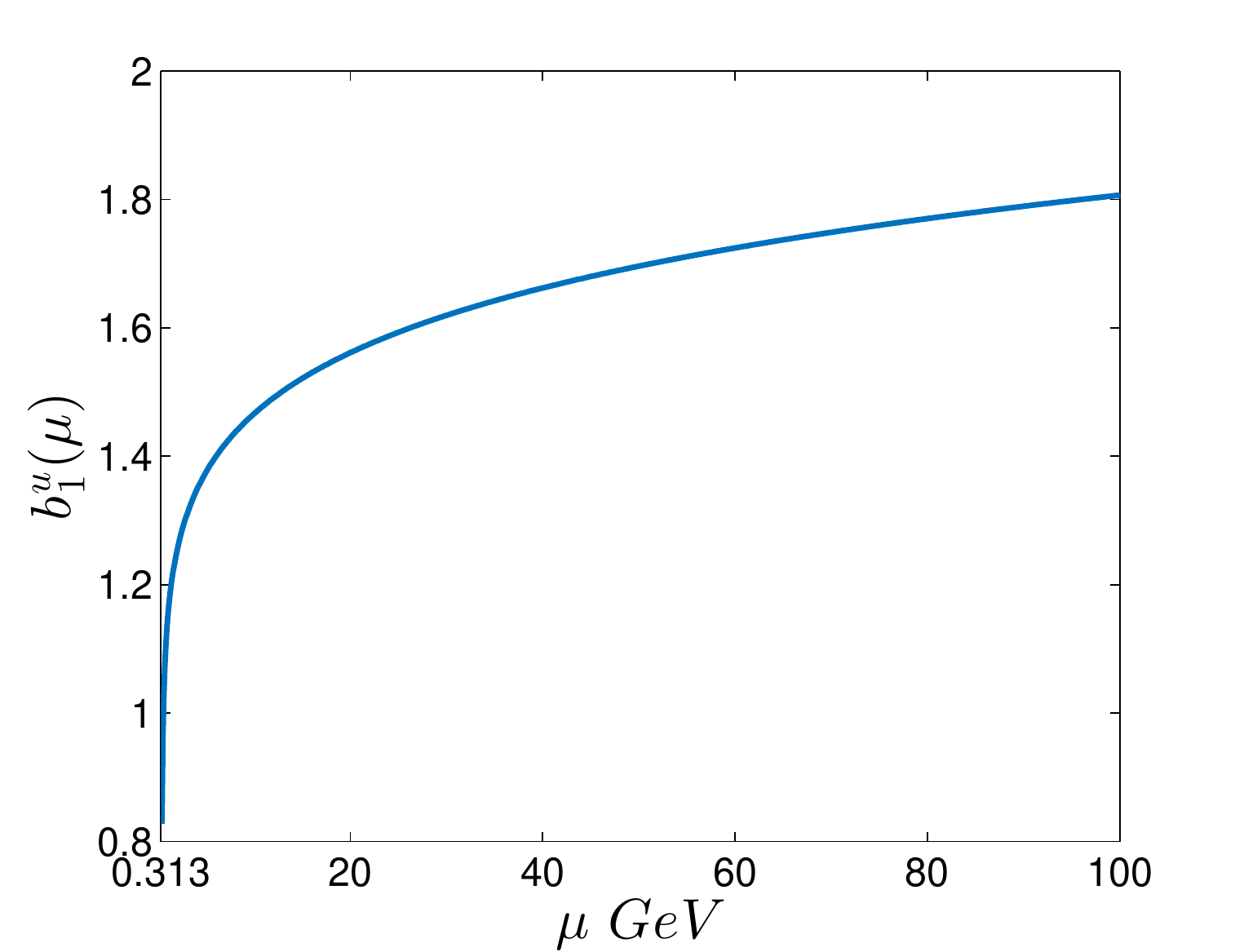}
\small{(d)}\includegraphics[width=7.5cm,clip]{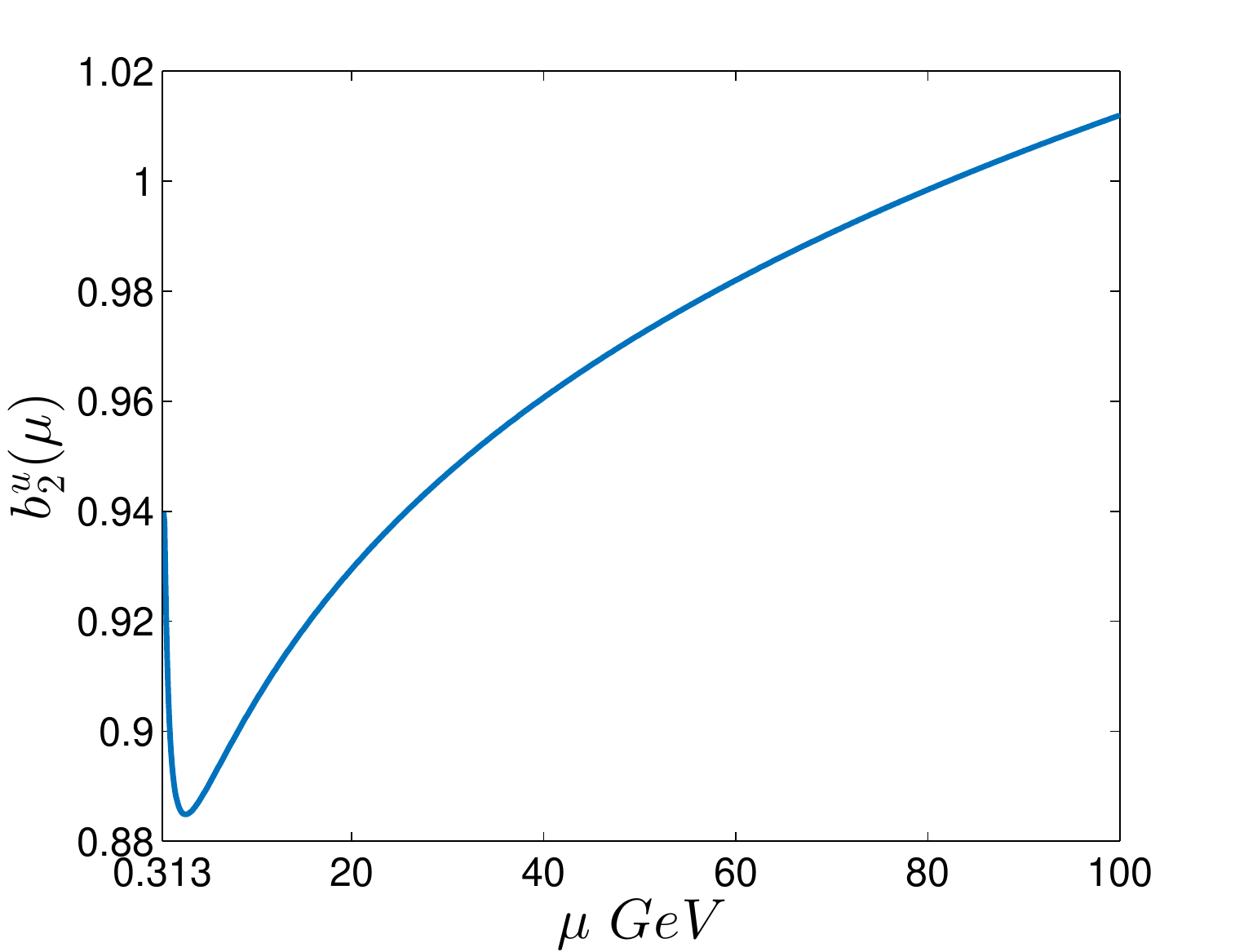}
\end{minipage}
\begin{minipage}[center]{0.98\textwidth}
\small{(a)}\includegraphics[width=7.5cm,clip]{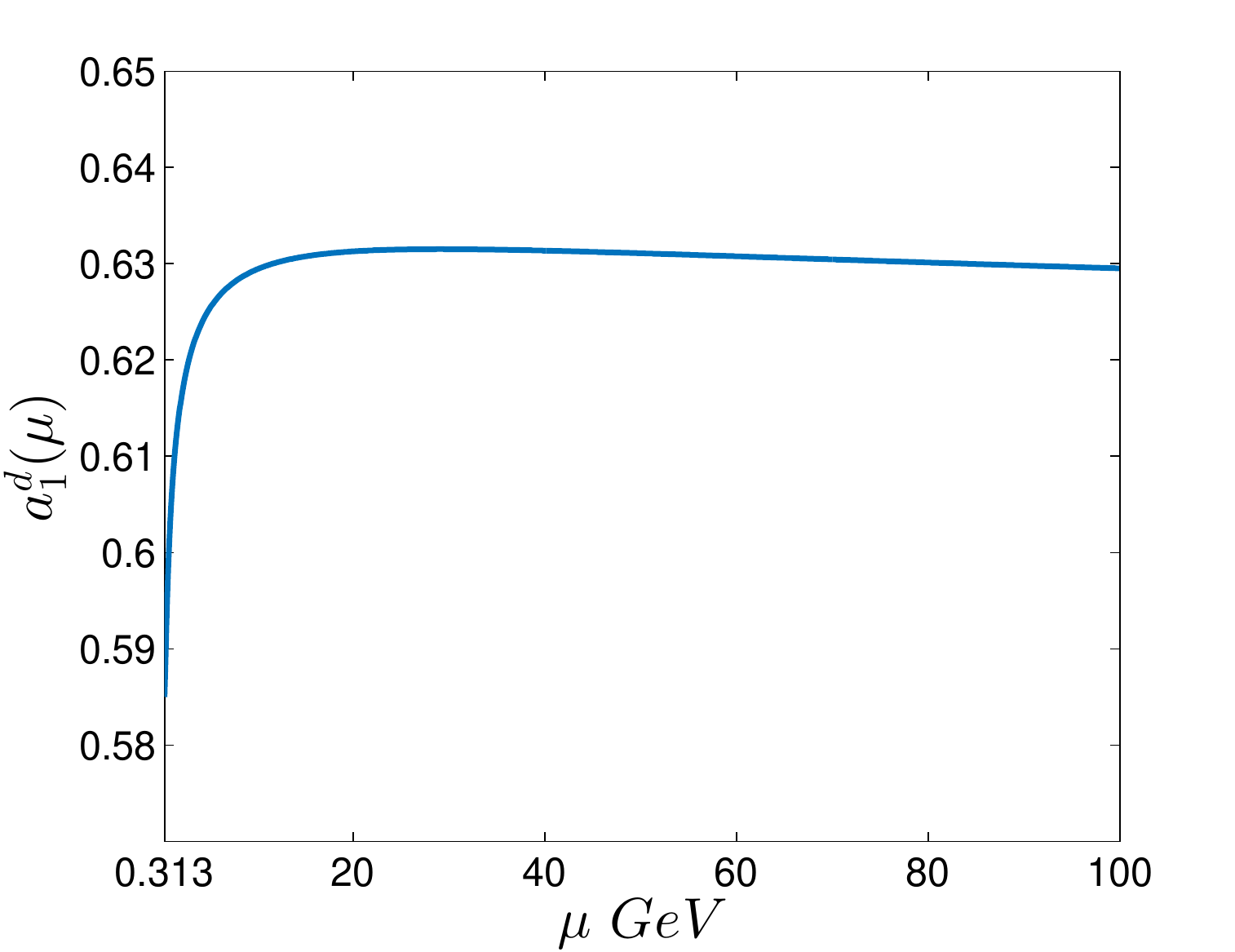}
\small{(b)}\includegraphics[width=7.5cm,clip]{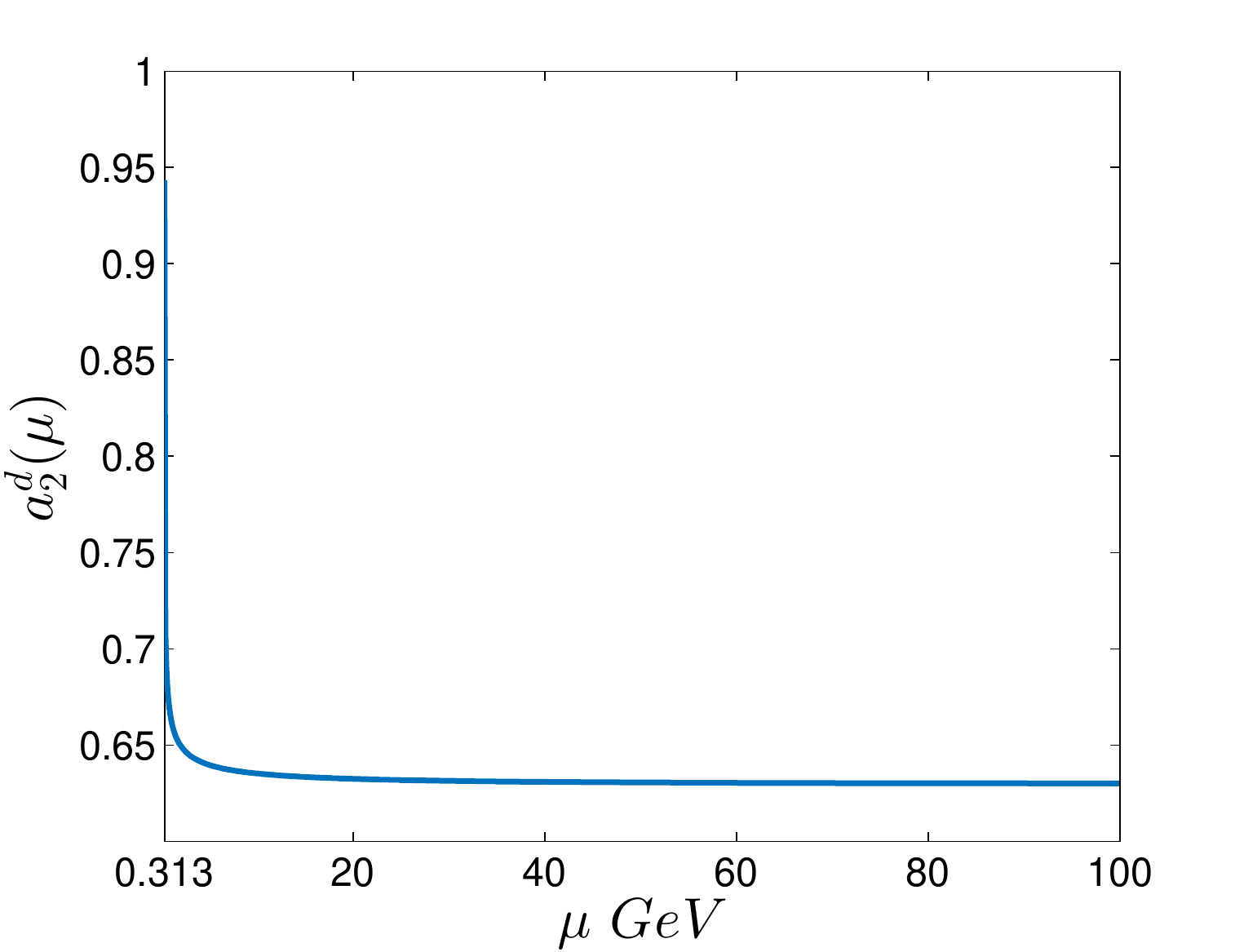}
\end{minipage}
\begin{minipage}[center]{0.98\textwidth}
\small{(c)}\includegraphics[width=7.5cm,clip]{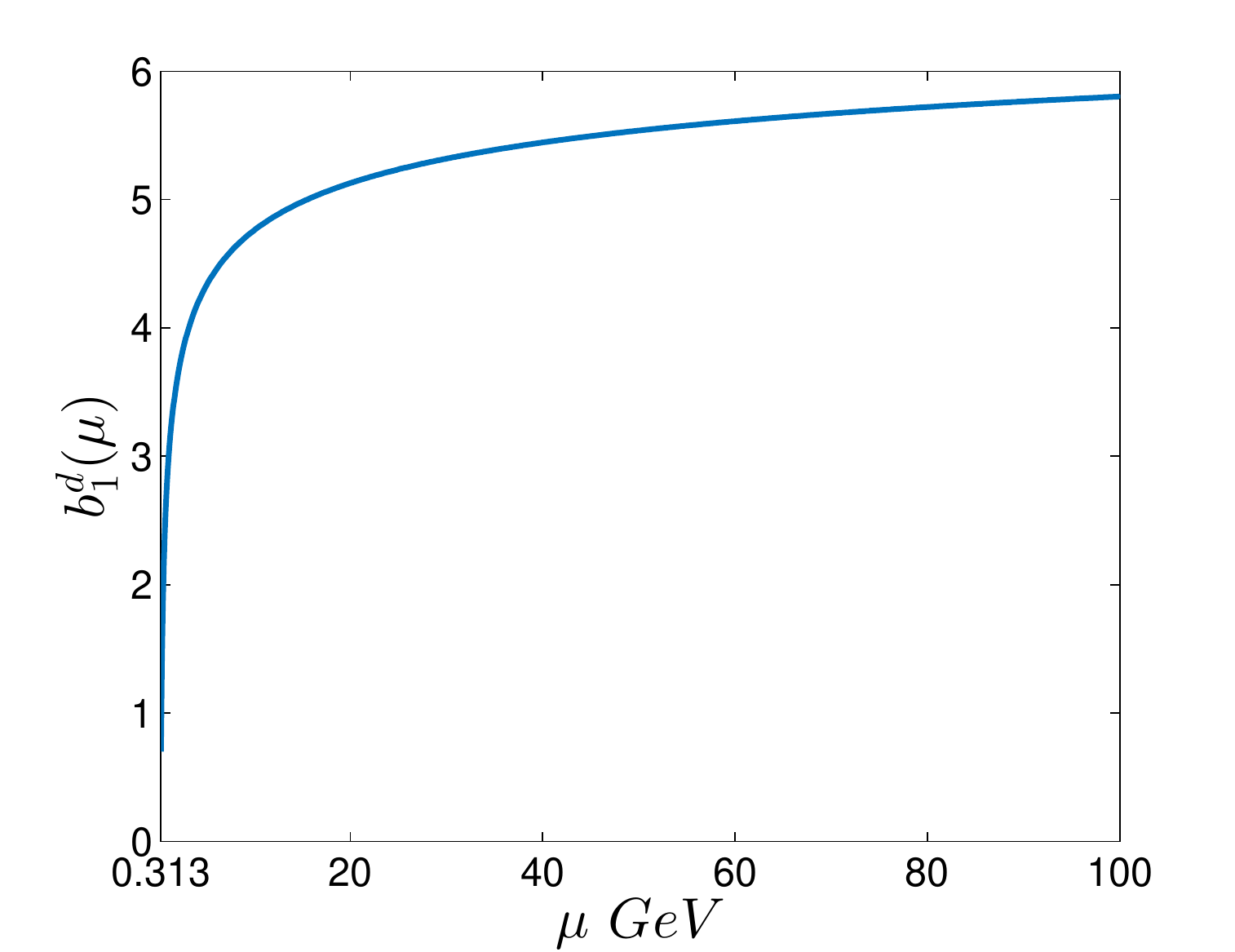}
\small{(d)}\includegraphics[width=7.5cm,clip]{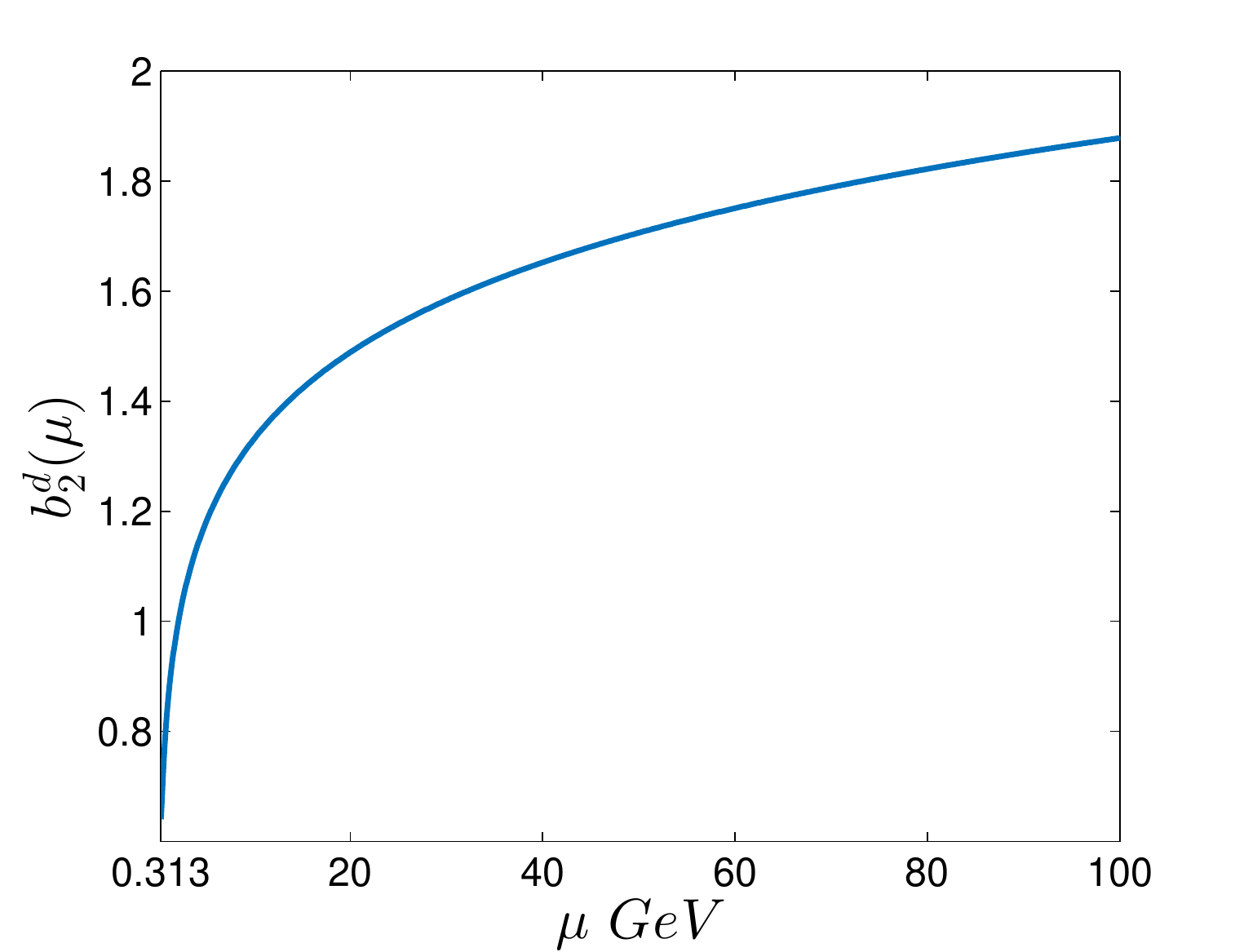}
\end{minipage}
\caption{\label{fig_ai_mu} Scale evolution of the parameters [Eq.(\ref{a_im}),(\ref{b_im})].}
\end{figure}

\begin{figure}[htbp]
\small{(a)}\includegraphics[width=7.5cm,clip]{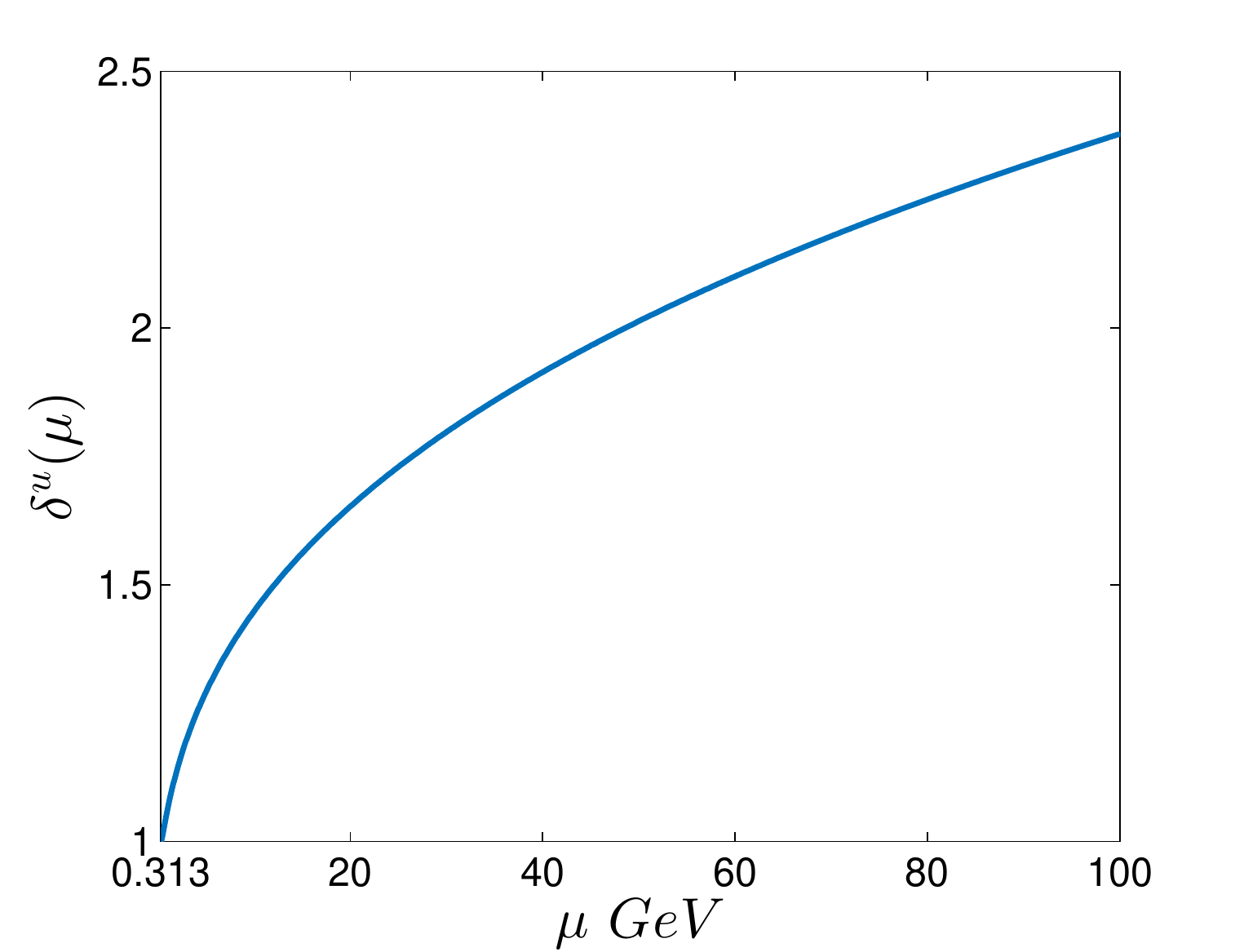}
\small{(b)}\includegraphics[width=7.5cm,clip]{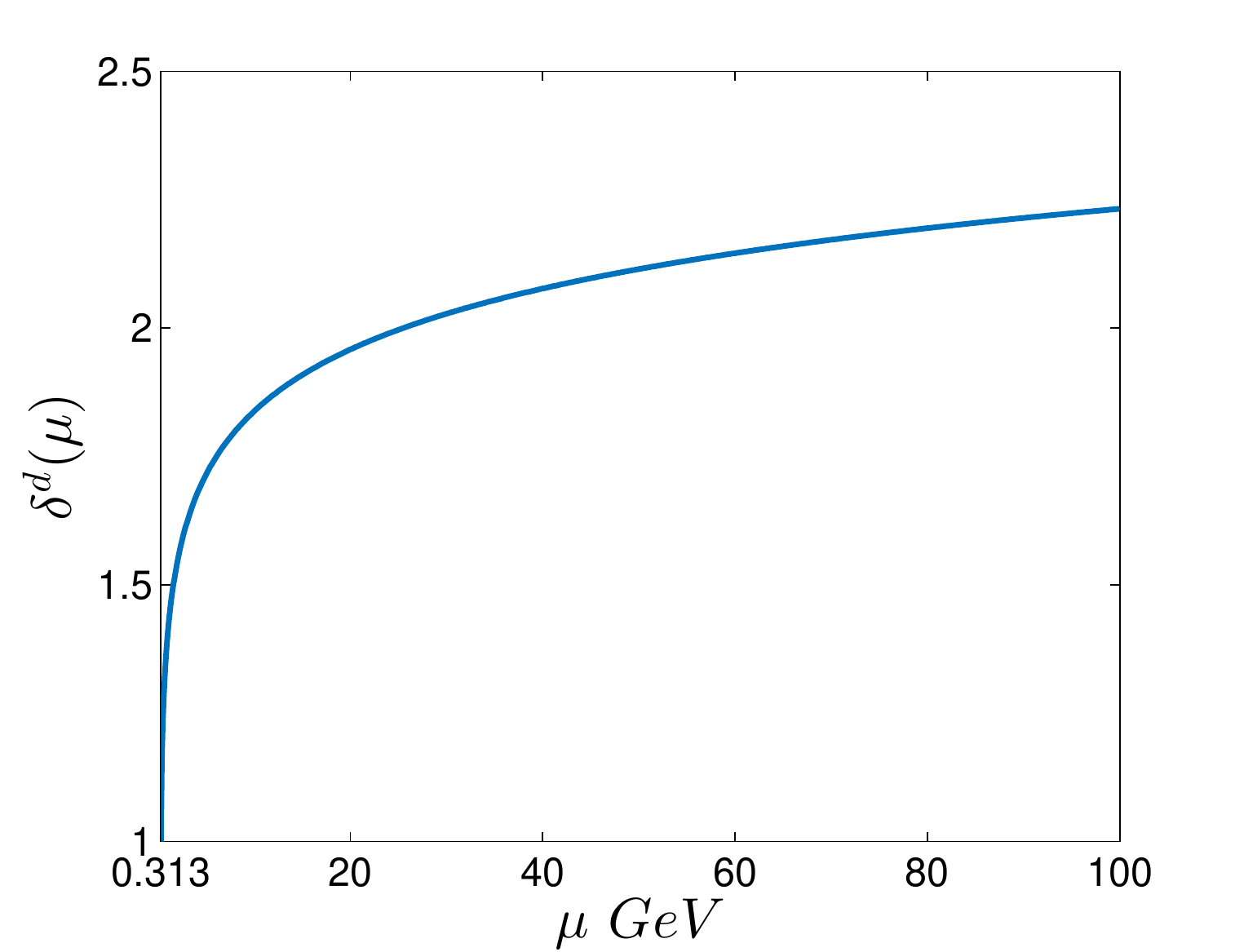}
\caption{\label{fig_DL}  Variation of $\delta^\nu$ with $\mu$ for $u$ and $d$ quark, see Eq.\ref{DL} }
\end{figure}

\begin{figure}[htbp]
\begin{minipage}[center]{0.98\textwidth}
\small{(a)}\includegraphics[width=7.5cm,clip]{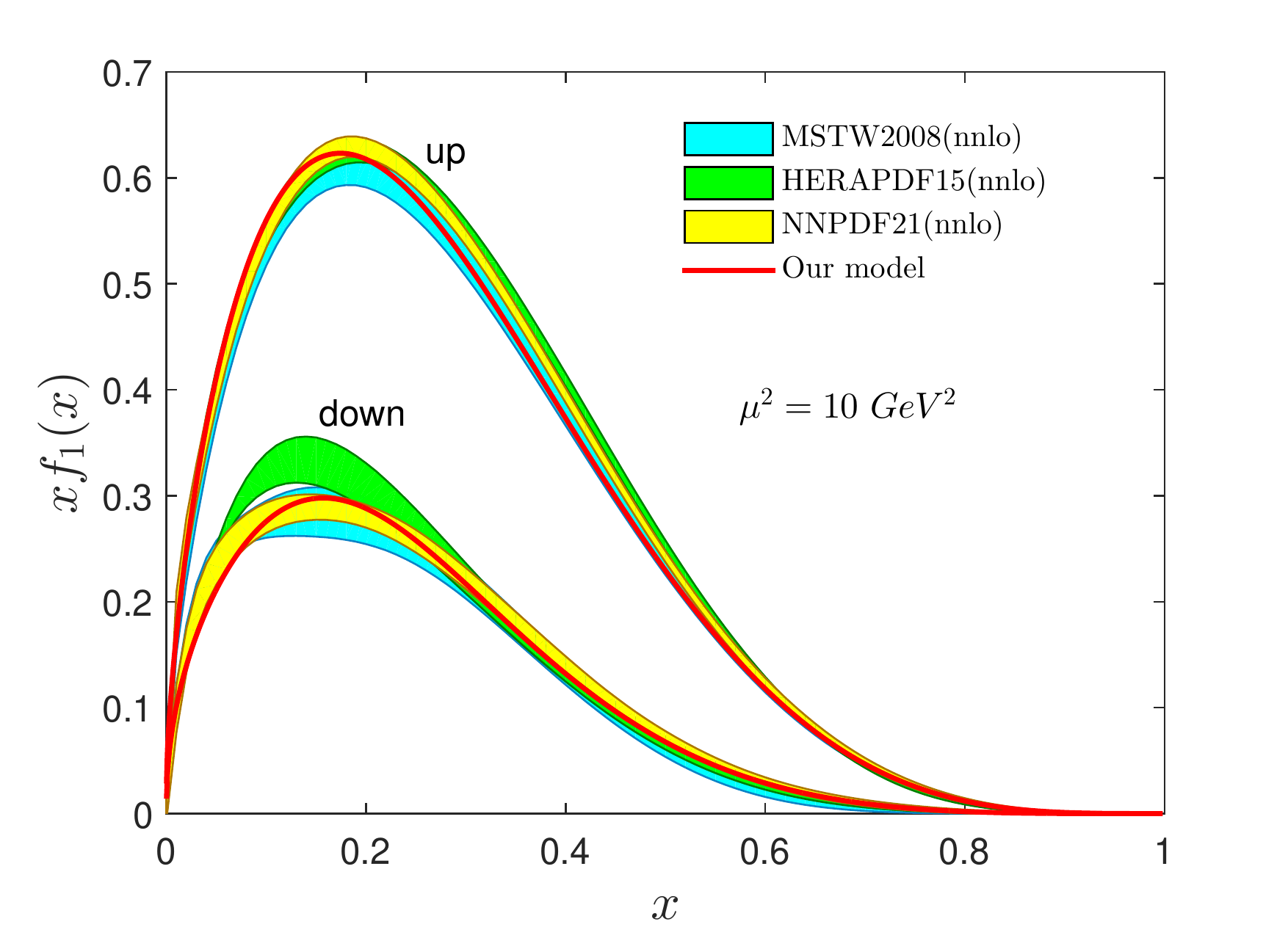}
\small{(b)}\includegraphics[width=7.5cm,clip]{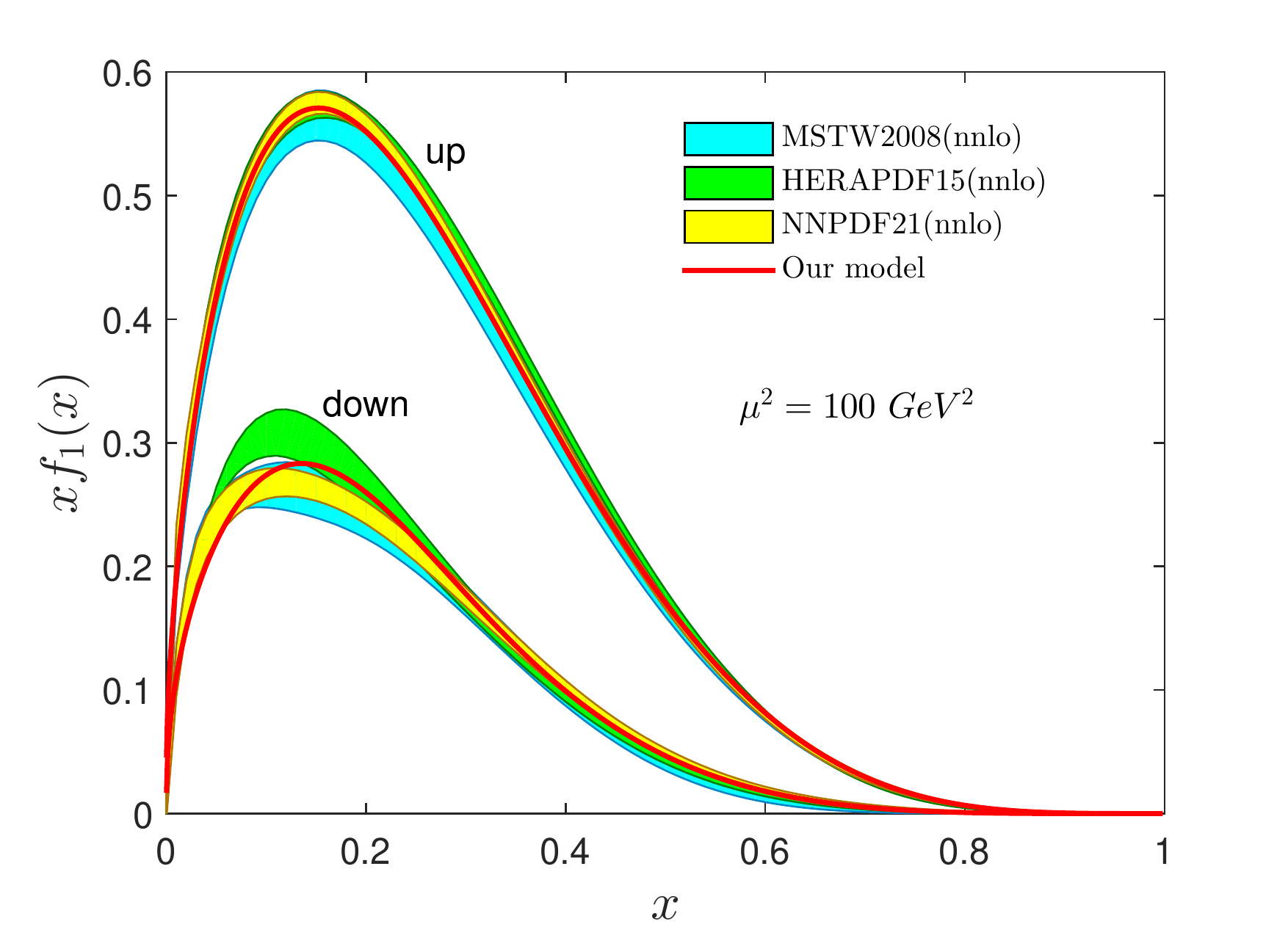}
\end{minipage}
\begin{minipage}[center]{0.98\textwidth}
\small{(c)}\includegraphics[width=7.5cm,clip]{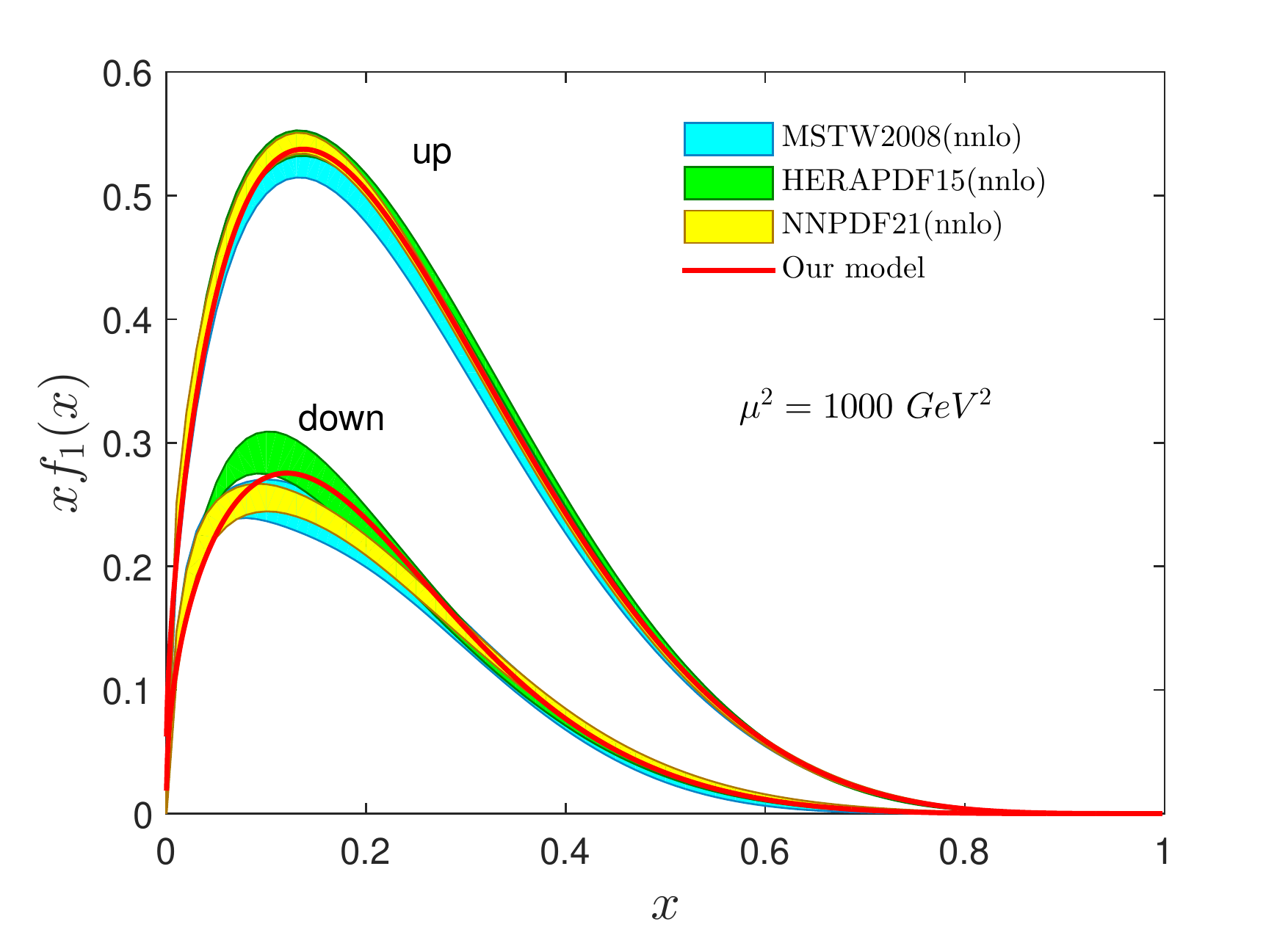}
\small{(d)}\includegraphics[width=7.5cm,clip]{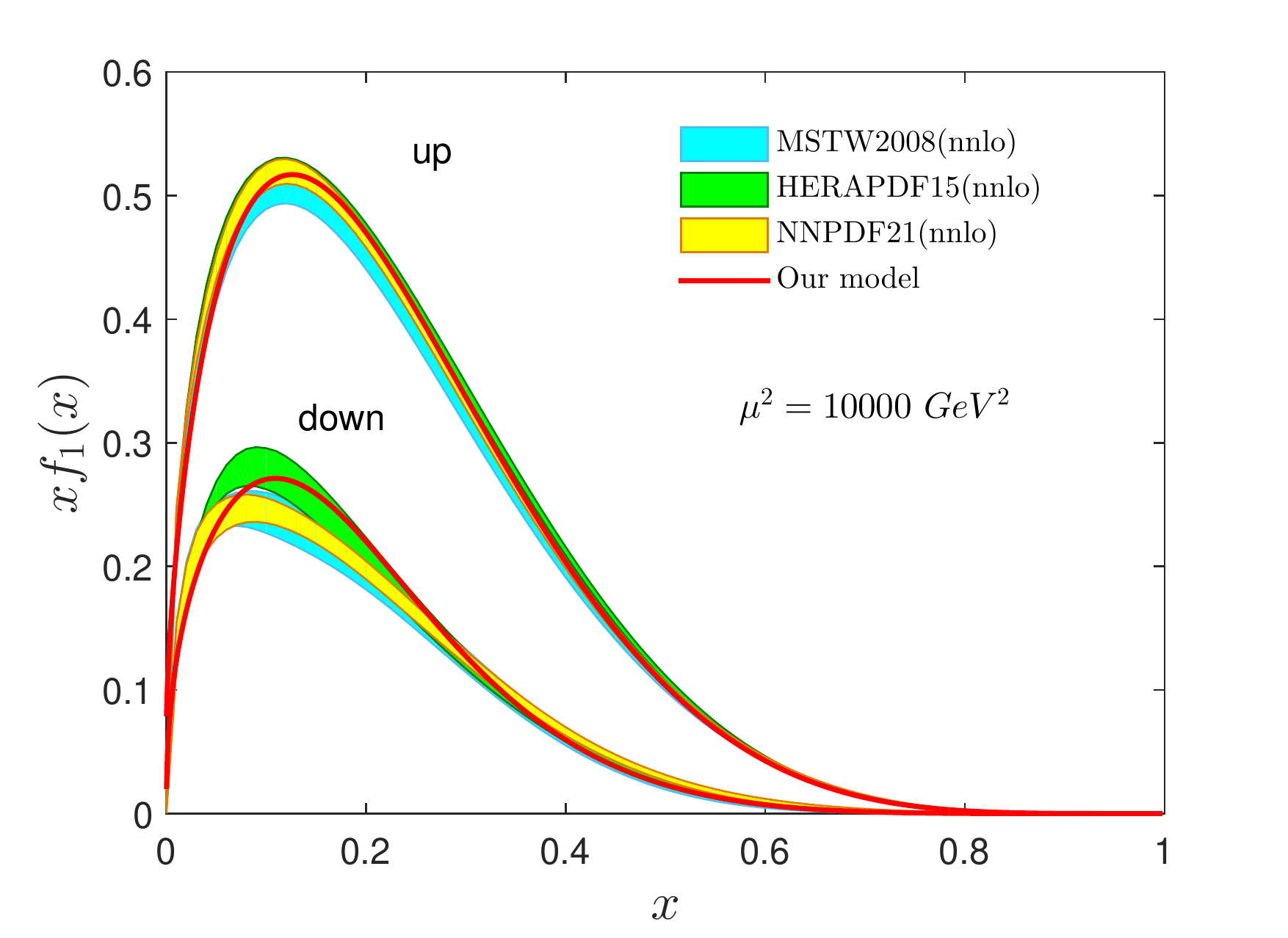}
\end{minipage}
\caption{\label{fig_PDFmu} Evolution of unpolarized PDF in this model at  $\mu^2=10,100,1000$ and $10000$ GeV$^2$ for both $u$ and $d$ quarks. Our model predictions are  compared with NNPDF21(NNLO)\cite{NNPDF}, HERAPDF15(NNLO)\cite{HERAPDF} and MSTW2008(NNLO)\cite{MSTW} results.}
\end{figure}

The PDF $f_1^\nu(x,\mu)$ at a scale $\mu$ can be written in our model as 
\be 
f^\nu_1(x,\mu)&=&  N^{(\nu)}\bigg[\frac{1}{\delta^\nu(\mu)} x^{2a_1^\nu(\mu)}(1-x)^{2b_1^\nu(\mu)+1}+ x^{2a_2^\nu(\mu)-2}(1-x)^{2b_2^\nu(\mu)+3}\frac{\kappa^2}{(\delta^\nu(\mu))^2 M^2\ln(1/x)}\bigg].\label{Eq_xf1mu}
\ee

The evolution of the parameters should be such that the PDF 
 satisfies the master evolution equation, Eq.(\ref{DGLAP_Eq}). 
The overall constants $N^{(u)}=(C_S^2N_s^2+C^2_V(\frac{1}{3} N_0^{(u)2}+\frac{2}{3} N_1^{(u)2}))$ and $N^{(d)}=C^2_{VV}(\frac{1}{3} N_0^{(d)2}+\frac{2}{3} N_1^{(d)2})$ for $u$ and $d$ quarks respectively. To fit the PDF data from NNPDF21(nnlo)\cite{NNPDF}, we find that the  scale dependence of the parameters can be written as
\be 
a_i^\nu(\mu)&=&a_i^\nu(\mu_0) + A^\nu_{i}(\mu), \label{a_im}\\
b_i^\nu(\mu)&=&b_i^\nu(\mu_0) - B^\nu_{i}(\mu)\frac{4C_F}{\beta_0}\ln\bigg(\frac{\alpha_s(\mu^2)}{\alpha_s(\mu_0^2)}\bigg),\label{b_im}\\
\delta^\nu(\mu)&=& \exp\bigg[\delta^\nu_1\bigg(\ln(\mu^2/\mu_0^2)\bigg)^{\delta^\nu_2}\bigg],\label{DL}
\ee
where the $a_i^\nu(\mu_0)$ and $b_i^\nu(\mu_0)$ are the parameters at $\mu=\mu_0$, given in Table.\ref{tab_para_mu0}. The parameter $\delta^\nu$ becomes unity at $\mu_0$ for both $u$ and $d$ quarks, as shown in Table.\ref{tab_para_mu0}.
The scale dependent parts $A^\nu_{i}(\mu)$ and $B^\nu_{i}(\mu)$  evolve as 
\be 
P^\nu_{i}(\mu)&=&\alpha^\nu_{P,i} ~\mu^{2\beta^\nu_{P,i}}\bigg[\ln\bigg(\frac{\mu^2}{\mu_0^2}\bigg)\bigg]^{\gamma^\nu_{P,i}}\bigg|_{i=1,2} ,\label{Pi_evolu}
\ee
where the subscript $P$ in the right hand side of the above equation stands for $P=A,B$ corresponding to $P^\nu_{i}(\mu)=A^\nu_{i}(\mu), B^\nu_{i}(\mu)$ respectively. Note that at $\mu=\mu_0$, $P^\nu_{i}(\mu_0)=0$. The unpolarized PDF data  are fitted  for $\mu^2=1,6,16,30,65$ and $150~ GeV^2$. 
The  evolution parameters $\alpha_{P,i}^\nu,~\beta_{P,i}^\nu$ and $\gamma_{P,i}^\nu$ are given in Table.\ref{tab_evopar} and the $\delta^\nu_i$ are given in Table.\ref{tab_DL} with the least $\chi^2$ per degrees of freedom ($\chi^2/d.o.f$) corresponding to the  PDF fit. 
The variation of $a_i^\nu(\mu)$, $b_i^\nu(\mu)$ and $\delta^\nu(\mu)$ with scale $\mu$ are shown in Fig.\ref{fig_ai_mu} and in Fig.\ref{fig_DL} respectively. At the initial scale $\mu_0$, the strong coupling  constant  is large $\alpha_s(\mu_0)/2\pi\sim 0.34$ and hence parameters evolve  very fast for  scales near the initial point. The rate of evolution decays down at higher scales where the coupling constant becomes small.  In appendix \ref{appa}, we have listed the parameters at different scales and shown the fitting of the parameters at the above mentioned scales.

With the fitted parameters, we predict the unpolarized PDF at other scales and compare with NNPDF21(nnlo), HERAPDF15(nnlo) and MSTW2008(nnlo) results in Fig.\ref{fig_PDFmu}.  Though to determine the evolution, we used up to $\mu^2=150 $ GeV$^2$, Fig.\ref{fig_PDFmu} shows that the model  reproduces the PDF quite accurately at very high scales.
 According to Drell-Yan-West relation\cite{DY70,West70}, the quark distribution should go like $(1-x)^p$ as $x \to 1$ at large $\mu^2$ and the Dirac form factor $F_1(Q^2) \sim 1/(Q^2)^{(p+1)/2}$ as $Q^2\to\infty$ where $p$ is related to the number of valence quark. For, proton, the number of valence quark is three which also gives $p=3$. In our model, we observe that for $u$-quark the unpolarized PDF at large $\mu^2$ goes as $f_1^u\sim (1-x)^{3.35}$  as $x\to1$ ( for $d$ quark $f_1^d\sim(1-x)^{3.09}$ ), and  $F_1(Q^2)\sim 1/(Q^2)^{2.16}$ at large $Q^2$ which are  consistent with Drell-Yan-West relation.


With these above PDF evolution parameters we can generate the scale evolution of the other distribution functions e.g. helicity distribution, transversity distribution, GPDs, TMDs etc. 
Now we show that the model can predict the helicity and transversity distributions at different scales.

\begin{table}[ht]
\centering 
\begin{tabular}{|c c c c c|}
 \hline
 $P_i^\nu(\mu)$~~&~~$\alpha_i^\nu$~~&~~$\beta_i^\nu$~~ & ~~$\gamma_i^\nu$~~ & ~~$\chi^2/d.o.f$~~  \\ \hline
$A_1^u$ &~~ $-0.2058\pm0.0187$ ~~&~~ $-0.0318\pm0.0209$ ~~&~~ $0.405\pm0.0937$ ~~&~~ 0.23\\
$B_1^u$ & $1.551^{+0.034}_{-0.035}$ & $0.0598\pm0.0057$ & $-0.4291\pm0.0242$ &0.02\\ 
$A_2^u$ & $-0.1637\pm0.0179$ & $-0.0066\pm0.0245$ & $0.3758\pm0.1111$ & 0.13\\ 
$B_2^u$ & $1.426\pm0.320$ & $0.07780.0605$ & $-0.7634^{+0.0241}_{-0.0251}$ &0.07\\ 
$A_1^d$ &~~ $0.0061\pm0.0098$ ~~&~~ $-0.1535\pm0.0257$ ~~&~~ $1.391^{+0.246}_{-0.245}$ ~~&~~ 0.14\\
$B_1^d$ & $2.072\pm0.0193$ & $-0.008^{+0.021}_{-0.022}$ & $0.1728\pm0.0972$ &0.10\\
$A_2^d$ & $-0.2493\pm0.0456$ & $-0.0116\pm0.0408$ & $0.1371\pm0.1783$ & 0.29\\ 
$B_2^d$ & $0.1399\pm0.0737$ & $0.0247\pm0.1086$ & $0.5733^{+0.0517}_{-0.0518}$ &0.10\\
 \hline
 \end{tabular} 
\caption{PDF evolution parameters with 95\% confidence bounds.} 
\label{tab_evopar} 
\end{table}
\begin{table}[ht]
\centering  
\begin{tabular}{|c c c c |}
 \hline
 $\delta^\nu(\mu)$~~~&~~~$\delta_1^\nu$~~~~&~~~~~$\delta_2^\nu$~~~ & ~~$\chi^2/d.o.f$~~  \\ \hline
$\delta^u$ & $0.015\pm0.008$ & $1.667\pm0.032$ & 1.16\\
$\delta^d$ & $0.212\pm0.0566$ & $~0.5444\pm0.1504$ & 0.81 \\ \hline
\end{tabular} 
\caption{PDF evolution parameter $\delta^\nu_1$ and $\delta^\nu_2$ for $\nu=u,d$. }  
\label{tab_DL}  
\end{table}
\section{Model predictions}\label{predicts}
\subsection{Helicity distributions and axial charges}
The polarized PDFs are evaluated as predictions of the model.  In terms of the light front wave functions,
 the helicity distribution $g_1(x)$  in the quark-diquark model at the initial scale $\mu_0$ is defined as 
\be 
g^{(S)}_1(x)&=& \int d^2\bfp \frac{1}{16\pi^3}\bigg[|\psi^{+(u)}_+(x,\bfp )|^2-|\psi^{+(u)}_-(x,\bfp)|^2\bigg],\\
g^{(A)}_1(x)&=& \int d^2\bfp \frac{1}{16\pi^3}\bigg[|\psi^{+(\nu)}_{++}(x,\bfp )|^2-|\psi^{+(\nu)}_{-+}(x,\bfp)|^2 \nonumber\\
&&~~~+ |\psi^{+(\nu)}_{+0}(x,\bfp )|^2-|\psi^{+(\nu)}_{-0}(x,\bfp)|^2\bigg].
\ee
for scalar and vector diquarks respectively. The scale evolutions of the polarized PDFs are simulated by the same scheme as the unpolarized PDF. The scale evolutions of the parameters are the same as given in Eqs.(\ref{a_im},\ref{b_im},\ref{DL},\ref{Pi_evolu}).
Thus, the flavour dependent helicity distributions at a scale $\mu$ are given by
\be 
g^{u}_1(x,\mu)&=&  \bigg(C_S^2N^{2}_S(\mu)+C^2_V\big(\frac{1}{3}N^{(u)2}_0(\mu)-\frac{2}{3}N^{(u)2}_1(\mu)\big)\bigg)\bigg[\frac{1}{\delta^u(\mu)} x^{2a_1^u(\mu)}(1-x)^{2b_1^u(\mu)+1}\nonumber\\
&&~~~~- x^{2a_2^u(\mu)-2}(1-x)^{2b_2^u(\mu)+3}\frac{\kappa^2}{(\delta^u(\mu))^2M^2\ln(1/x)}\bigg],\\
g^{d}_1(x,\mu)&=&  C^2_{VV}\bigg(\frac{1}{3}N^{(d)2}_0(\mu)-\frac{2}{3}N^{(d)2}_1(\mu)\bigg)\bigg[\frac{1}{\delta^d(\mu)} x^{2a_1^d(\mu)}(1-x)^{2b_1^d(\mu)+1}\nonumber\\
&&~~~~~- x^{2a_2^d(\mu)-2}(1-x)^{2b_2^d(\mu)+3}\frac{\kappa^2}{(\delta^d(\mu))^2M^2\ln(1/x)}\bigg].
\ee

\begin{figure}[htbp]
\begin{minipage}[c]{0.98\textwidth}
\small{(a)}\includegraphics[width=7.5cm,clip]{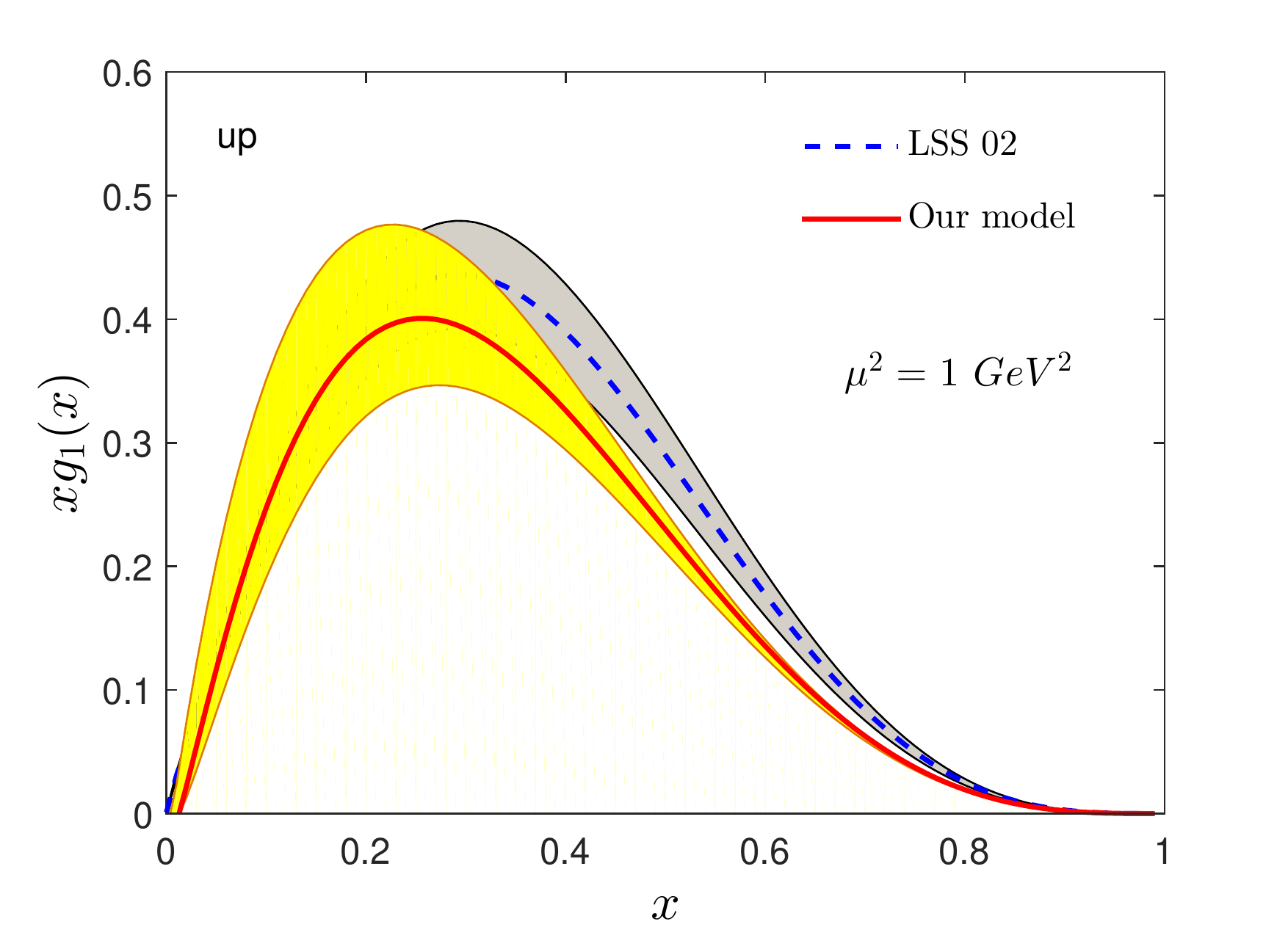}
\small{(b)}\includegraphics[width=7.5cm,clip]{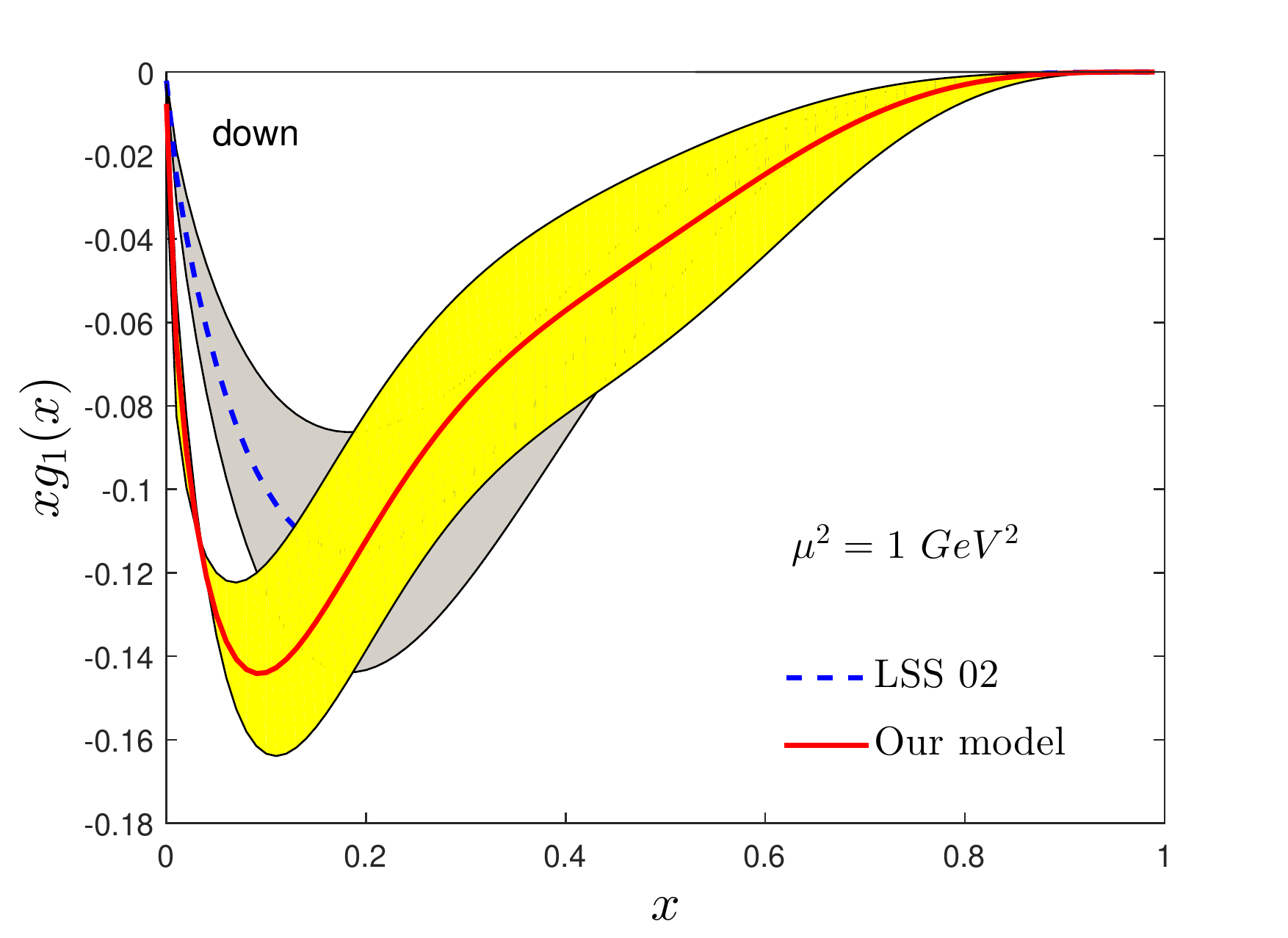}
\end{minipage}
\caption{\label{fig_g1} Helicity PDF at $\mu^2=1~ GeV^2$ compared with ref.\cite{LSS02}. The error bands(yellow) in our model come from the errors in the evolution parameters given in Table.\ref{tab_evopar}.}
\end{figure}

Helicity PDF $g_1(x)$ are shown in Fig.\ref{fig_g1}, at scale $\mu^2=1~GeV^2$, for $u$ and $d$ quarks.  Following  \cite{Bacc08,Hirai04}, we include a constant relative error of $10\%$ to $g^u_1$ and $25\%$ to $g^d_1$ in the data taken from\cite{LSS02}. The errors in the model predictions are due to the uncertainties in the parameters as listed in Tab.\ref{tab_gA}. The model predicts  the helicity PDF for $u$ quark quite well.

\begin{table}[ht]
\centering 
\begin{tabular}{|c|c|c|c|c|c|c|}
 \hline
 & $g^u_A$& $g^d_A$&~$g_A$ & $g^{u(1)}_A$ & $g^{d(1)}_A$ & $g^{(1)}_A$\\ \hline
 Our result & $0.71\pm0.09$ & $-0.54^{+0.19}_{-0.13}$ & $1.25^{+0.28}_{-0.22}$ & $0.18\pm0.15$ & $-0.052^{+0.003}_{-0.007}$ & $0.23^{+0.15}_{-0.16}$ \\ 
  Measured Data & $0.82\pm 0.07$ & $-0.45\pm 0.07$ & $1.27\pm0.14$ & $0.19\pm0.07$ &  $-0.06\pm0.07$ & $0.25\pm0.14$ \\
\hline
 \end{tabular} 
\caption{Axial charge and second moment of helicity distribution at the scale $\mu^2=1~ GeV^2$ and compared with LSS fit to experimental data\cite{Lead10}.} 
\label{tab_gA} 
\end{table} 
\begin{figure}[htbp]
\begin{minipage}[c]{0.98\textwidth}
\small{(a)}\includegraphics[width=7.5cm,clip]{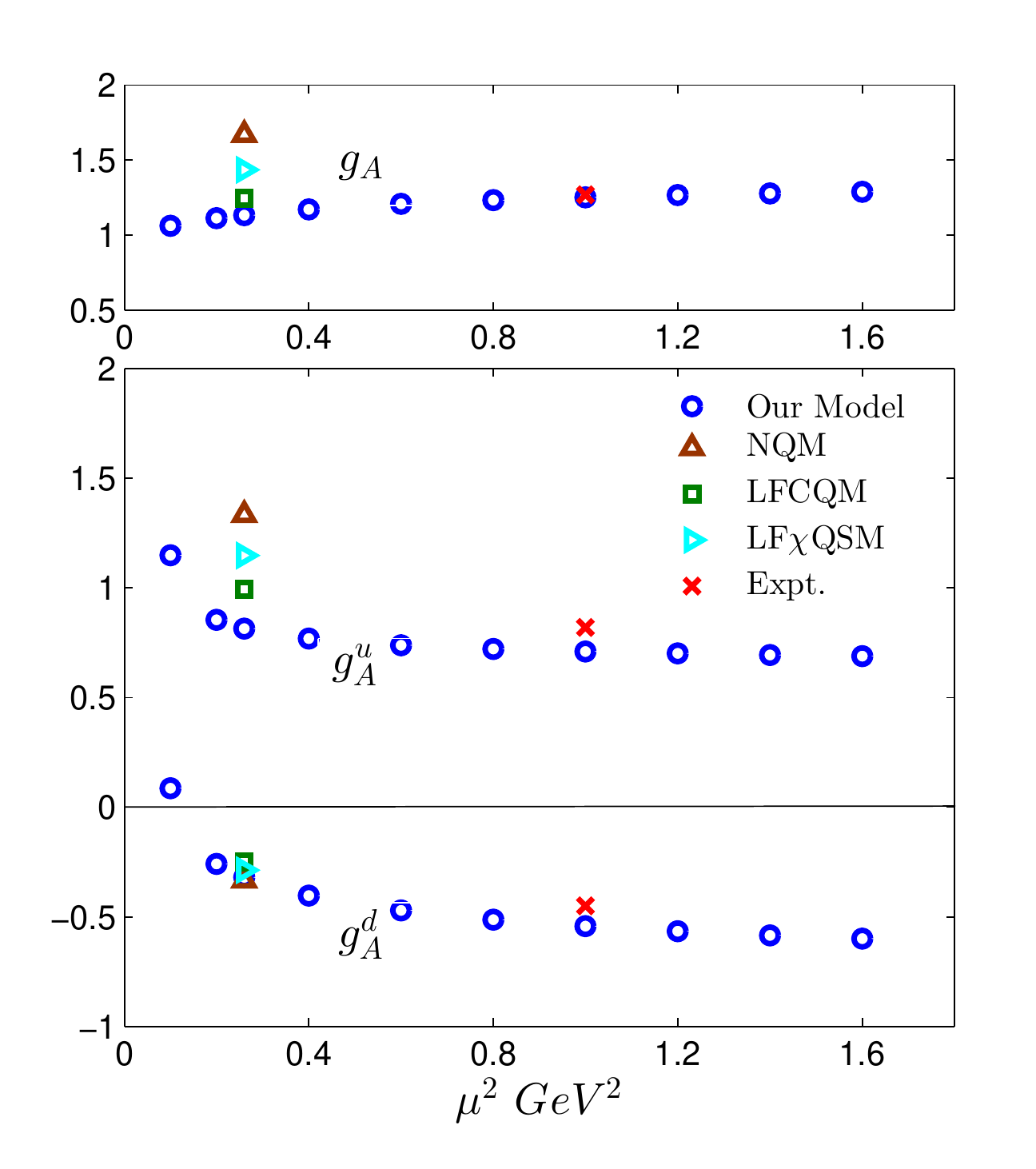}
\small{(b)}\includegraphics[width=7.5cm,clip]{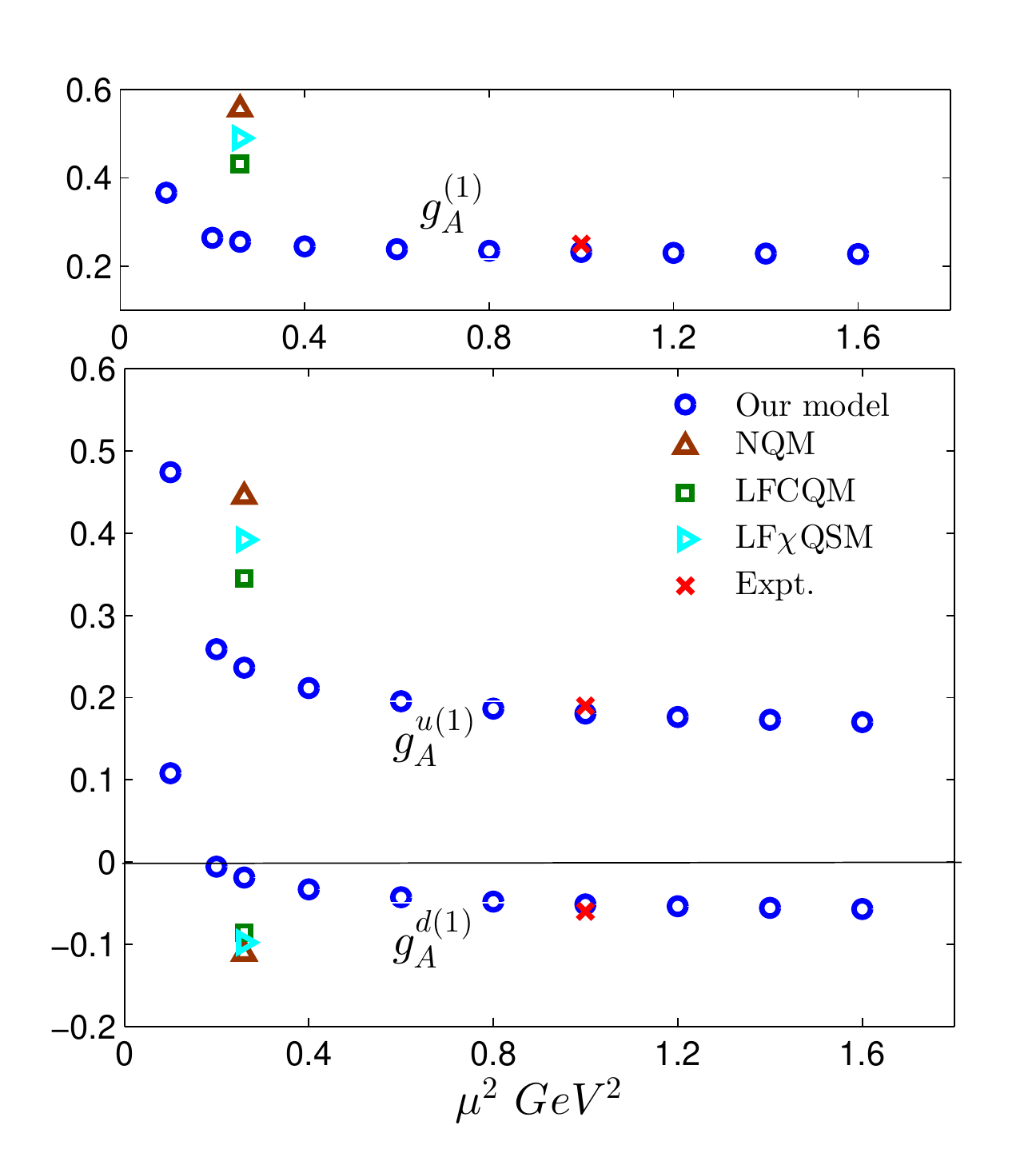}
\end{minipage}
\caption{\label{fig_gA} Scale evolution of axial charge in the range $\mu^2=0.1 ~to~1.6 ~GeV^2$ is shown in(a). We compare our result with other models e.g. NQM, LFCQM, LF$\chi$QSM\cite{Lorce11} for $\mu^2=0.26$ GeV$^2$ and also with experimental value at $\mu^2=1.0~GeV^2$\cite{Lead10}. The second moment of helicity and a comparison with measured value\cite{Lead10} at $\mu^2=0.26,1.0~GeV^2$ are shown in (b).  The top panels in the plots represent the total proton axial charge (a) and second moment of the total helicity distribution (b). }
\end{figure}
The axial charges which are obtained from the first moment of the helicity distributions are given in Table.\ref{tab_gA} and compared with the measured data\cite{Lead10}. The axial charge of proton is  defined as 
\be  g_A=g_A^u-g_A^d. \ee
The model prediction is in excellent agreement with the experimental data.
 The second  moment of the helicity distributions are also presented in the table where $g_A^{\nu(1)}=\int_0^1 dx x g_1^\nu(x)$ and  $g_A^{(1)}$ is defined as 
\be g_A^{(1)}=\int_0^1 dx x (g_1^u(x)-g_1^d(x)).
\ee
 In Fig.\ref{fig_gA}(a),  the scale evolution of the axial charges for $u$ and $d$ quarks are shown. The top panels in Fig.\ref{fig_gA}(a) represents the axial charge of  the proton,  $g_A$.  Results from other models and experimental data are shown in the same plot for comparison. Results from other models e.g., NQM, LFCQM, LF$\chi$QSM\cite{Lorce11} at $\mu^2=0.26$ GeV$^2$ are in agreement with our model prediction and again our model predicts the experimental data at $\mu^2=1~\rm{GeV}^2$ (shown in red in Fig.\ref{fig_gA}) quite well. In Fig.\ref{fig_gA}(b), the scale evolution of the second moment of the helicity distributions for both $u$ and $d$ quarks are shown and compared with other model predictions and experimental data. The top panel in the plot represents   $g_A^{(1)}$.
 Our model predictions  show excellent agreement with  the experimental data available at $\mu^2=1$ GeV$^2$.  

\subsection{Transversity distributions and tensor charges}
\begin{figure}[htbp]
\begin{minipage}[c]{0.98\textwidth}
\small{(a)}\includegraphics[width=7.5cm,clip]{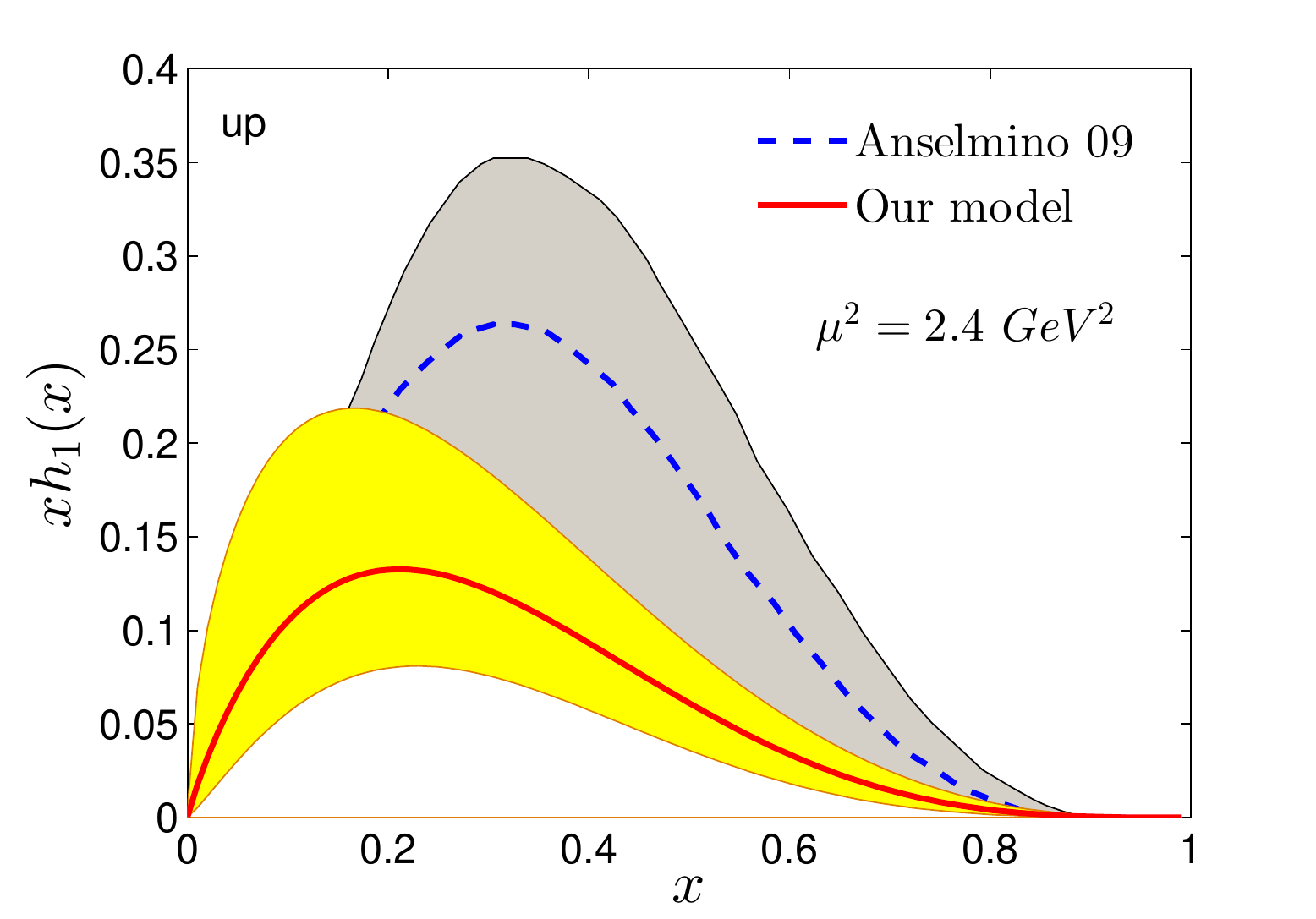}
\small{(b)}\includegraphics[width=7.5cm,clip]{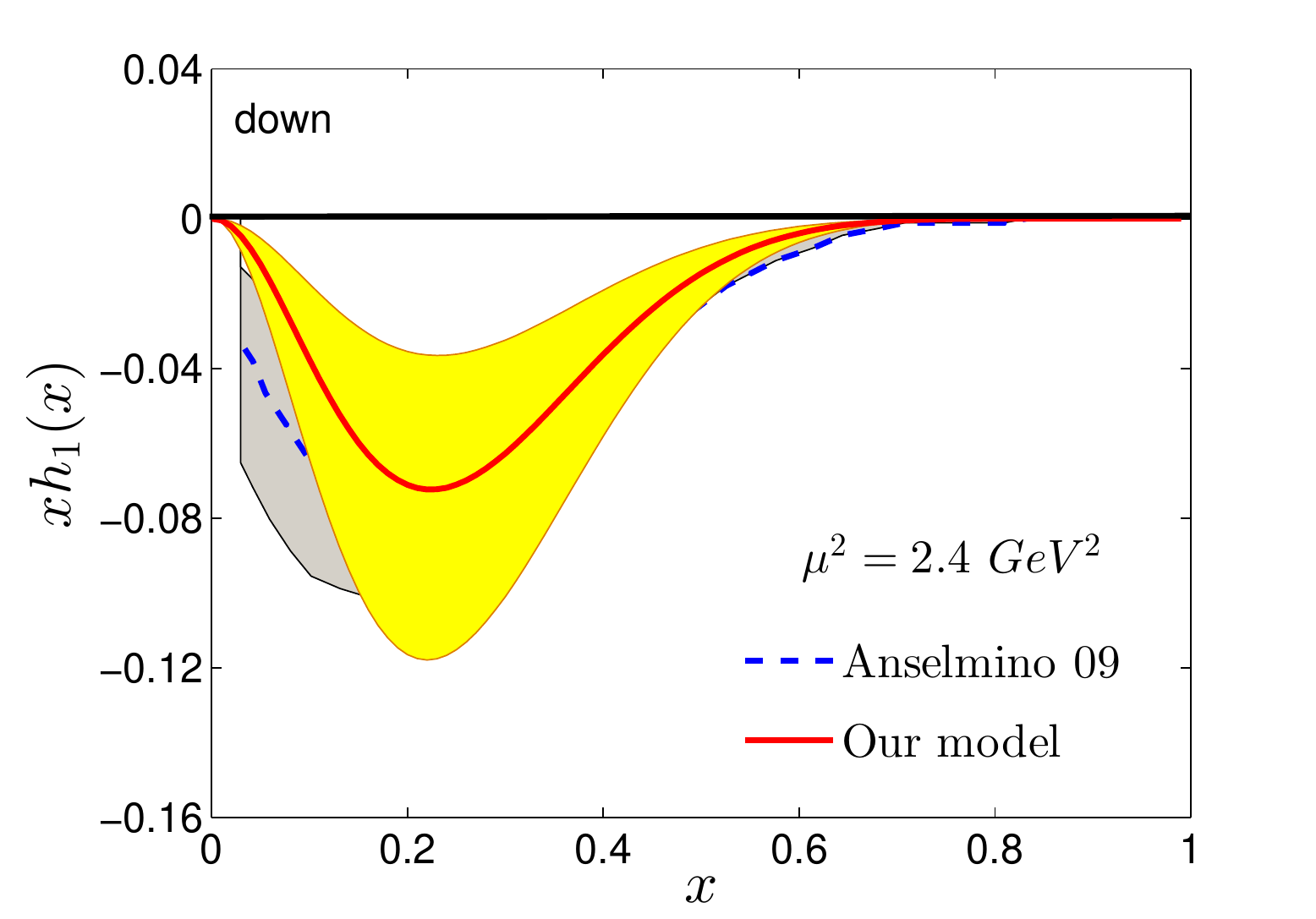}
\end{minipage}
\caption{\label{fig_h1} The transversity distribution at $\mu^2=2.4~GeV^2$ for $u$ quark(a) and $d$ quark(b). Our results are compared with Anselmino et. al.\cite{Anse09}.}
\end{figure}
\begin{figure}[htbp]
\small{(a)}\includegraphics[width=7.5cm,clip]{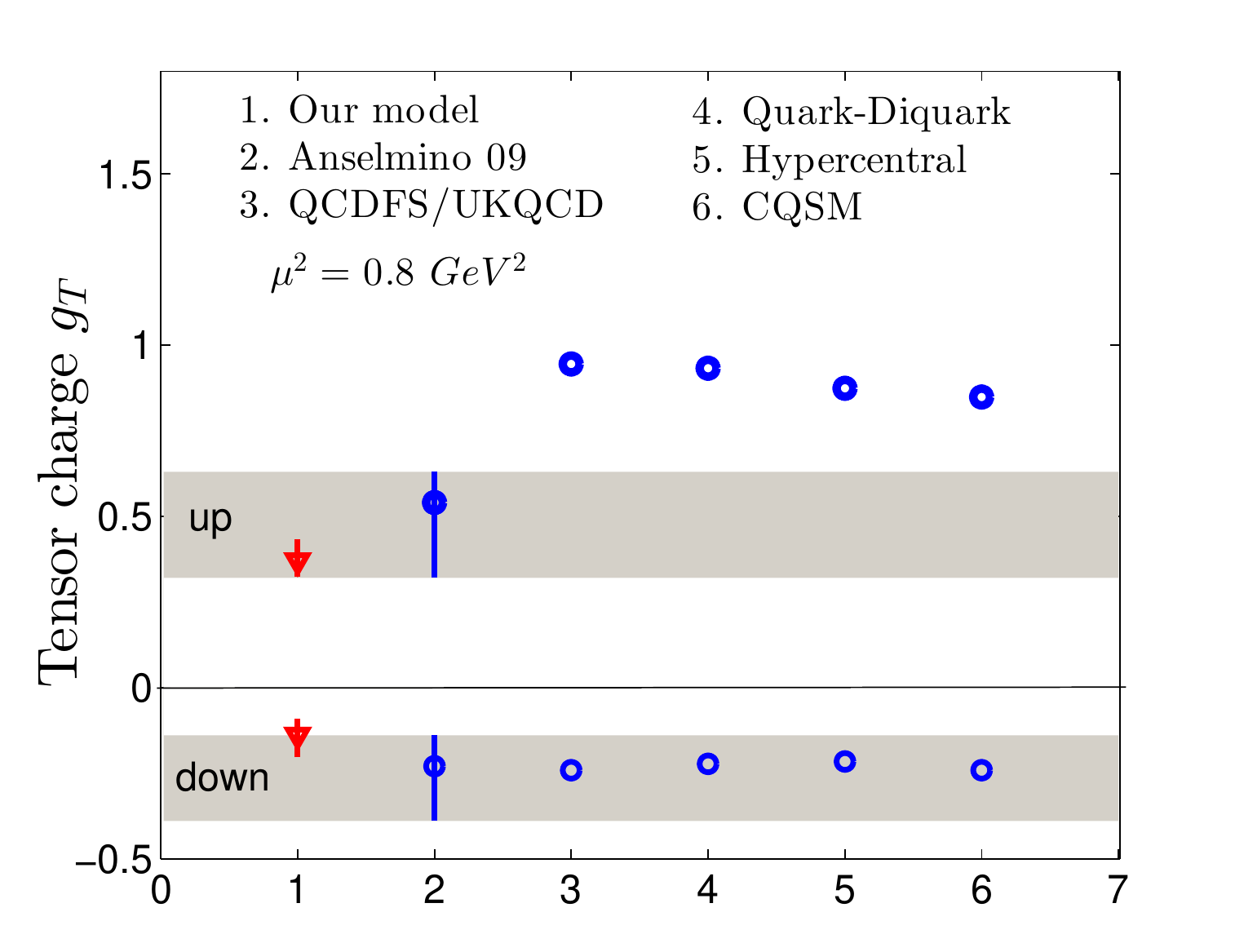}
\small{(b)}\includegraphics[width=7.5cm,clip]{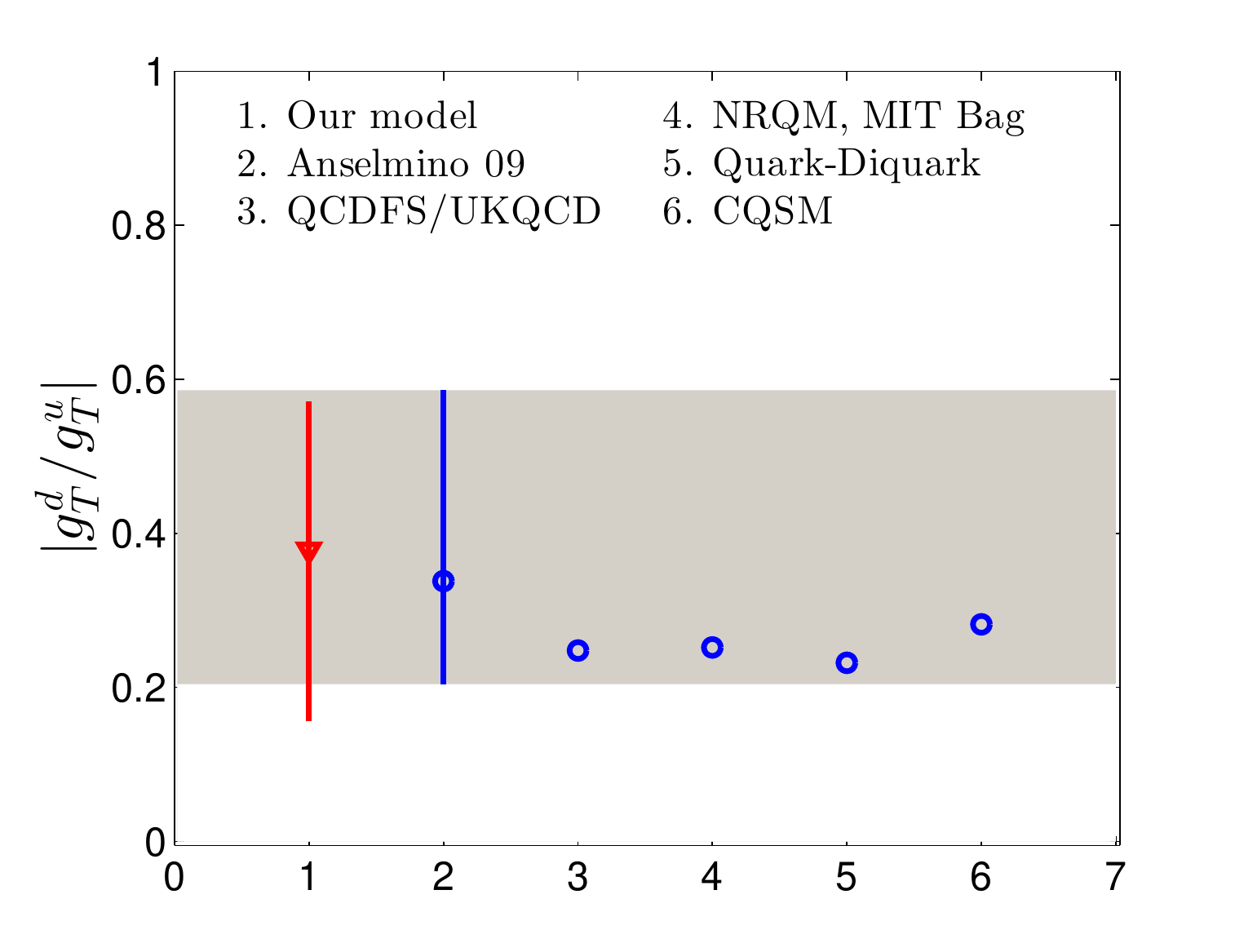}\\
\caption{\label{fig_gT_comp} Comparisons of tensor charges $gT$ with the experimental data fit \cite{Anse09} and other model predictions \cite{Gock05,Cloet08,Pasq05,Waka07} are shown in (a),  for $u$ and $d$ quarks. The shaded regions are the error band for $u$ and $d$ quarks in the experimental extraction of $g_T$\cite{Anse09}.  Our results of tensor charges for both  $u$ and $d$ quarks are shown in red.  The ratio of the tensor charges $|g^d_T/g^u_T|$ is shown in (b). Our results are compared with other models e.g. Lattice QCD\cite{Gock05}, MIT Bag model, Quark-Diquark model\cite{Cloet08}, CQSM\cite{Waka07}. } 
\end{figure}

The transversity distributions in this model reads as
\be 
h^{u}_1(x,\mu)&=&\bigg(C_S^2N^{2}_S(\mu)-C^2_V\frac{1}{3}N^{(u)2}_0(\mu)\bigg)\frac{1}{\delta^u} x^{2a^u_1(\mu)}(1-x)^{2b^u_1+1},\\
h^{d}_1(x,\mu)&=&-C^2_{VV}\frac{1}{3}N^{(d)2}_0(\mu) \frac{1}{\delta^d} x^{2a^d_1(\mu)}(1-x)^{2b^d_1+1}.
\ee
Transversity PDF $h_1(x)$ are shown in Fig.\ref{fig_h1}, at scale $\mu^2=1~GeV^2$, for $u$ and $d$ quarks.   The model predictions are shown to agree with the experimental data\cite{Anse09}. The first moment of the transversity distribution gives the tensor charge $g_T$. The model again predicts the tensor charges quite accurately as shown in
  Table.\ref{tab_gT}. For both $u$ and $d$ quarks, we have $\mid g_T^\nu\mid  < \mid g_A^\nu\mid$.
  In Fig.\ref{fig_gT_comp}(a), we have compared our model predictions of tensor charge for both $u$ and $d$ quarks with other models along with the phenomenological fit of experimental data\cite{Anse09}.  Similar  comparisons are studies by Anselmino et. al.\cite{Anse09} and Wakamatsu\cite{Waka09}. Our predictions fall within the uncertainty bands of the phenomenological fits for both $u$ and $d$ quarks.  The ratio of the two tensor charges $\mid {g_T^d}/ {g_T^u}\mid $ is totally scale independent and a better quantity to compare with other models.  Fig.\ref{fig_gT_comp}(b),  shows the comparison of that ratio with other model predictions. Our model predicts $\mid {g_T^d}/ {g_T^u}\mid =0.38$ which is very close to the phenomenological prediction.

\begin{table}[ht]
\centering 
\begin{tabular}{|c|c|c|c|c|c|c|}
 \hline
 ~~ ~~&~~ $g^u_T$~~&~~ $g^d_T$~~&~~$g_T$ \\ \hline
 ~~ Our result ~~&~~ $0.37^{+0.06}_{-0.05}$ ~~&~~ $-0.14^{+0.05}_{-0.06}$~~&~~ $0.51^{+0.12}_{-0.11}$ ~~ \\ 
 ~~ Measured Data\cite{Anse07}   ~~&~~ $0.59^{+0.14}_{-0.13}$ ~~&~~ $-0.20^{+0.05}_{-0.07}$~~&~~ $0.79^{+0.19}_{-0.20}$ ~~\\
\hline
 \end{tabular}
\caption{Tensor charge at the scale $\mu^2=0.8~ GeV^2$. Our results are compared with measured data\cite{Anse07}.} 
\label{tab_gT} 
\end{table} 

\begin{figure}[htbp]
\begin{minipage}[c]{0.98\textwidth}
\small{(a)}\includegraphics[width=7.5cm,clip]{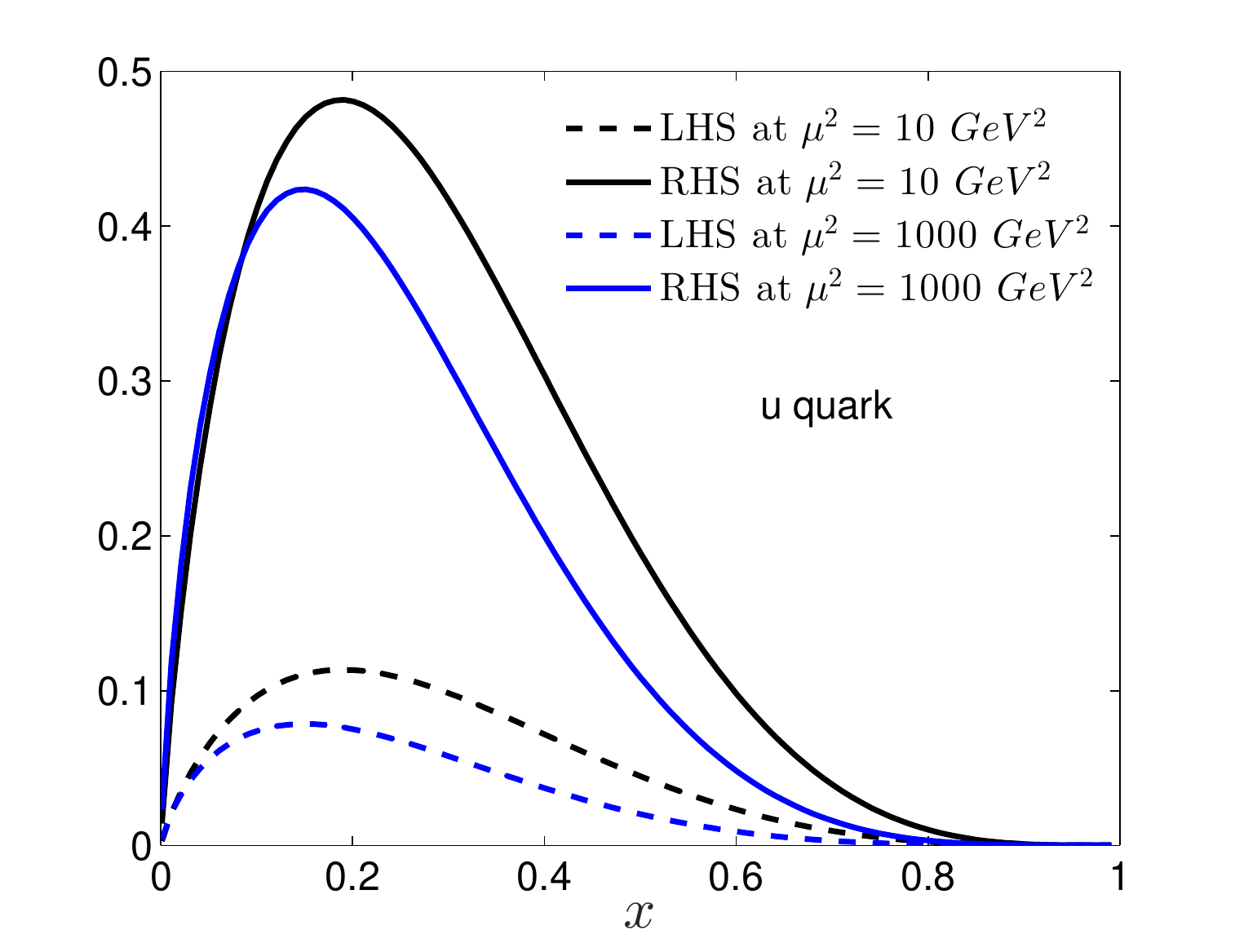}
\small{(b)}\includegraphics[width=7.5cm,clip]{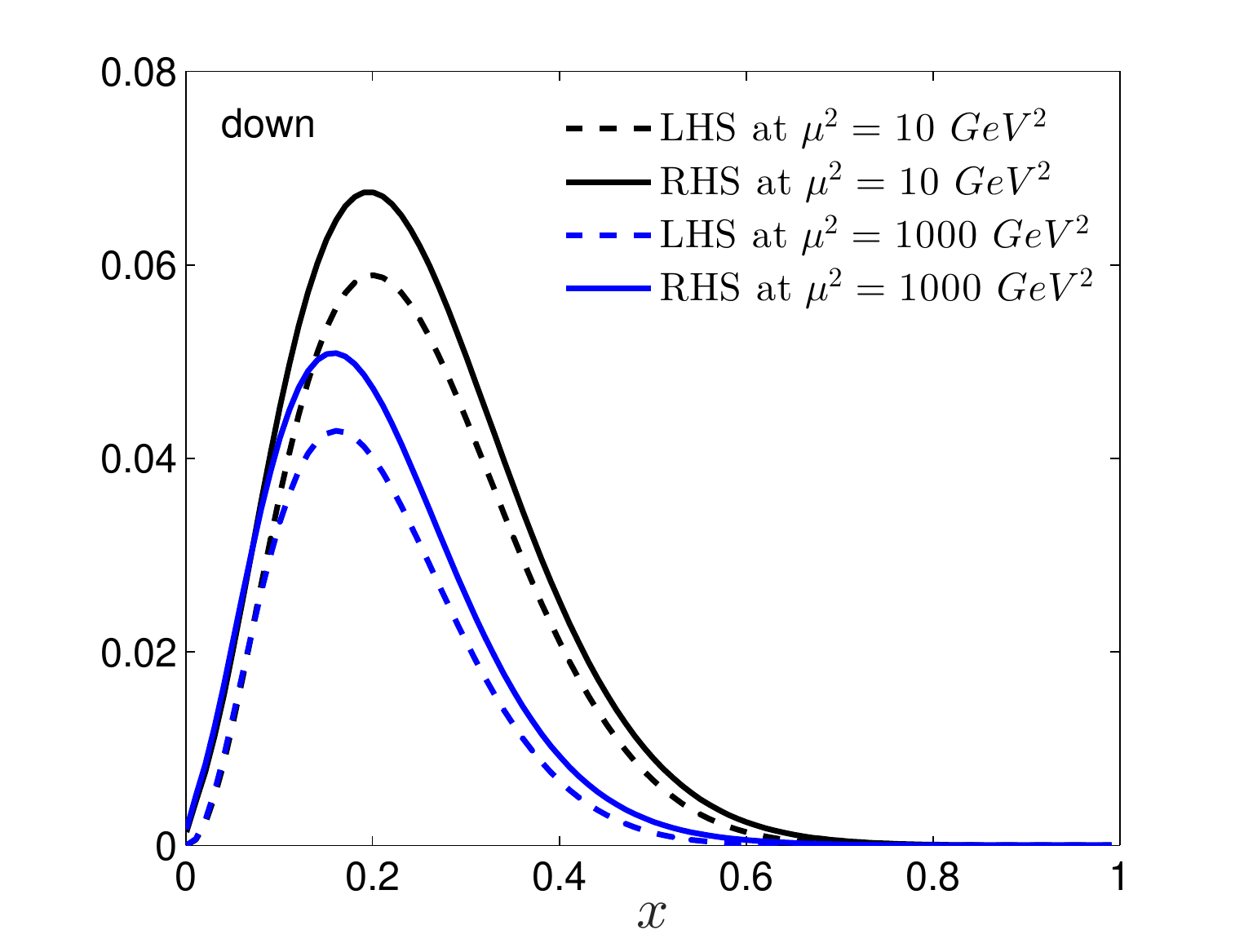}
\end{minipage}
\caption{\label{fig_soff} The Soffer bound at  $\mu^2=10$ GeV$^2$, and $1000$ GeV$^2$ for  (a) $u$  quark and (b) $d$ quark. Where the LHS and RHS stand for the left hand side and right hand side of Eq.(\ref{soff}) multiplied by $x$. }
\end{figure}

This model also satisfies the  Soffer bound which at an arbitrary scale $\mu$ is defined as \cite{Soff95}
\be 
|h^\nu_1(x,\mu)|\leq \frac{1}{2}\bigg[ f^\nu_1(x,\mu)+g^\nu_1(x,\mu) \bigg].\label{soff}
\ee
In Fig.\ref{fig_soff} we show the left (LHS) and right hand sides (RHS) of the the above equation multiplied by $x$ for both $\nu=u,d$ and at  both low and high  scales.
According to Soffer bound, LHS should always lie below RHS which can be easily seen in Fig.\ref{fig_soff}.

\begin{table}[ht]
\centering  
\begin{tabular}{|c | c c c c c c|}
 \hline
 $\mu^2~(GeV^2)$~~&~~$0.2$~~&~~0.6~~ &~~1.0 ~~&~~7.0~~&~~10~~&~~20~~  \\ \hline
$P_q$ ~~&~~0.60~~&~~0.48~~&~~0.45~~&~~0.38[0.55]~~&~~0.37~~&~~0.35\\ \hline
\end{tabular} 
\caption{Proton momentum fraction carried by valence quarks $P_q$ with the scale $\mu$. The value given within square bracket is the ZEUS result \cite{ZEUS03} at the scale $\mu^2=7~GeV^2$. }  
\label{tab_Pq}  
\end{table}
Proton momentum fraction carried by valence quarks can be estimated from the unpolarized PDFs at different scale $\mu$  
\be
P_q(\mu)&=&\int dx x[C_S^2 f^{(s)}_1(x) + C_V^2 f^{(V)}_1(x) + C_{VV}^2 f^{(VV)}_1(x)],\\ \nonumber
&=&\int dx x[f^{(u)}_1(x,\mu) + f^{(d)}_1(x,\mu)].
\ee
The values of $P_q$ with the scale are shown in Table.\ref{tab_Pq}. The  momentum carried by the valence quarks decreases as the value of $\mu^2$ increases.


\section{Summary and conclusion}\label{concl}
Light front Ads/QCD has predicted many interesting nucleon properties. Light front AdS/QCD predicts a particular form of wave function for a two body bound state\cite{BT}. In this paper, we have developed a quark-diquark model for proton  where the light front wave functions are constructed from the AdS/QCD predictions. The model is consistent with the quark counting rule and Drell-Yan-West relation.
 The model has $SU(4)$ spin-flavor structure and includes the contributions from scalar($S=0$) and axial vector($S=1$)  diquarks.    The evolution of  $f_1(x)$ is simulated by introducing scale dependence parameters in the PDF.    The scale evolutions of the parameters are  determined by satisfying the PDF evolution in the range $\mu^2=0.09 GeV^2$ to $150 GeV^2$.  We have given the explicit  scale evolution of each parameter in the model, so  the distributions can be calculated at any arbitrary scale. Though PDF data  up to $\mu^2=150$ GeV$^2$ are used to determine the scale evolution,   we have shown that our model can  accurately predict the PDF evolution up to a very high scale ($\mu^2=10^4~ GeV^2$).  The helicity and transversity PDFs are calculated as predictions of the model and are shown to satisfy Soffer bound and  have good agreement with the available data. Our model reproduces the experimental values of axial and tensor charges quite well. It will be interesting to study the other proton properties like GPDs, TMDs, Wigner distributions and GTMDs etc  and their scale evolutions in this model and compare with  other model predictions.

 {\bf Acknowledgements:} DC thanks Stan Brodsky, Guy de Teramond  for  many insightful discussions. We also  thank Chandan Mondal for many useful discussions.

\appendix
\section{Parameter fitting for PDF evolution\label{appa}}
\begin{figure}[htbp]
\begin{minipage}[c]{0.98\textwidth}
\small{(a)}\includegraphics[width=6.5cm,clip]{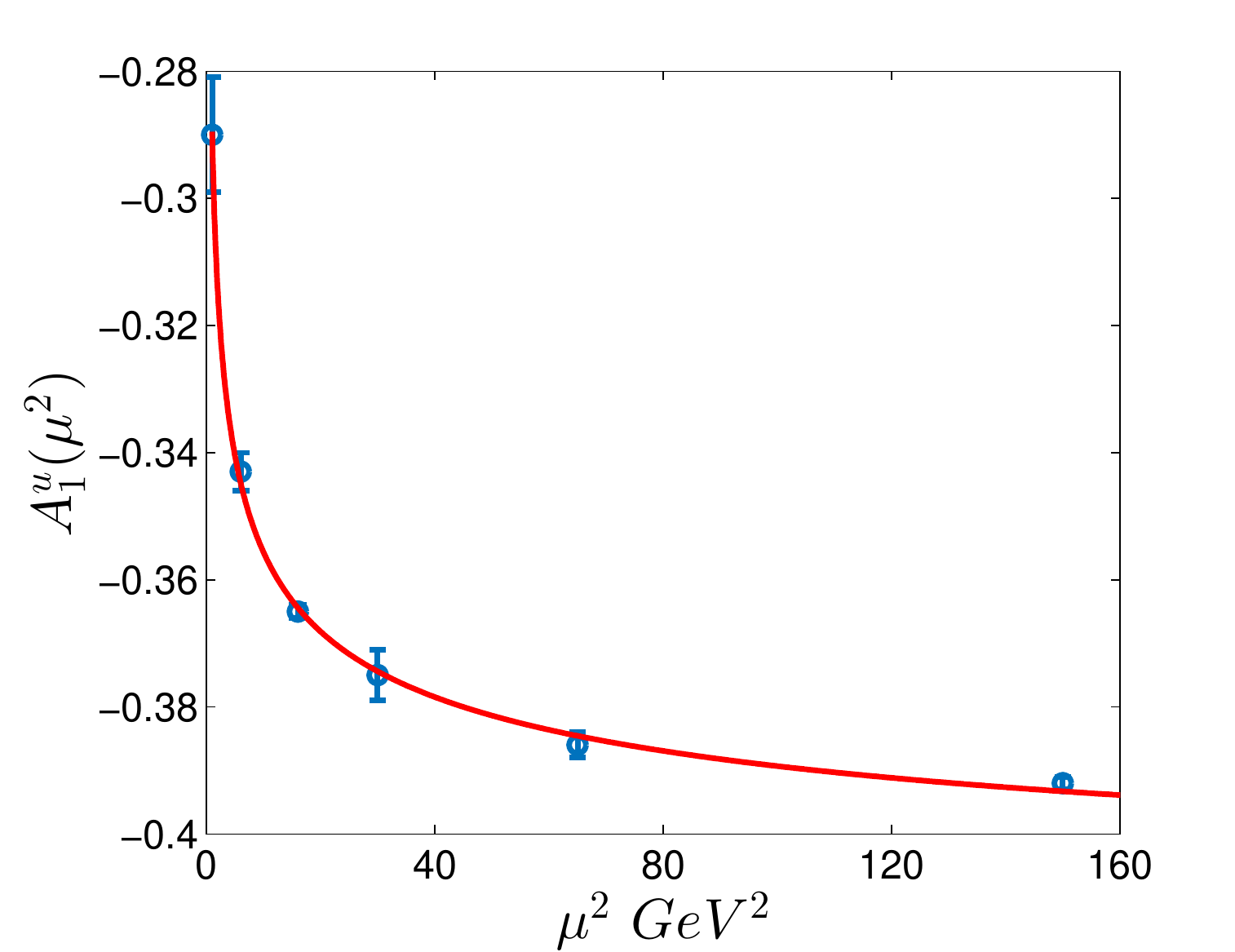}
\small{(b)}\includegraphics[width=6.5cm,clip]{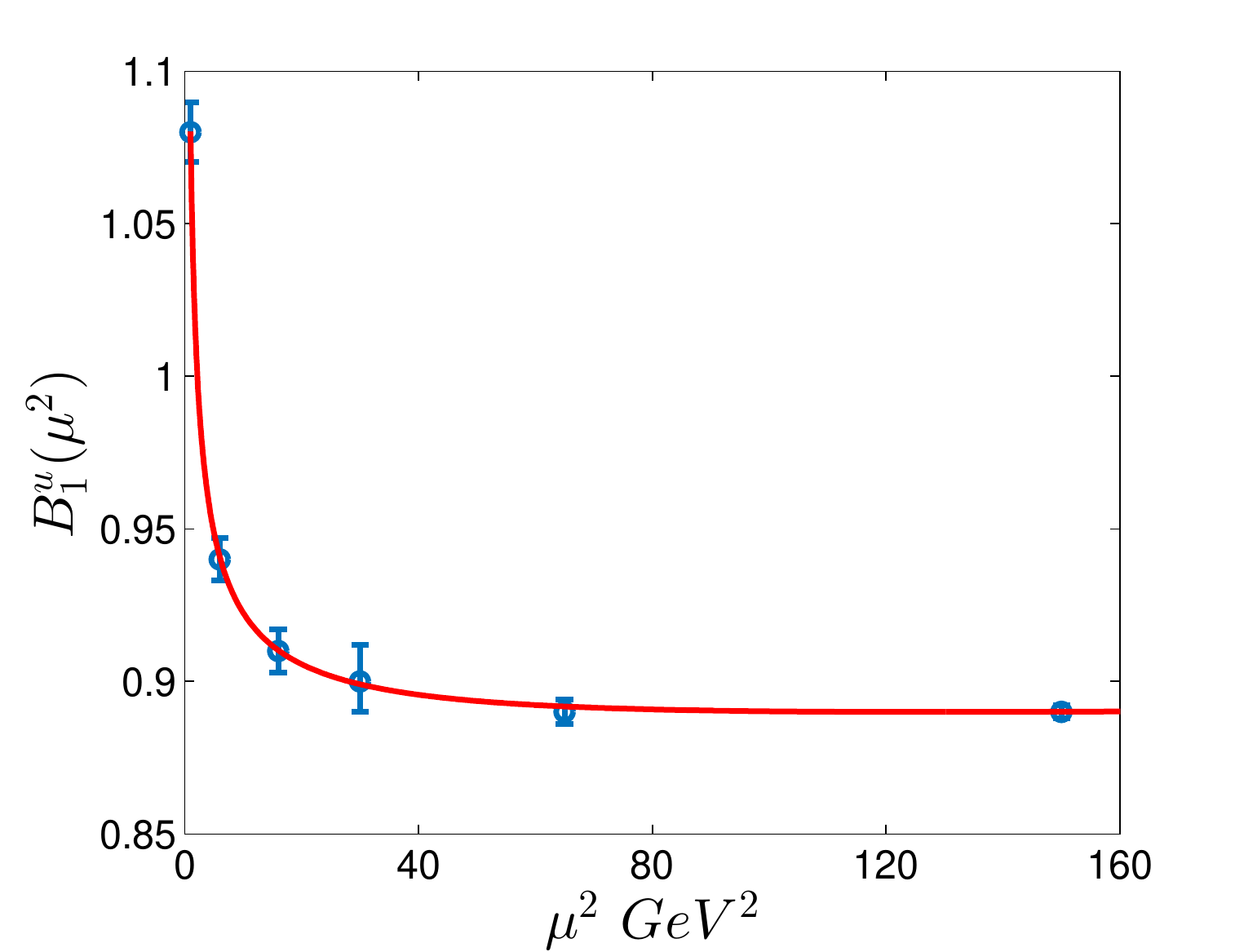}
\end{minipage}
\begin{minipage}[c]{0.98\textwidth}
\small{(c)}\includegraphics[width=6.5cm,clip]{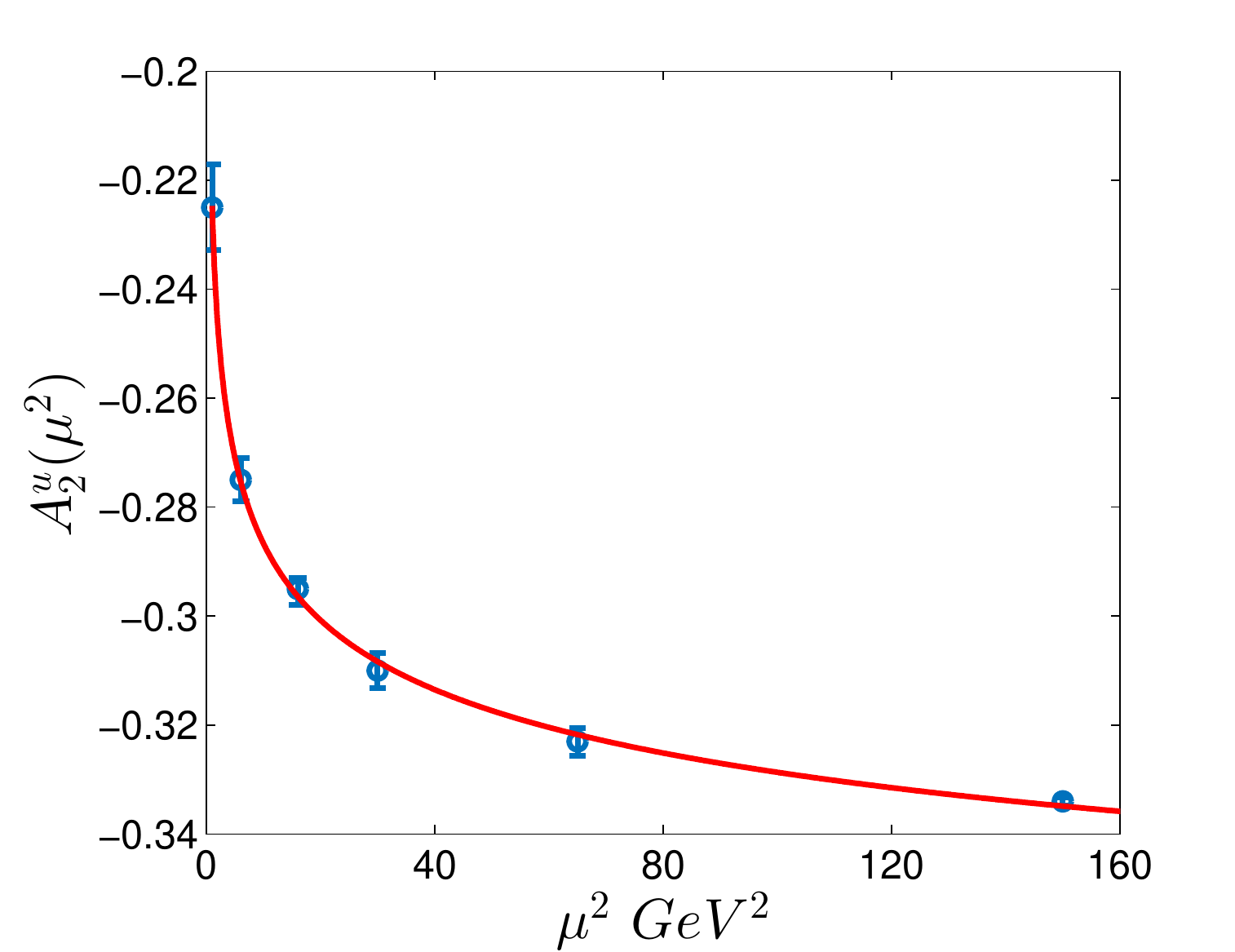}
\small{(d)}\includegraphics[width=6.5cm,clip]{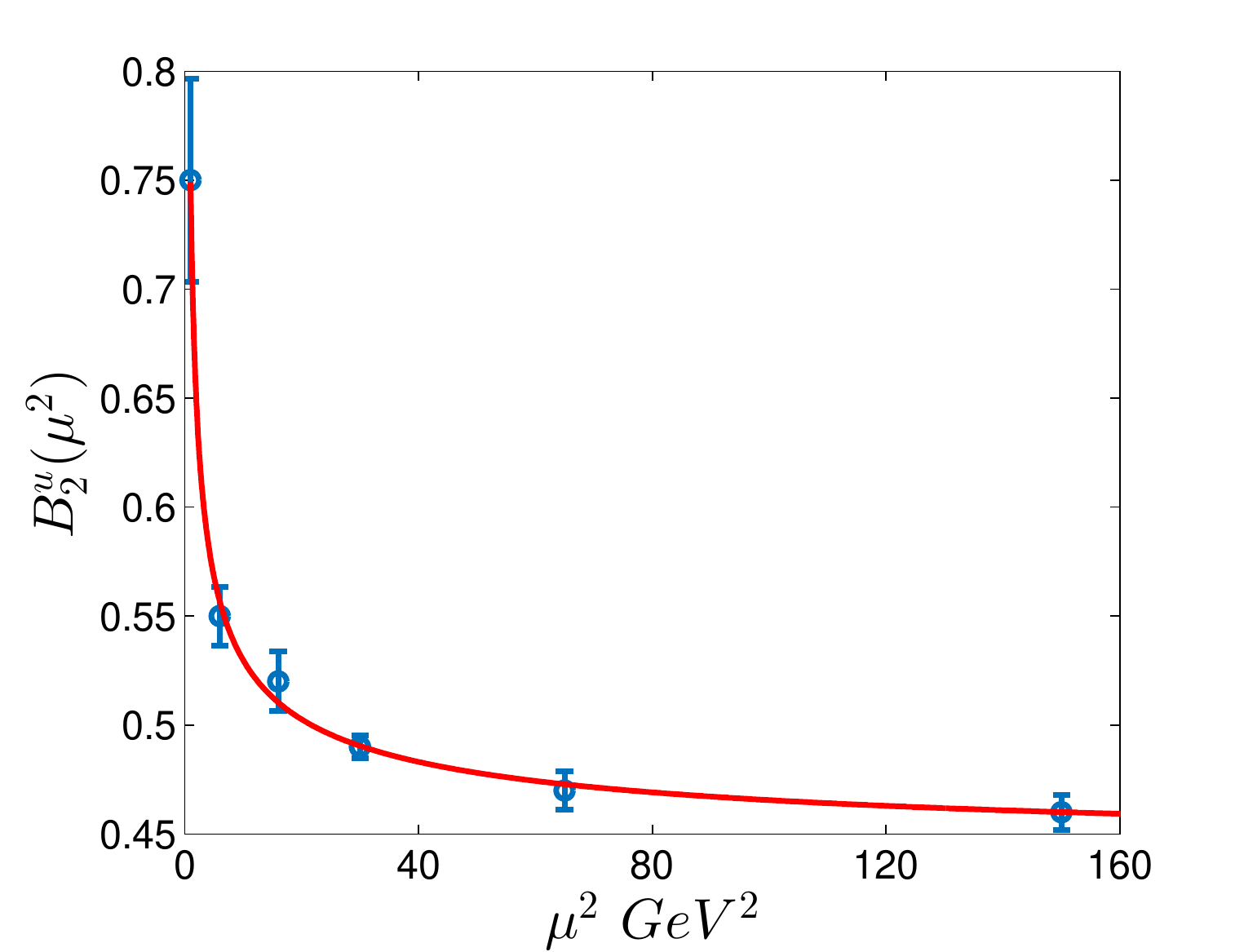}
\end{minipage}
\begin{minipage}[c]{0.98\textwidth}
\small{(e)}\includegraphics[width=6.5cm,clip]{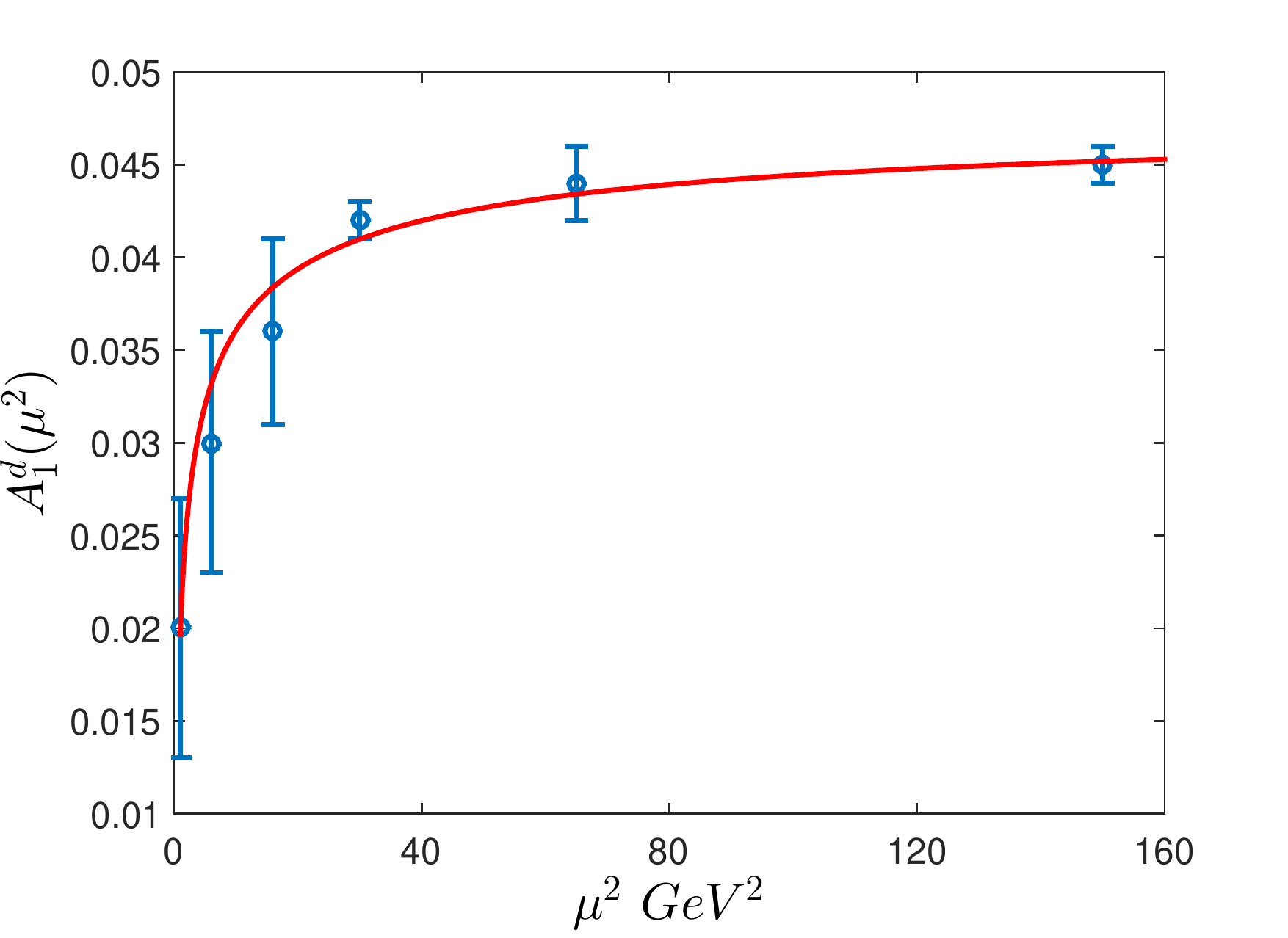}
\small{(f)}\includegraphics[width=6.5cm,clip]{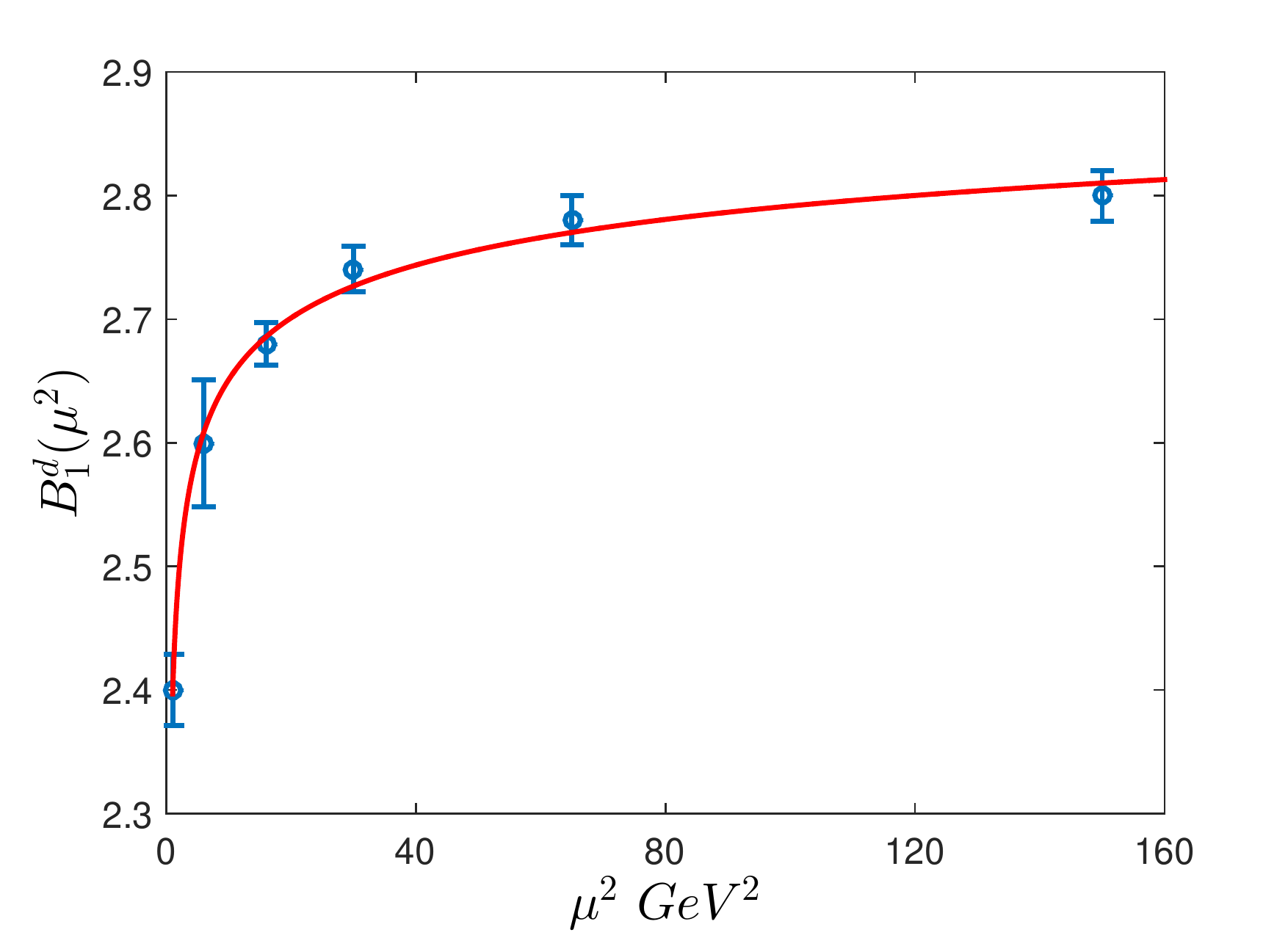}
\end{minipage}
\begin{minipage}[c]{0.98\textwidth}
\small{(g)}\includegraphics[width=6.5cm,clip]{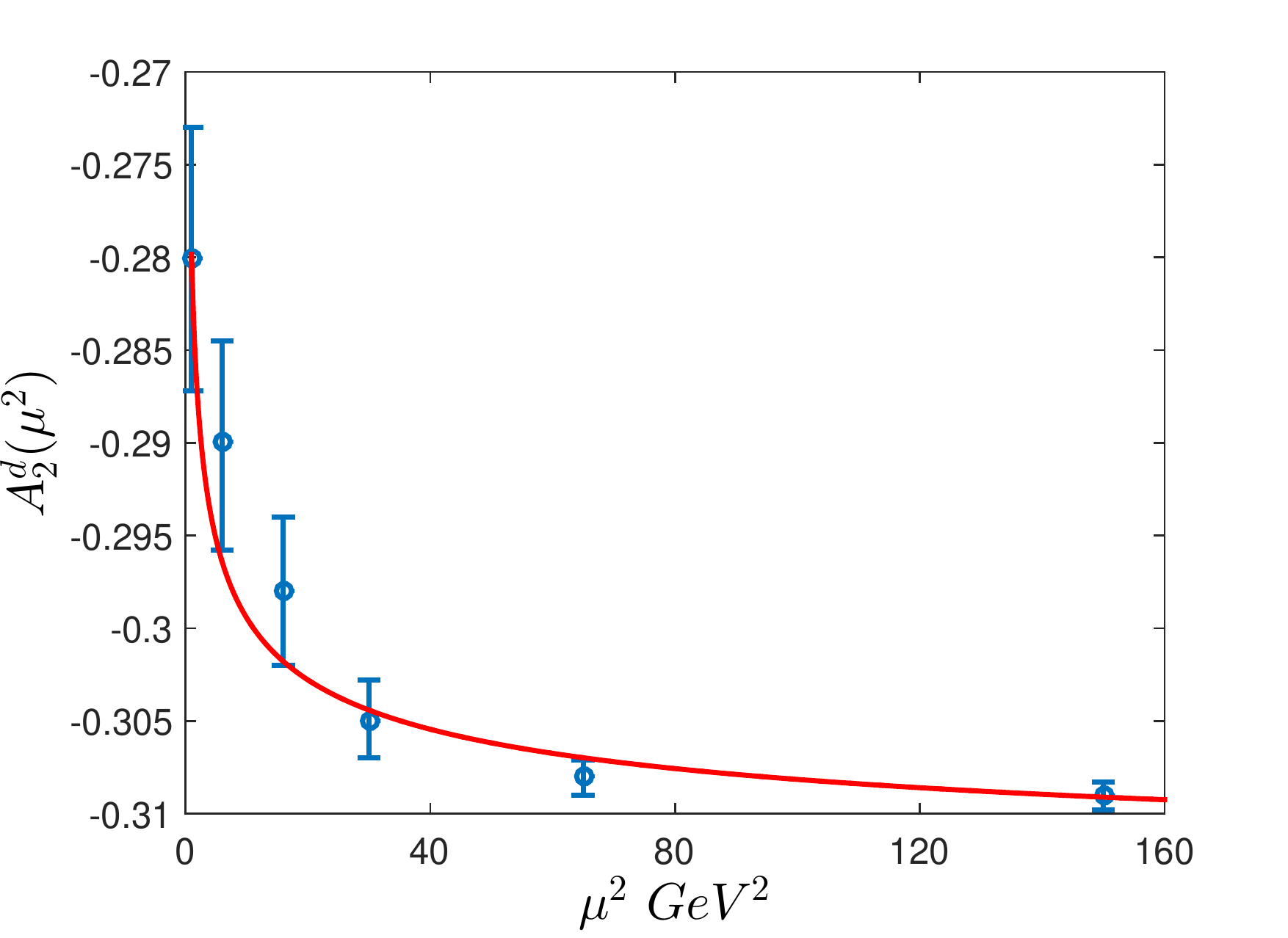}
\small{(h)}\includegraphics[width=6.5cm,clip]{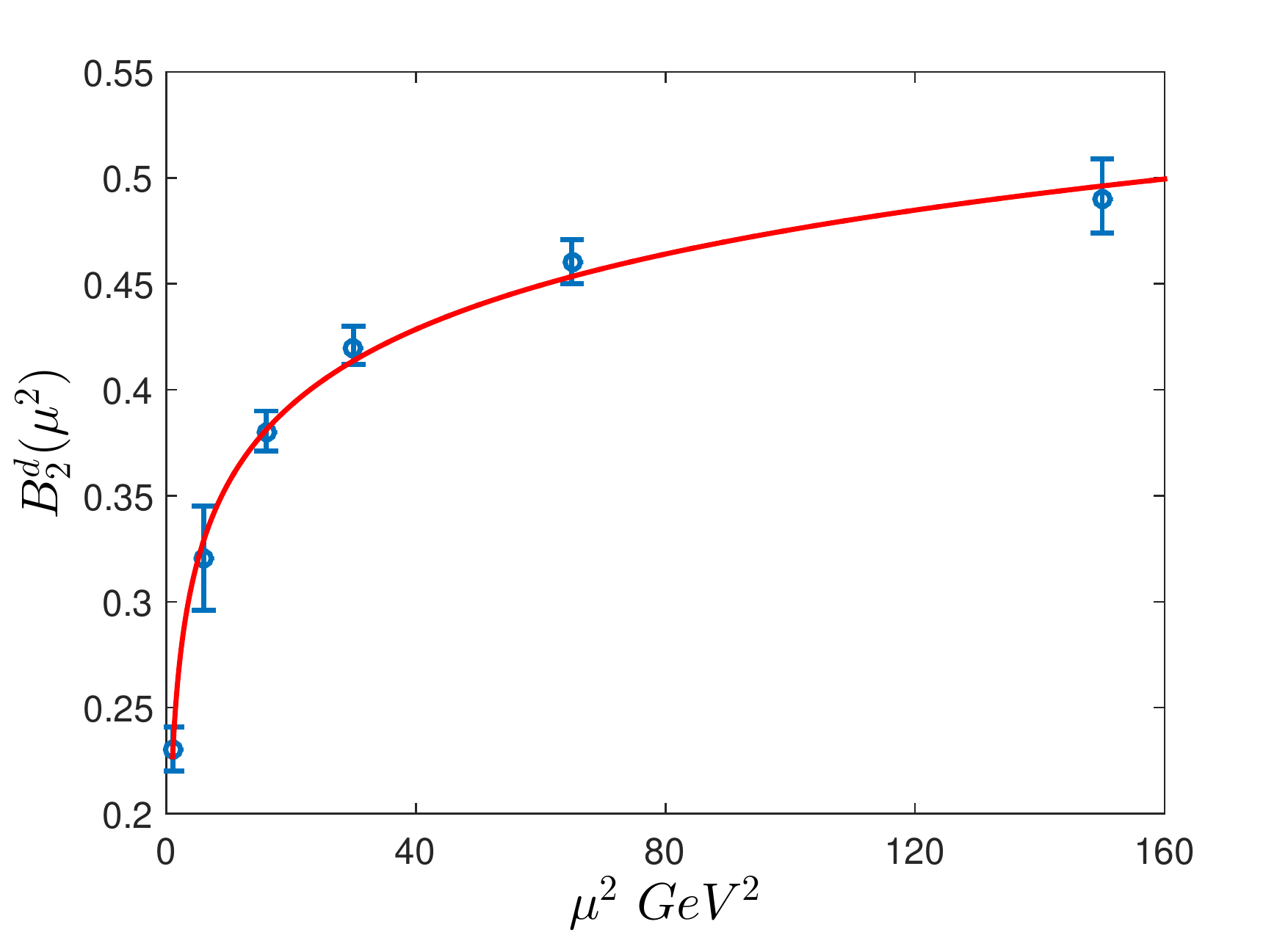}
\end{minipage}
\caption{\label{fig_ABud} Using Eq.(\ref{Pi_evolu}), the data of Table.\ref{tab_ABu} are fitted by varying evolution parameters $\alpha^\nu_{P,i},\beta^\nu_{P,i} ~and~ \gamma^\nu_{P,i}$, for $u$ quark ($(a) - (d)$). Similar data fitting plots for $d$ quark(Table.\ref{tab_ABd}) are shown 
in $(e) - (h)$.} 
\end{figure}

\begin{figure}[htbp]
\begin{minipage}[c]{0.98\textwidth}
\small{(a)}\includegraphics[height=5.0cm,clip]{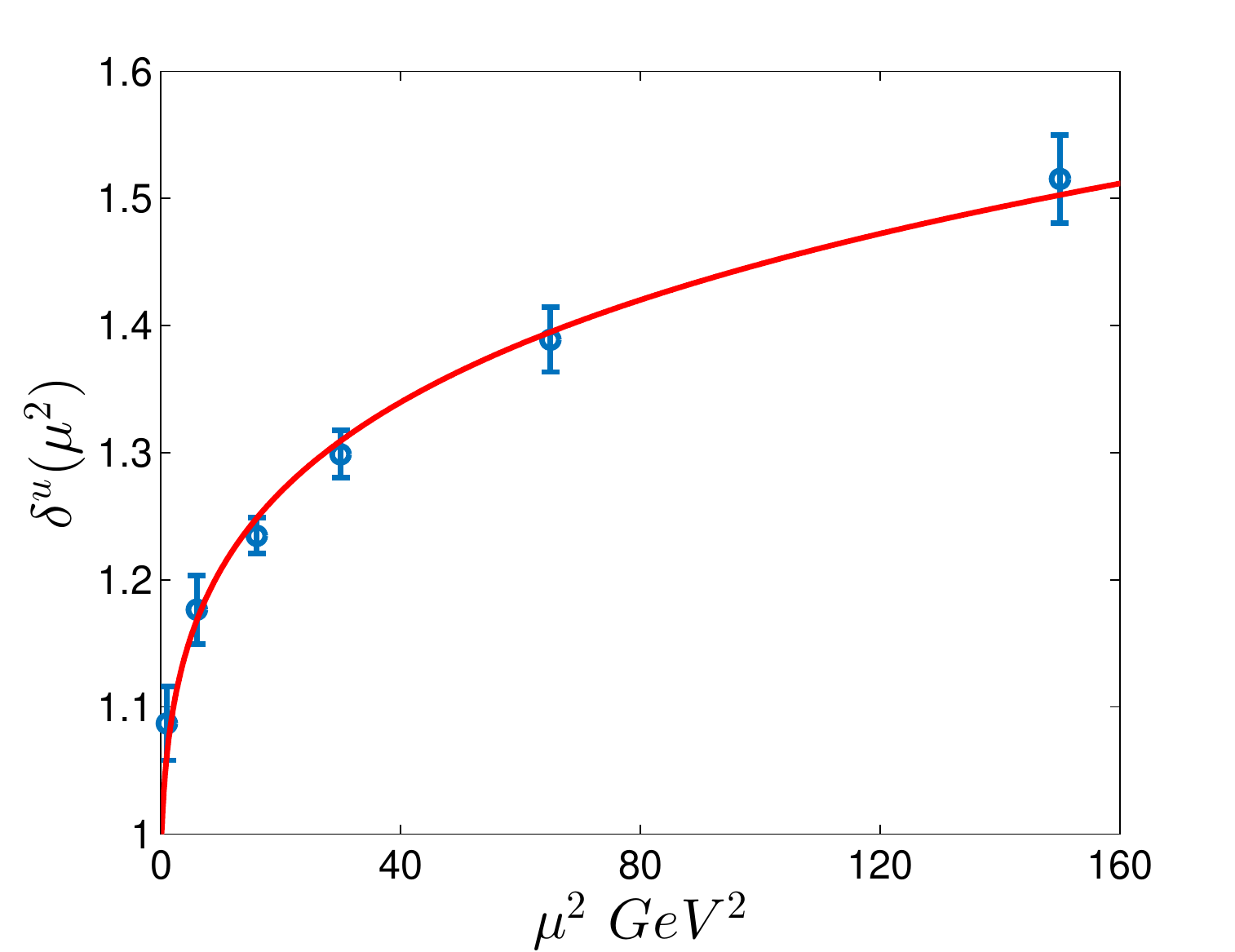}
\small{(b)}\includegraphics[height=5.0cm,clip]{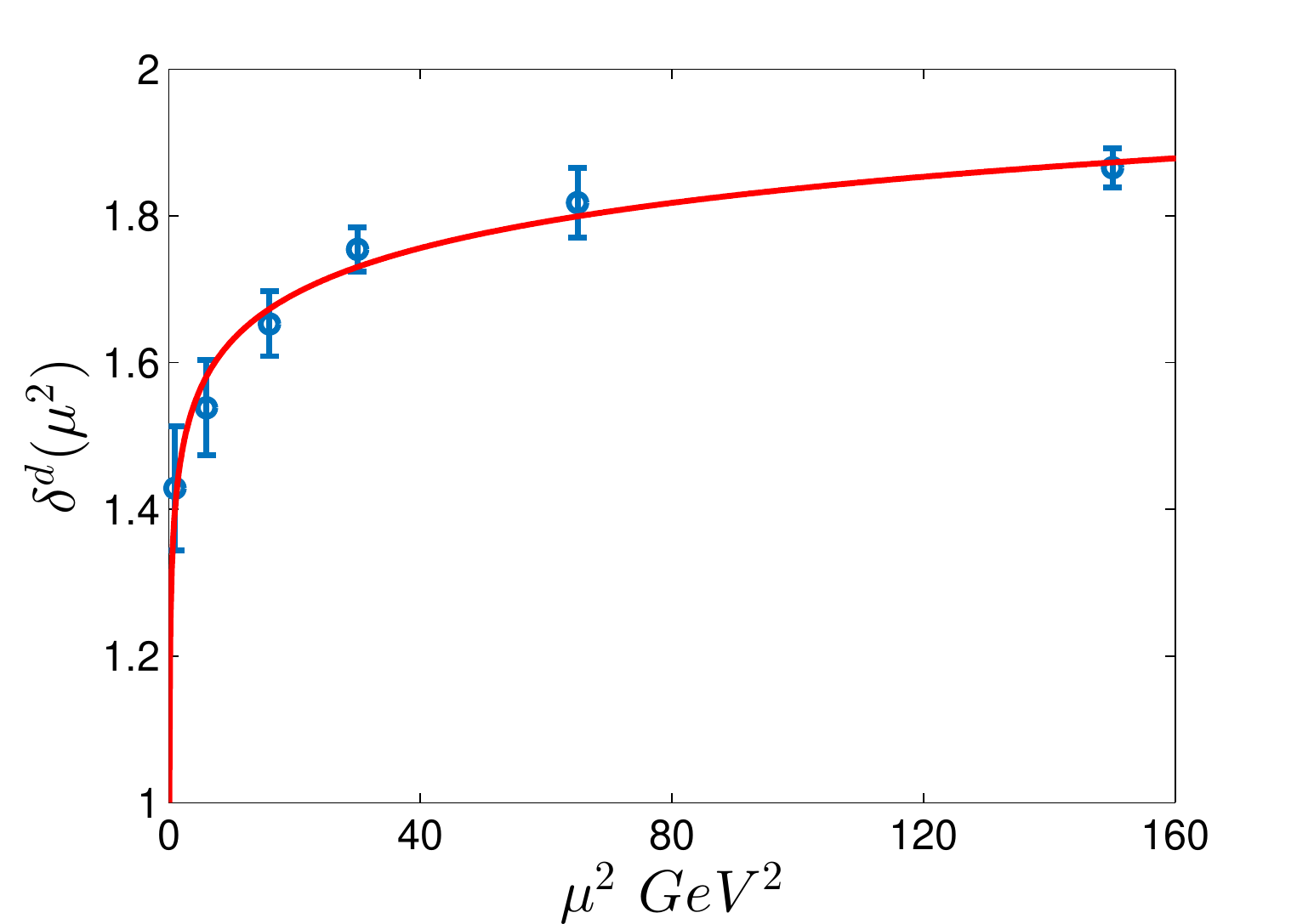}
\end{minipage}
\caption{\label{fig_del}  The data of Table.\ref{tab_DL} are fitted by varying evolution parameters $\delta^\nu_1$ and $\delta^\nu_2$, for (a)  $u$  quark and (b)  $d$ quark.}
\end{figure}

Scale evolutions of  $A_i^\nu$ and $B_i^\nu$ are parameterized by the parameters $\alpha_i^\nu,~\beta_i^\nu$, and $\gamma_i^\nu$ and that of $\delta^\nu$ is parameterized by $\delta_1^\nu$ and $\delta_2^\nu$.  $f_1(x,\mu)$ is given by Eq.(\ref{Eq_xf1mu}) along with Eqs.(\ref{a_im},\ref{b_im},\ref{DL}) and Eq.(\ref{Pi_evolu}). Here we list the parameters$A_i^\nu,~ B_i^\nu$ and $\delta^\nu$ fitted at different scales $\mu^2$ in Table \ref{tab_ABu} and Table \ref{tab_ABd}.   The last column in each  table indicates the least $\chi^2$  error in the PDF estimation. Each  $\chi^2/d.o.f$ has been evaluated  from 100 data points for different $x~(0<x<1)$) i.e., for the five parameter fit, we have  (100-5)=95 degrees of freedom. The fitting of the parameters at $\mu^2=1,6,16,30,65$ and $150$ GeV$^2$ are shown in Fig.\ref{fig_ABud} and Fig.\ref{fig_del}. The data points are extracted from the PDF data. 
The error bars shown in the plots are the errors in  the extracted values of the  parameters  due to the uncertainties in the PDF data.
\begin{table}[ht]
\centering 
\begin{tabular}{|c | c c c c c c |}
 \hline
 $\mu^2~GeV^2$~~&~~$A_1^u$~~ & ~~$B_1^u$~~ & ~~$A_2^u$~~ & ~~$B_2^u$~~ & ~~$\delta^u$~~ &~~$\chi^2/d.o.f$~~ \\ \hline
1&~$-0.29\pm0.009$~&~$1.08\pm0.009$~&~$-0.225\pm0.008$~&~$0.75^{+0.047}_{-0.046}$~&$1.087\pm0.029$&~1.41\\
6&$-0.343\pm0.003$&$0.94\pm0.007$&$-0.275\pm0.004$&$0.55\pm0.013$&$1.176\pm0.027$&4.8 \\
16&$-0.365\pm0.001$&$0.91\pm0.007$&$-0.295^{+0.003}_{-0.002}$&$0.52^{+0.013}_{-0.014}$&$1.234\pm0.014$&1.8\\
30&$-0.375\pm0.004$&$0.9^{+0.010}_{-0.012}$&$-0.31^{+0.003}_{-0.002}$&$0.49\pm0.005$&$1.298\pm0.018$&1.2 \\
65&$-0.386\pm0.002$&$0.89\pm0.004$&$-0.323^{+0.002}_{-0.003}$&$0.47\pm 0.009$&$1.389\pm0.025$&0.54\\
150&$-0.392\pm0.001$&$0.89\pm0.002$&$-0.334^{+0.0008}_{-0.001}$&$0.46\pm0.008$&$1.515\pm0.034$&0.29\\ \hline
 \end{tabular} 
\caption{Fitting of the PDF $f_1(x)$ at various scales for $u$ quark.  } 
\label{tab_ABu} 
\end{table} 
\begin{table}[ht]
\centering 
\begin{tabular}{|c | c c c c c c |}
 \hline
 $\mu^2~GeV^2$~~&~~$A_1^d$~~ & ~~$B_1^d$~~ & ~~$A_2^d$~~ & ~~$B_2^d$~~ & ~~$\delta^d$~~&~~$\chi^2/d.o.f$ \\ \hline
 1&~$0.02\pm0.007$~&~$2.4^{+0.029}_{-0.029}$~&~$-0.28^{+0.012}_{-0.017}$~&~$0.23^{+0.010}_{-0.011}$~&$1.43\pm0.085$&~0.21\\
 6&$0.03\pm0.007$&$2.6^{+0.052}_{-0.051}$&$-0.29^{+0.006}_{-0.005}$&$0.32^{+0.024}_{-0.025}$&$1.54\pm0.065$&0.38\\
 16&$0.036\pm0.005$&$2.68^{+0.017}_{-0.017}$&$-0.298\pm0.001$&$0.38^{+0.009}_{-0.010}$&$1.65\pm0.044$&0.40\\
 30&$0.042\pm0.001$&$2.74^{+0.018}_{-0.019}$&$-0.305\pm0.002$&$0.42^{+0.008}_{-0.010}$&$1.75\pm0.030$&0.49\\
 65&$0.044\pm0.002$&$2.78^{+0.020}_{-0.020}$&$-0.308\pm0.001$&$0.46\pm0.010$&$1.82\pm0.048$&0.59\\
 150&$0.045\pm0.001$&$2.8^{+0.021}_{-0.020}$&$-0.309\pm0.0008$&$0.49^{+0.016}_{-0.019}$&$1.86\pm0.036$&0.77\\ \hline
 \end{tabular}
\caption{ Fitting of the PDF $f_1(x)$ at various scales for $d$ quark. \label{tab_ABd}.} 
\end{table}


\vskip2in

\end{document}